\documentclass[12pt]{article}

\usepackage{amsmath}
\usepackage{graphicx}
\usepackage{amsfonts}
\usepackage{amssymb}
\usepackage{epsfig}
\usepackage{color}
\usepackage{psfrag}

\newcommand{\EQ}{\begin{equation}}
\newcommand{\EN}{\end{equation}}

\newcommand{\hs}{\hspace{1mm}}

\begin{document}
\setcounter{page}{0} \topmargin 0pt \oddsidemargin 5mm
\renewcommand{\thefootnote}{\arabic{footnote}}
\newpage
\setcounter{page}{0}

\begin{titlepage}

\begin{flushright}
\end{flushright}
\vspace{0.5cm}
\begin{center}
{\Large {\bf Neutral Bound States in Kink-like Theories}}\\

\vspace{2cm}
{\large G. Mussardo$^{a,b,1}$
\footnotetext[1]{On leave of absence from International School for Advanced Studies, Trieste (Italy)}}
\\
\vspace{0.5cm} $^a${\em Ecole Normale Superieure de Lyon}\\
{\em Laboratoire de Physique} \\
\vspace{0.3cm} $^b${\em Istituto Nazionale di Fisica Nucleare}\\
{\em Sezione di Trieste}

\end{center}
\vspace{1cm}

\begin{abstract}
\noindent
In this paper we present an elementary derivation of the semi-classical spectrum 
of neutral particles in a field theory with kink excitations. In the non-integrable cases, 
we show that each vacuum state cannot generically support more than two stable 
particles, since all other neutral exitations are resonances, which will eventually decay. 
A phase space estimate of these decay rates is also given. This shows that there may be 
a window of values of the coupling constant where a particle with higher mass is more 
stable than the one with lower mass. We also discuss the crossing symmetry properties 
of the semiclassical form factors and the possibility of extracting the elastic part of 
the kink $S$-matrix below their inelastic threshold. We present the analysis of theories 
with symmetric and asymmetric wells, as well as of those with symmetric or asymmetric kinks. 
Illustrative examples of such theories are provided, among others, by the Tricritical Ising Ising,  
the Double Sine Gordon model and by a class of potentials recently introduced by Bazeira et al.

\end{abstract}

\end{titlepage}

\newpage

\section{Introduction}\label{intro}

The aim of this paper is to explain in the easiest possible way the 
presence of neutral bound states in two dimensional field theories with 
kink topological excitations. The theories that we will consider are those 
described by a scalar real field $\varphi(x)$, with a Lagrangian density 
\EQ
{\cal L} \,=\,\frac{1}{2} (\partial_{\mu} \varphi)^2 - U(\varphi) \,\,\,, 
\label{Lagrangian}
\EN 
where the potential $U(\varphi)$ possesses several degenerate minima at 
$\varphi_a^{(0)}$ ($a =1,2,\ldots,n$), as the one shown in Figure 1. These 
minima correspond to the different vacua $\mid \,a\,\rangle$ of 
the associate quantum field theory. 

\vspace{5mm}

\begin{figure}[ht]
\hspace{25mm}
\vspace{10mm}
\psfig{figure=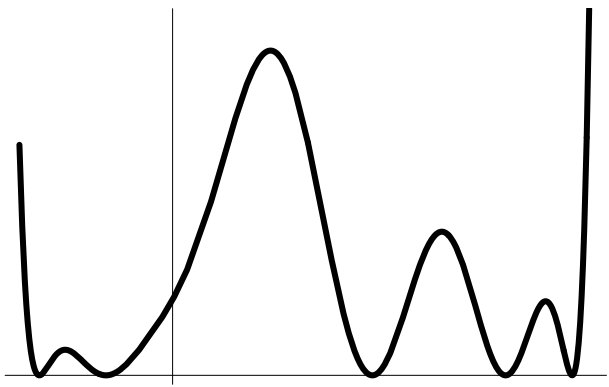,height=5cm,width=8cm}
\vspace{1mm}
\caption{{\em Potential $U(\varphi)$ of a quantum field theory with kink excitations.}}
\label{potential}
\end{figure}

\noindent
The basic excitations of this kind of models are kinks and anti-kinks, i.e. topological configurations 
which interpolate between two neighbouring vacua. Semiclassically they correspond to the static 
solutions of the equation of motion, i.e. 
\EQ
\partial^2_x \,\varphi(x) \,=\,U'[\varphi(x)] \,\,\,, 
\label{static}
\EN
with boundary conditions $\varphi(-\infty) = \varphi_a^{(0)}$ and $\varphi(+\infty)= \varphi_{b}^{(0)}$, 
where $b = a \pm 1$. This equation can be equivalently expressed in terms of a first order differential 
equation  
\EQ
\frac{d\varphi}{d x} \,=\,\pm \sqrt{2 U(\varphi)} \,\,\,. 
\label{kinkequation}
\EN 
Denoting by $\varphi_{ab}(x)$ the solutions of this equation, their classical energy density 
is given by 
\EQ
\epsilon_{ab}(x) \,=\,\frac{1}{2} \left(\frac{d\varphi_{ab}}{d x}\right)^2 + U(\varphi_{ab}(x))  \,\,\,,
\EN 
and its integral provides the classical expression of the kink masses 
\EQ
M_{ab} \,=\,\int_{-\infty}^{\infty} \epsilon_{ab}(x) \,\,\,.
\label{integralmass}
\EN   
As a rule of thumb, it is useful to notice that the classical masses of the kinks $\varphi_{ab}(x)$ are 
simply proportional to the heights of the potential between the two minima 
$\varphi_a^{(0)}$ and $\varphi_b^{(0)}$. 

\begin{figure}[h]
\hspace{10mm}
\vspace{8mm}
\psfig{figure=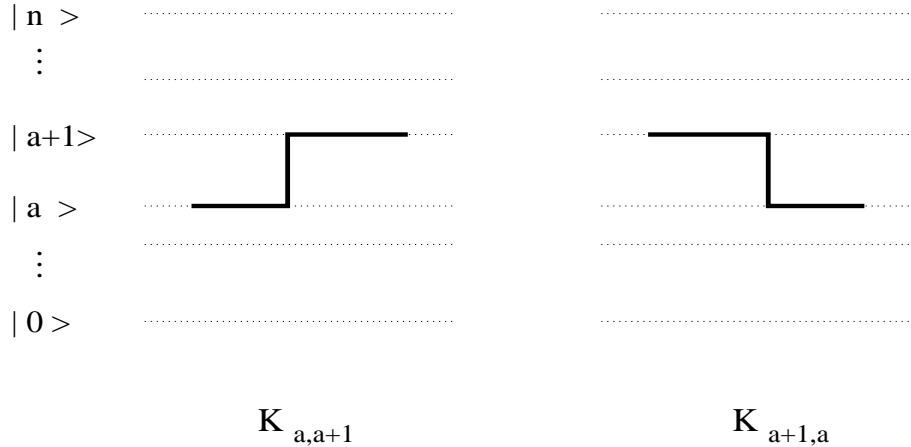,height=6cm,width=12cm}
\caption{{\em Kink and antikink configurations.}}
\label{step}
\end{figure}

The classical solutions can be set in motion by a Lorentz transformation, i.e. 
$\varphi_{ab}(x) \rightarrow \varphi_{ab}\left[(x \pm v t)/\sqrt{1-v^2}\right]$. 
In the quantum theory, these configurations describe the kink states $\mid K_{ab}(\theta)\,\rangle$, 
where $a$ and $b$ are the indices of the initial and final vacuum, respectively. The quantity $\theta$ 
is the rapidity variable which parameterises the relativistic dispersion relation of these excitations, i.e. 
\EQ
E = M_{ab}\,\cosh\theta
\,\,\,\,\,\,\,
,
\,\,\,\,\,\,\,
P = M_{ab} \,\sinh\theta
\,\,\,.
\label{rapidity}
\EN 
Conventionally $\mid K_{a,a+1}(\theta) \,\rangle$ denotes the {\it kink} between the pair 
of vacua $\left\{\mid a\,\rangle ,\mid a+1\,\rangle\right\}$ while $\mid K_{a+1,a}\,\rangle$ 
is the corresponding {\it anti-kink}. For the kink configurations it may be useful 
to adopt the simplified graphical form shown in Figure \ref{step}\footnote{Although very convenient, 
one should keep in mind that this graphical representation oversimplifies the exponential approaching 
to the vacua, which can be different for the one to the right and for the one to the left.}.  

\vspace{3mm}

\noindent
The multi-particle states are given by a string of these excitations, with the adjacency condition 
of the consecutive indices for the continuity of the field configuration
\EQ
\mid K_{a_1,a_2}(\theta_1) \,K_{a_2,a_3}(\theta_2)\,K_{a_3,a_4}(\theta_3) \ldots \rangle 
\,\,\,\,\,\,\,\,
,
\,\,\,\,\,\,\,\, (a_{i+1} = a_i \pm 1)
\EN 
In addition to the kinks, in the quantum theory there may exist other excitations in the guise of 
ordinary scalar particles (breathers). These are the neutral excitations $\mid B_c(\theta)\,\rangle_a$ 
($c=1,2,\ldots$) around each of the vacua $\mid a\,\rangle$. For a theory based on a Lagrangian of 
a single real field, these states are all non-degenerate: in fact, there are no extra quantities 
which commute with the Hamiltonian and that can give rise to a multiplicity of them. The only exact 
(alias, unbroken) symmetries for a Lagrangian as (\ref{Lagrangian}) may be the discrete ones, like the 
parity transformation $P$, for instance, or the charge conjugation ${\cal C}$. However, since they 
are neutral excitations, they will be either even or odd eigenvectors of ${\cal C}$. 
 
\begin{figure}[h]
\hspace{10mm}
\vspace{8mm}
\psfig{figure=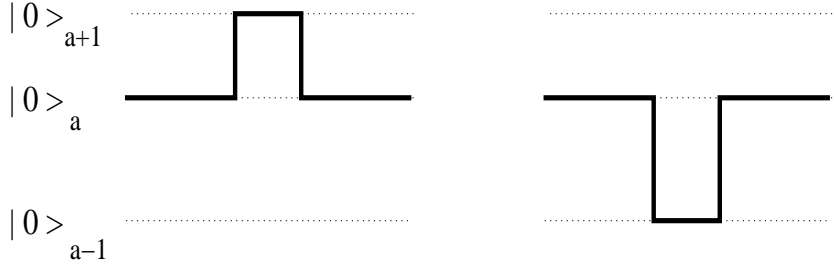,height=35mm,width=11cm}
\caption{{\em Kink-antikink configurations which may give rise to a bound state nearby the vacuum 
$\mid 0\,\rangle_a$.}}
\label{tooth}
\end{figure}

The neutral particles must be identified as the bound states of the kink-antikink configurations that 
start and end at the same vacuum $\mid a\,\rangle$, i.e. $
\mid K_{ab}(\theta_1) \,K_{ba}(\theta_2)\,\rangle
$, with the ``tooth'' shapes shown in Figure\,\,\ref{tooth}.

\vspace{3mm}

\noindent
If such kink states have a pole at an imaginary value 
$i \,u_{a b}^c$ within the physical strip $0 < {\rm Im}\, \theta < \pi$ of their rapidity difference 
$\theta = \theta_1 - \theta_2$, then their bound states are defined through the factorization formula 
which holds in the vicinity of this singularity 
\EQ
\mid K_{ab}(\theta_1) \,K_{ba}(\theta_2) \,\rangle \,\simeq \,i\,\frac{g_{ab}^c}{\theta - i u_{ab}^c}
\,\mid B_c\,\rangle_a \,\,\,.
\label{factorization}
\EN 
In this expression $g_{ab}^c$ is the on-shell 3-particle coupling between the kinks and the neutral particle. 
Moreover, the mass of the bound states is simply obtained by substituing the resonance value $i \,u_{ab}^c$ 
within the expression of the Mandelstam variable $s$ of the two-kink channel 
\EQ
s = 4 M^2_{ab} \,\cosh^2\frac{\theta}{2} 
\,\,\,\,\,\,
\longrightarrow 
\,\,\,\,\,\, 
m_c \,=\,2 M_{ab} \cos\frac{u_{ab}^c}{2} \,\,\,.
\label{massboundstate}
\EN 
Since the kink $\varphi_{a-1,a}(x)$ that interpolates to the vacuum on the left of $\mid a\,\rangle$ 
may be different from the kink $\varphi_{a,a+1}(x)$ which interpolates to the vacuum on the right, 
a-priori there could be two different towers of breathers that pile up in each vacuum: the first 
coming from the poles of 
$\mid K_{a,a-1}(\theta_1) K_{a-1,a}(\theta)\,\rangle$ while the second one from the poles of 
$\mid K_{a,a+1}(\theta) K_{a+1,a}(\theta_2) \rangle$. However, as we will discuss in the following, 
this situation cannot occur. 

Concerning the vacua themselves, as well known, in the infinite volume their classical degeneracy 
is removed by selecting one of them, say  $\mid k \,\rangle$, out of the $n$ available. This happens 
through the usual spontaneously symmetry breaking mechanism, even though -- stricly speaking -- there may 
be no internal symmetry to break at all. This is the case, for instance, of the potential shown in Figure 1, 
which does not have any particular invariance. In the absence of a symmetry which connects the various 
vacua, the world -- as seen by each of them -- may appear very different: they can have, indeed, 
different particle contents. The problem we would like to examine in this paper concerns 
the neutral excitations around each vacuum, in particular the question of the existence of 
such particles and of the value of their masses. 

The answer provided by the perturbation theory to this question is straightforward: after 
defining $\eta(x) \equiv \varphi(x) - \varphi_a^{(0)}$ and making a Taylor expansion 
of $U(\varphi)$ near $\varphi_a^{(0)}$ 
\EQ
U(\varphi_a^{(0)} + \eta) \,=\,\frac{1}{2} \omega_a^2 \,\eta^2 + \frac{1}{3} \lambda_3 \,\eta^3 + 
\frac{1}{4} \lambda_4 \,\eta^4 + \cdots 
\EN 
one identifies the mass $m$ of the fundamental particle around the vacuum $\mid a\,\rangle$ 
and $\omega_a$, while the rest of the expansion with its interaction terms. The quantity $\omega_a$ is 
of course the zero-order value of $m$ of such a particle, but the crucial point is another one: according 
to the perturbation theory, as far as the potential has a quadratic curvature at its minimum, there is always a 
neutral excitation above the corresponding vacuum state. This conclusion is, unfortunately, false.    

A famous counter-example is given by the Sine-Gordon model, i.e. the quantum field theory associated 
to the potential 
\EQ
U_{SG}(\varphi)\,=\,\frac{m_0^2}{\beta^2} \left[1 - \cos(\beta \varphi)\right] \,\,\,,
\label{SG}
\EN 
where $m_0$ is a mass-like parameter and $\beta$ is a coupling constant. Such a theory has an infinite 
number of degenerate vacua $\mid a\,\rangle$ ($a=0,\pm 1,\pm 2,\ldots$), localised at $\varphi_a^{(0)} 
\,=\,2 \pi \,a/\beta$, each of them with the same curvature $\omega^2 = m_0^2$. Through the above 
perturbative argument, one would conclude that each vacuum has always, at least, one neutral 
excitation. On the other hand, the exact $S$-matrix of this model \cite{zamzam} shows that the situation is 
rather different: indeed, such a particle does not exist if $\beta^2 > 4 \pi$.

As in the case of the Sine-Gordon model, the knowledge of the exact $S$-matrix of a quantum problem  
would obviously provide a clear cut answer to the question of the particle content of a theory: 
a proper identification of its poles gives its spectrum. This has been amply proved by the large 
number of the exact $S$-matrices associated to the integrable deformations of Conformal Field Theories 
\cite{Zam} (for a review, see \cite{GM}). But, what happens if the theory is not integrable? How can we  
proceed if the exact $S$ matrix is not known? 

\section{A semiclassical formula}

The particle content of certain non-integrable models can be studied by using the so-called 
Form Factor Perturbation Theory \cite{DMS,DM,CM}. As shown in \cite{decay}, this approach can be also 
extended to compute, in a reliable way, the decay widths of the unstable particles of the theory\footnote{
The analitic prediction of the paper \cite{decay} for the decay widths of the unstable 
particles of the Ising model has been confirmed by the numerical analysis done in \cite{ungheresi1}.}. 
In addition to this approach, another interesting route for investigating the non-integrable models comes 
from semi-classical methods. Originally proposed by Dashen-Hasslacher-Neveu \cite{DHN} 
and by Goldstone-Jackiw \cite{GJ}, this approach has been recently applied either to study quantum 
field theories on a finite volume \cite{FFvolume,volume,MRSD} or to obtain their spectrum at the 
semiclassical level \cite{doubleSG}.
 
The main difference between the two approaches is the following. The Form Factor Perturbation Theory 
is a formalism based on the $S$-matrix theory. More precisely, it moves its first steps with the 
exact scattering amplitudes of the integrable models, reached as a limit of the non-integrable ones. 
On the contrary, the semi-classical methods build their analysis on the Lagrangian density of the model, 
irrespectively whether it describes an integrable system or not. As shown in \cite{doubleSG}, the two 
approaches sometimes coincide while, in some other cases, they complement each other, i.e. one needs 
both methods to recover the whole spectrum of the theory. 

\begin{figure}[h]
\hspace{55mm}
\vspace{1mm}
\psfig{figure=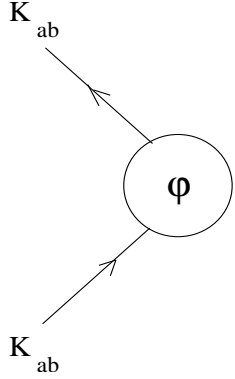,height=5cm,width=3cm}
\caption{{\em Matrix element between kink states.}}
\label{formfactor}
\end{figure}

For the problem that concerns this paper -- to find the neutral spectrum of the Lagrangian theory 
(\ref{Lagrangian}) -- the semiclassical methods are the obvious choice. Their correct implementation 
may need, though, important pieces of information coming from the $S$-matrix theory or from the 
Form Factor Perturbation Theory. The starting point of our analysis is a remarkably simple formula due 
to Goldstone-Jackiw \cite{GJ}. In its refined version, given in \cite{JW} and rediscovered in 
\cite{FFvolume}, 
it reads as follows\footnote{The matrix element of the field $\varphi(y)$ is easily obtained by using 
$\varphi(y) = e^{-i P_{\mu} y^{\mu}} \,\varphi(0) \,e^{i P_{\mu} y^{\mu}}$ and by acting with the 
conserved energy-momentum operator $P_{\mu}$ on the kink state. Moreover, for the semiclassical 
matrix element $F_{ab}^{\cal G}(\theta)$ of the operator $G[\varphi(0)]$, one should employ 
$G[\varphi_{ab}(x)]$. For instance, the matrix element of $\varphi^2(0)$  
are given by the Fourier transform of $\varphi_{ab}^2(x)$.} (Figure \ref{formfactor}) 
\EQ
f_{ab}^{\varphi}(\theta) \,=\,\langle K_{ab}(\theta_1) \,\mid \varphi(0) \,\mid \, K_{ab}(\theta_2) \rangle 
\,\simeq \,\int_{-\infty}^{\infty} dx \,e^{i M_{ab} \,\theta\,x} \,\,
\varphi_{ab}(x) \,\,\,,
\label{remarkable1}
\EN  
where $\theta = \theta_1 - \theta_2$. 

\vspace{3mm}

\noindent
Notice that, if we substitute in the above formula 
$\theta \rightarrow i \pi - \theta$, the corresponding expression may be interpreted as 
the following Form Factor 
\EQ
F_{ab}^{\varphi}(\theta) \,=\, f(i \pi - \theta) \,=\,\langle a \,\mid \varphi(0) \,\mid \,K_{ab}(\theta_1) \,
K_{ba}(\theta_2) \rangle \,\,\,.
\label{remarkable2}
\EN    
In this matrix element, it appears the neutral kink states around the vacuum $\mid a \rangle$ we 
are interested in.

By following the references \cite{GJ,JW,FFvolume,raj}, let's firstly recall the main 
steps that lead to this formula and let's make the first comments on its content. 
Denoting the adimensional coupling constant of the theory generically by $g$, we will assume that the 
mass of the kink becomes arbitrarly large when $g \rightarrow 0$, say as $M_{ab} \simeq 1/g$. 
Consider now the Heisenberg equation of motion satisfied by the field $\varphi(x)$ 
\EQ
\left(\partial^2_{t} - \partial^2_x\right) \,\varphi(x,t) \,=\,-U'[\varphi(x,t)] \,\,\,,
\EN 
and sandwich it between the kink states of momentum $p_1$ and $p_2$. By using 
\EQ
\langle K_{ab}(p_1) \,\mid \varphi(x,t) \,\mid \,K_{ab}(p_2) \,\rangle \,=\,
e^{-i (p_1 - p_2)_{\mu} \,x^{\mu}} \,\,\langle K_{ab}(p_1) \,\mid \varphi(0) \,\mid \,K_{ab}(p_2) \,\rangle 
\label{decomposition}
\EN 
we have 
\begin{eqnarray}
&& \left[-(p_1 - p_2)_{\mu} \,(p_1 - p_2)^{\mu} \right] \,\,\langle K_{ab}(p_1) \,\mid \varphi(0) \,\mid 
K_{ab}(p_2) \,\rangle \nonumber \\
&& \,\,\,\,=\,- \langle K_{ab}(p_1) \,\mid U'\left[\varphi(0)\right]\,\mid K_{ab}(p_2) \,\rangle 
\,\,\,.
\label{intermediate}
\end{eqnarray}
Once it has been extracted the $x^{\mu}$-dependence of the matrix element (\ref{decomposition}), 
the remaining expression $ \langle K_{ab}(p_1) \,\mid \varphi(0) \,\mid \,K_{ab}(p_2) \,\rangle $
should depend on the relativistic invariants of the channel of the two kinks. 
Since these invariants can be expressed in terms of difference of the rapidities of the 
two kinks, this suggests to adopt the rapidity variables and write eq.\,(\ref{intermediate}) as 
\begin{eqnarray}
&& 2\,M_{ab}^2 (\cosh\theta -1) \,\,\langle K_{ab}(\theta_1) \,\mid \varphi(0) \,\mid \, K_{ab}(\theta_2)  
\,\rangle
\nonumber \\
&& \,\,\,\,=\, - 
\langle K_{ab}(\theta_1) \,\mid U'[\varphi(0)] \,\mid \, K_{ab}(\theta_2) \,\rangle \,\,\,,
\label{intermediate2}
\end{eqnarray}
where $\theta = \theta_1 - \theta_2$. Let's now assume that the kinks are sufficiently slow, so that 
their dispersion relations can be approximated by the non-relativistic expressions 
\EQ
E \,=\, M \,\cosh\theta \simeq M\,\left(1 + \frac{\theta^2}{2}\right)
\,\,\,\,\,\,\,\,
,
\,\,\,\,\,\,\,\,
P \,=\,M\,\sinh\theta \simeq M \,\theta \,\ll M \,\,\,.
\EN 
In this quasi-static regime, we can define the matrix element of the field $\varphi(0)$ between 
the kink states as the Fourier transform with respect to the Lorentz invariant difference 
$\theta = \theta_1 - \theta_2$
\EQ
f_{ab}^{\varphi}(\theta)\,=\,
\langle K_{ab}(\theta_1) \,\mid \varphi(0) \,\mid \, K_{ab}(\theta_2) \rangle 
\,\simeq \,\int_{-\infty}^{\infty} dx \,\,e^{i M_{ab}\, \theta\,x} \,\,
\hat f(x) \,\,\,,
\label{FF}
\EN
with the inverse Fourier transform given by 
\EQ
\hat f(x) \,=\,\int \frac{d\theta}{2\pi} \,e^{-i M_{ab}\, \theta \,x} 
\,f_{ab}^{\varphi}(\theta)
\,\,\,.
\EN 
In the quasi-static limit, the left hand side of equation (\ref{intermediate2}) becomes 
\EQ
M_{ab}^2 \,\theta^2 \,f_{ab}^{\varphi}(\theta)  \,\,\,,
\EN 
which, in real space, corresponds to
\EQ
-\frac{d^2}{d x^2} \,\hat f_{ab}(x) \,\,\,.
\EN  
Concerning the right hand side, let's assume that $U'[\phi(0)]$ can be expressed in terms of 
powers of the field $\varphi(x)$ (either as a finite sum or an infinite series) 
\EQ
U'[\varphi] \,=\,\sum_{n=1} \alpha_n \,\varphi^n \,\,\,.
\EN 
Consider now the matrix elements of the generic term of this expression between the kink states 
\EQ
\langle K_{ab}(\theta_1) \,\mid \varphi^n(0) \,\mid \, K_{ab}(\theta_2) \rangle\,\,\,.
\EN 
By inserting $(n-1)$ times a complete set of state, we have  
\EQ
\sum_{m_1,\ldots m_{n-1}} 
\langle K_{ab}(\theta_1) \,\mid \varphi(0) \,\mid \, m_1 \rangle \,\langle m_1 \mid \,\varphi(0) 
\mid \,m_2 \rangle \ldots \langle m_{n-1} \mid \varphi(0) \mid K_{ab}(\theta_2) \rangle\,\,\,.
\label{completeset}
\EN
The only states which are involved in the above sums are those having the same topological charge of 
the kink $K_{ab}$, with the lowest mass states given precisely by the kinks $K_{ab}(\theta_i)$ 
themselves. By truncating the sums just on these states and using the definition (\ref{FF}), we have 
then 
\EQ
\langle K_{ab}(\theta_1) \,\mid \varphi^n(0) \,\mid \, K_{ab}(\theta_2) \rangle\,
\simeq \,\int_{-\infty}^{\infty} dx \,e^{i M_{ab} \,\theta\,x} \left(\hat f(x)\right)^n 
\,\,\,.
\EN 
Hence, at the leading order $1/g$, the function $\hat f(x)$ satisfies the same differential equation 
(\ref{static}) satisfied by the static kink solution, i.e.  
\EQ
\frac{d^2}{d x^2} \hat f_{ab}(x) \,=\,U'[f_{ab}(x)]  \,\,\,, 
\EN 
arriving then to the result (\ref{remarkable1}). 

\begin{figure}[h]
\hspace{15mm}
\vspace{10mm}
\psfig{figure=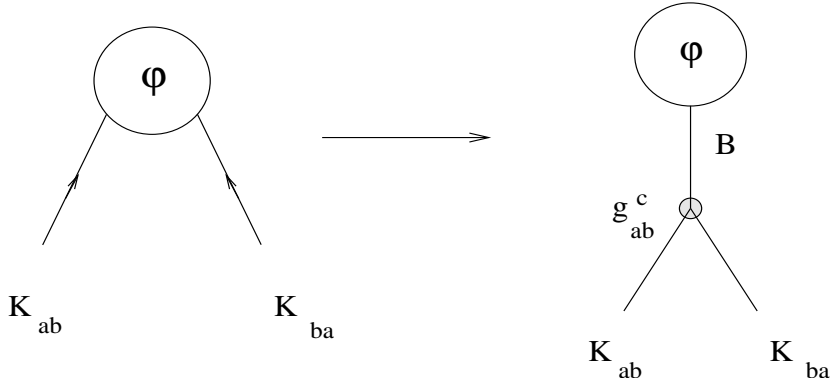,height=5cm,width=11cm}
\vspace{1mm}
\caption{{\em Residue equation for the matrix element on the kink states.}}
\label{residuef}
\end{figure}

The appealing aspect of the formula (\ref{remarkable1}) stays in the relation between the 
Fourier transform of the {\em classical} configuration of the kink, -- i.e. the solution $\varphi_{ab}(x)$  
of the differential equation (\ref{kinkequation}) -- to the {\em quantum} matrix element 
of the field $\varphi(0)$ between the vacuum $\mid a\,\rangle$ and the 2-particle kink state 
$\mid K_{ab}(\theta_1) \,K_{ba}(\theta_2)\,\rangle$. Once the solution of eq.\,(\ref{kinkequation}) 
has been found and its Fourier transform has been taken, the poles of $F_{ab}(\theta)$ within 
the physical strip of $\theta$ identify the neutral bound states which couple to $\varphi$. Then, 
their mass can be extracted by using eq.\,(\ref{massboundstate}), while the on-shell 3-particle coupling 
$g_{ab}^c$ can be obtained from the residue at these poles (Figura \ref{residuef})  
\EQ
\lim_{\theta \rightarrow i \,u_{ab}^c} (\theta - i u_{ab}^c)\, F_{ab}(\theta)
\,=\,i \,g_{ab}^c \,\,\langle a \,\mid \varphi(0) \,\mid \,B_c \,\rangle \,\,\,.
\label{residue}
\EN  

It is important to stress that, for a generic theory, the classical kink configuration $\varphi_{ab}(x)$ is 
not related in a simple way to the anti-kink configuration $\varphi_{ba}(x)$. It is precisely for this reason 
that neighbouring vacua may have a different spectrum of neutral excitations, as shown in the examples 
discussed in the following sections.   

It is also worth noting that this procedure for extracting the bound states masses permits in many cases 
to avoid the semiclassical quantization of the breather solutions \cite{DHN}, making their derivation much 
simpler. The reason is that, the classical breather configurations depend also on time and have, in general, 
a more complicated structure than the kink ones. Yet, in non--integrable theories these configurations do 
not exist as exact solutions of the partial differential equations of the field theory. On the contrary, in 
order to apply eq.\,(\ref{remarkable1}), one simply needs the solution of an {\em ordinary} differential 
equation, the one given by (\ref{kinkequation}): in the absence of an exact expression, one could even 
conceive the idea of employing the solution extracted by a numerical integration of the equation 
(\ref{kinkequation}), an operation which requires few seconds on any laptop\footnote{One might doubt about 
the possibility to locate the poles of $F_{ab}(\theta)$ by performing a Fourier transform of the numerical 
solution. However, their position can be easily determined in a different way, i.e. by looking at the 
exponential behavior of the solutions at $x \rightarrow \pm \infty$, as discussed in the next section. 
This behavior can be quickly extracted by numerical solutions and it is also, analitically, 
perfectly under control.}.

Let's now add a remark of technical nature: the Fourier transform (\ref{kinksolphi4}) has always 
a singular part, due to the constant asymptotic behaviours of $\varphi_(x)$ at $x \rightarrow \pm \infty$. 
This piece can be easily isolated by splitting the integral as follows 
\begin{eqnarray}
f_{ab}(k) & \,=\,& 
\int_{-\infty}^{\infty} dx \, e^{i k x} \varphi_{ab}(x) \,=\,\int_{-\infty}^0 dx \,e^{i k x} \,\varphi_{ab}(x) + 
\int_{0}^{\infty} dx \,e^{i k x} \,\varphi_{ab}(x)  \nonumber \\ 
& = &  \int_{-\infty}^0 dx \,e^{i kx} \left[\varphi_{ab}(x) - \varphi_a^{(0)}\right] + 
\int_{0}^{\infty} dx \,e^{i k x} \,\left[\varphi_{ab}(x) - \varphi_b^{(0)}\right] \\  
&& + \varphi_a^{(0)} \int_{-\infty}^0 dx \,e^{i k x} +  
\varphi_b^{(0)} \,\int_0^{\infty} dx \,e^{i k x} \,\,\,.\nonumber 
\end{eqnarray}
The singular part is enconded in the last two terms, for which we have  
\[
\varphi_{a}^{(0)} \,\int_{-\infty}^{0} dx \,e^{i k x} + 
\varphi_{b}^{(0)} \,\int_0^{\infty} dx \,e^{i k x} \,=\,i (\varphi_a^{(0)} - \varphi_b^{(0)}) 
\,P\left(\frac{1}{k}\right) 
+ \pi \,(\varphi_a^{(0)} + \varphi_b^{(0)}) \,\delta(k) \,\,\,,
\] 
where $P$ stays for the Cauchy principal value. Since these singular terms do not contain information on 
the pole structure of the matrix elements, they will always be discarded from now on. Concering the regular 
part, it can be computed by means of the derivative of the kink solutions, by using 
${\cal F}\left(\frac{d f}{dx}\right) = - i k \,{\cal F}(f)$, where 
\[
{\cal F}(f) \,=\,\int_{-\infty}^{\infty} d k \,e^{i k x} \,f(x) \,\,\,.
\]  
Additional comments on the nature of the poles of the semiclassical form factors will be given in 
the next section, with the help of an explicit example.

The range of validity of the formula (\ref{remarkable1}) is a more delicate issue. 
As discussed above, its derivation relies on the basic hypothesis that the kink momentum is very small 
compared to its mass, and also on the possibility of neglecting intermediate higher particle 
contributions \cite{GJ,JW,FFvolume}. These two assumptions usually translate into the combined condition 
$\theta \simeq {\cal O}(g) \ll 1$, where $g$ is the adimensional coupling constant of the theory. This 
authorises, for instance, to substitute in the result of the Fourier transform, $\theta \rightarrow \sinh\theta$ 
(since $\theta$ is infinitesimal), but keeping untouched all expressions containing $\theta/g$. 
But, the above constraint may result in a different level of accuracy on various physical 
quantities, with the precision that may also depend on the model under investigation. 

Consider, for instance, the Sine-Gordon model, a theory that will be studied in details in 
Section \ref{SGsection}. 
Its semiclassical mass formula turns out to be an exact expression, i.e. valid at all orders in the 
coupling constant till its critical value $\beta^2_c = 4 \pi$, beyond which, the bound states disappear.
Moreover, the numerical values of the semiclassical Form Factors do not significatively differ from the 
exact ones for {\em all real} values of $\theta$. On the wave of these success, one may dare to 
extract the $S$-matrix by using the semiclassical Form Factors. If one tries to do so, 
the outcoming expression turns out to be remarkably close to the exact one but, it always fails to 
meet an important crossing symmetric factor. The same happens in other models too, as it will be explained 
later. 

All this to say that, some caution is necessary in handling the results obtained by using the formula 
(\ref{remarkable1}). Morally it has the same status of the WKB approximation in quantum mechanics:   
this usually provides very accurate results for the discrete states, remaining nevertheless   
a poor approximation for those of the continuum. But even for the bound states, the formula (\ref{remarkable1}) 
seems sometimes to lead to some puzzling conclusions. This happens, for instance, in the case of the Double 
Sine-Gordon model \cite{doubleSG,ungheresi2,ungheresi3}. Fortunately, there is a way out from 
the possible paradox that involves the bound states, the solution of this problem being one of 
the main motivations of this paper. In summary, properly handled, the semiclassical formalism remains 
one of the most powerful method for extracting the spectrum of the bound states in kink-like theories. 

In the next two sections we will first analyse a class of theories with only two vacua, which 
can be either symmetric or asymmetric ones. After the analysis of the Sine-Gordon model, 
we will proceed to discuss the interesting case of a vacuum state, in communication 
through asymmetric kinks, with two of its neighbouring ones. The simplest example of this kind of 
situation is provided by the Double Sine Gordon model. As we shall see, the conclusions drawn 
from all the above cases enable us to address the study of the most general theories with kink 
excitations, as shown by the examples discussed in the last section.

\section{Symmetric wells}\label{phi4section}

A prototype example of a potential with two symmetric wells is the $\varphi^4$ theory in its 
broken phase. The potential is given in this case by 
\EQ
U(\varphi) \,=\,\frac{\lambda}{4} \left(\varphi^2 - \frac{m^2}{\lambda}\right)^2 \,\,\,.
\EN 
Let us denote with $\mid \pm 1 \,\rangle$ the vacua corresponding to the classical 
minima $\varphi_{\pm}^{(0)} \,=\,\pm \frac{m}{\sqrt{\lambda}}$. By expanding around them, 
$\varphi \,=\,\varphi_{\pm}^{(0)} + \eta$, we have 
\EQ
U(\varphi_{\pm}^{(0)} + \eta) \,=\, m^2 \,\eta^2 \pm m \sqrt{\lambda}\,\eta^3 + \frac{\lambda}{4}
\eta^4 \,\,\,.
\label{potentialphi4}
\EN 
Hence, perturbation theory predicts the existence of a neutral particle for each of the two vacua,  
with a bare mass given by $m_b = \sqrt{2} m$, irrespectively of the value of the coupling $\lambda$. 
Let's see, instead, what is the result of the semiclassical analysis. 

The kink solutions are given in this case by\footnote{In the following we will always discard 
the integration constant $x_0$ on which is localised the kink solution.} 
\EQ
\varphi_{-a,a}(x) \,=\,a \,\frac{m}{\sqrt{\lambda}} \,\tanh\left[\frac{m x}{\sqrt{2}}\right]
\,\,\,\,\,\,\,
,
\,\,\,\,\,\,\, a = \pm 1 
\label{kinksolphi4}
\EN 
and their classical mass is 
\EQ
M_0\,=\,\int_{-\infty}^{\infty} \epsilon(x) \,dx \,=\,\frac{2 \sqrt{2}}{3} \,
\frac{m^3}{\lambda}  
\,\,\,.
\EN 
The value of the potential at the origin, which gives the height of the barrier between the two vacua, 
can be expressed as 
\EQ
U(0) \,=\,\frac{3 m}{8 \sqrt{2}} \,M_0 \,\,\,,
\EN 
and, as noticed in the introduction, is proportional to the classical mass of the kink. 

If we take into account the contribution of the small oscillations around the classical static 
configurations, the kink mass gets corrected as \cite{DHN}
\EQ
M \,=\,\frac{2 \sqrt{2}}{3} \,\frac{m^3}{\lambda} - m 
\left(\frac{3}{\pi \sqrt{2}} - \frac{1}{2 \sqrt{6}}\right) 
+ {\cal O}(\lambda) \,\,\,.
\label{mass1phi4}
\EN 
It is convenient to define 
\[
c =  \left(\frac{3}{2\pi} - \frac{1}{4 \sqrt{3}}\right) > 0 \,\,\,,
\]
and also the adimensional quantities
\EQ
g = \frac{3 \lambda}{2 \pi m^2}
\,\,\,\,\,\,\,\,
;
\,\,\,\,\,\,\,\,
\xi \,=\,\frac{g}{1 - \pi c g} \,\,\,.
\label{definitiong}
\EN 
In terms of them, the mass of the kink can be expressed as 
\EQ
M \,=\,\frac{\sqrt{2} m}{\pi \,\xi}\,=\,\frac{m_b}{\pi \,\xi}\,\,\,.
\label{newmassphi4}
\EN
Since the kink and the anti-kink solutions are equal functions (up to a sign), their Fourier transforms 
have the same poles. Hence, the spectrum of the neutral particles will be the same on both vacua, in 
agreement with the $Z_2$ symmetry of the model. For explicitly computing it, let's first consider 
\EQ
\left(\frac{d\varphi}{d x}\right)_{-a,a} \,=\, a\,
\frac{m^2}{\sqrt{2 \lambda}} \,\frac{1}{\cosh^2\frac{m x}{\sqrt{2}}} 
\,\,\,,
\EN 
and then apply the prescription for the Fourier transform given in the previous section. As a result, 
we have  
\EQ
f_{-a,a}(\theta) \,= \, \int_{-\infty}^{\infty} 
dx \,e^{i M  \theta\,x} \varphi_{-a,a}(x)  \nonumber \, =\, 
 i\,a \sqrt{\frac{2}{\lambda}}\,\frac{1}{\sinh\left(\frac{\pi M}{\sqrt{2} m} \theta\right)}\,\,\,.
\EN
By making now the analitical continuation $\theta \rightarrow i \pi - \theta$ and using the above definitions 
(\ref{definitiong}), we arrive to   
\EQ
F_{-a,a}(\theta) \,=\,
\langle a\,\mid \varphi(0)\,\mid K_{-a,a}(\theta_1) K_{a,-a}(\theta_2) \rangle 
\,=\, A_a \,  
\frac{1}{\sinh\left(\frac{(i \pi - \theta)}{\xi}\right)} \,\,\,,
\label{FFphi4}
\EN 
where $A_a$ is a constant, given by  
\[
A_a \,=\,i \,a \,m\,\left(\frac{3 (1 + \pi \,c \,\xi)}{(\pi\, \xi)}\right)^{1/2} \,\,\,.
\]
The poles of the above expression are located at 
\begin{equation}
\theta_{n}\,=\,i \pi \left(1 - \xi \,n\right)
\,\,\,\,\,\,\,
,
\,\,\,\,\,\,\,
n = 0,\pm 1,\pm 2,\ldots
\label{polesphi4}
\end{equation}
and, if 
\EQ
\xi \geq 1 \,\,\,,
\label{conditionphi4}
\EN 
none of them is in the physical strip $0 < {\rm Im}\,\theta < \pi$. Consequently, in the 
range of the coupling constant  
\EQ
\frac{\lambda}{m^2} \geq \frac{2 \pi}{3} \,\frac{1}{1 + \pi c} \,=\,1.02338...
\label{criticalg}
\EN 
the theory does not have any neutral bound states, neither on the vacuum to the right nor on the one to 
the left. Viceversa, if $\xi < 1$, there are $n = \left[\frac{1}{\xi}\right]$ neutral bound states, 
where $[ x ]$ denote the integer part of the number $x$. Their semiclassical masses are 
given by 
\begin{equation}
m_{b}^{(n)} \,=\, 2 M\,\sin\left[n\frac{\pi \xi}{2}\,\right]\,=\,
n\,\,m_b\left[1 - \frac{3}{32}\,\frac{\lambda^2}{m^4} 
\,n^{2} +...\right]\,.
\label{massphi4}
\end{equation}
Note that the leading term is given by multiples of the mass of the elementary boson $\mid B_1\rangle$.  
Therefore the $n$-th breather may be considered as a loosely bound state of $n$ of it, with the binding 
energy provided by the remaining terms of the above expansion. But, for the non-integrability of the 
theory, all particles with mass $m_n > 2 m_1$ will eventually decay. It is easy to see that, if there 
are at most two particles in the spectrum, it is always valid the inequality $m_2 < 2 m_1$. However, 
if $\xi < \frac{1}{3}$, for the higher particles one always has 
\EQ
m_k > 2 m_1
\,\,\,\,\,\,\,
,
\,\,\,\,\,\,\,
{\makebox for}\,\, k=3,4,\ldots n \,\,\,.
\EN 
According to the semiclassical analysis, the spectrum of neutral particles of $\varphi^4$ theory is then 
as follows: (i) if $\xi > 1$, there are no neutral particles; (ii) if $\frac{1}{2} < \xi < 1$, there is one 
particle; (iii) if $\frac{1}{3} < \xi < \frac{1}{2}$ there are two particles; (iv) if $\xi < \frac{1}{3}$ 
there are $\left[\frac{1}{\xi}\right]$ particles, although only the first two are stable, because the others 
are resonances.  
\vspace{5mm}

\begin{figure}[h]
\hspace{45mm}
\vspace{3mm}
\psfig{figure=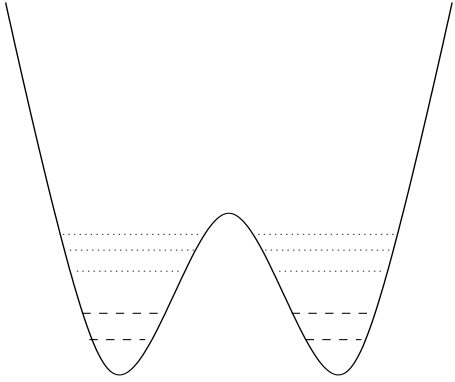,height=5cm,width=6cm}
\vspace{1mm}
\caption{{\em Neutral bound states of $\varphi^4$ theory for $g < 1$. The lowest two lines are 
the stable particles whereas the higher lines are the resonances.}}
\label{doublewell}
\end{figure}

The decay processes of the higher particles, $B_k \rightarrow r \,B_1 + s \,B_2$, where $r$ and $s$ 
are all those integers which satisfy  
\EQ
m_k \,\geq \,r \,m_1 + s \, m_2 
\,\,\,\,\,\,\,
,
\,\,\,\,\,\,\,
r+s = n  
\EN 
can be computed by Fermi golden rule
\EQ
d\Gamma \,=\,
\,(2\pi)^2 \,\delta^2(P - p_1 -  \cdots - p_n) 
\,\mid T_{fi}\mid^2 \,\frac{1}{2 E} \,\prod_{i=1}^n 
\frac{d p_i}{(2\pi) 2 E_i} \,\,\,.
\label{ratedecay}
\EN  
In this formula, $P$ denote the $2$-momentum of the decay particle whereas the amplitude $T_{fi}$ is 
given by the matrix element 
\EQ
T_{fi} \,=\,\langle B_k(P) \mid B_1(p_1) \ldots B_1(p_s) \,B_2(p_{s+1}) \ldots B_2(p_n) \,\rangle 
\,\,\,.
\EN 
At the moment it is difficult to have control of this matrix element, so the best we can do is to estimate 
some universal ratios of the decay rates by phase space alone, assuming that the matrix elements are of the 
same order of magnitude for the various processes. For instance, taking $\xi < \frac{1}{5}$, we 
can estimate the decay rates 
\EQ
\begin{array}{lll}
\Gamma^5_{11} & \cdots \cdots & B_5 \rightarrow B_1 + B_1 \\
\Gamma^5_{12} & \cdots \cdots & B_5 \rightarrow B_1 + B_2 \\
\Gamma^5_{22} & \cdots \cdots & B_5 \rightarrow B_2 + B_2 
\end{array}
\EN 
with respect to the decay rate $\Gamma^3_{11}$ of the process $B_3 \rightarrow B_1 + B_1$. We have 
\begin{eqnarray}
A & \,=\,& \frac{\Gamma^5_{11}}{\Gamma^3_{11}} \,\simeq \,
\frac{m_3}{m_5} \,\sqrt{\frac{m_3^2 - 4 m_1^2}{m_5^2 - 4 m_1^2}} \,\,\,,\nonumber \\
B & \,=\,& \frac{\Gamma^5_{12}}{\Gamma^3_{11}} \,\simeq \,
\,\sqrt{\frac{m_3^2 \,(m_3^2 - 4 m_1^2)}{[ m_5^2 - (m_1 - m_2)^2] [m_5^2 -(m_1 + m_2)^2]}}
\,\,\, , \\
C & \,=\,& \frac{\Gamma^5_{22}}{\Gamma^3_{11}} \,\simeq \,
\frac{m_3}{m_5} \,\sqrt{\frac{m_3^2 - 4 m_1^2}{m_5^2 - 4 m_2^2}} \,\,\,.\nonumber 
\end{eqnarray}

\begin{figure}[h]
\begin{tabular}{p{8cm}p{8cm}}
\psfig{figure=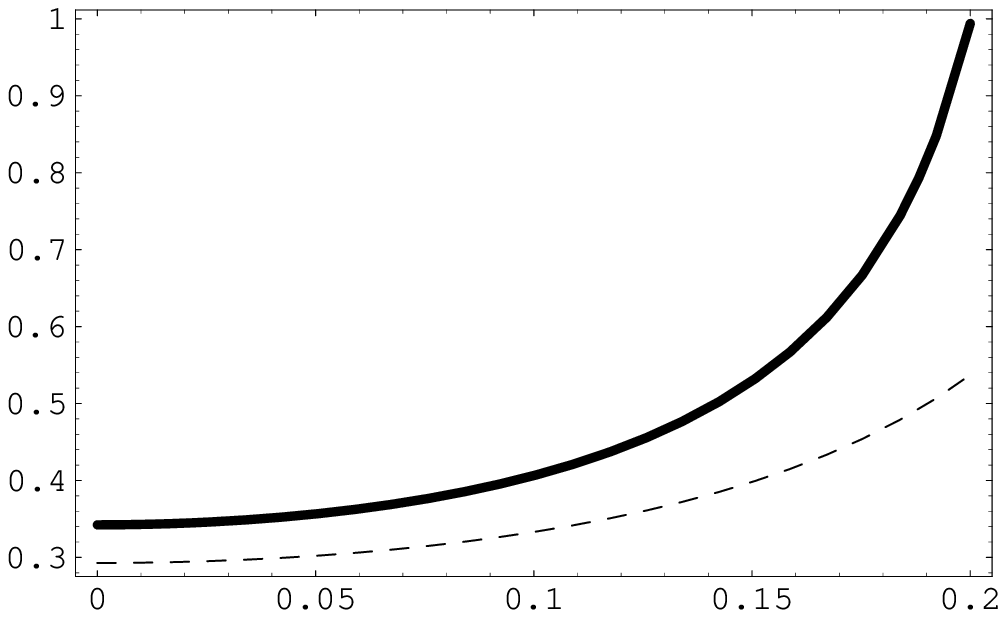,
height=5cm,width=7cm} \vspace{0.2cm}&
\psfig{figure=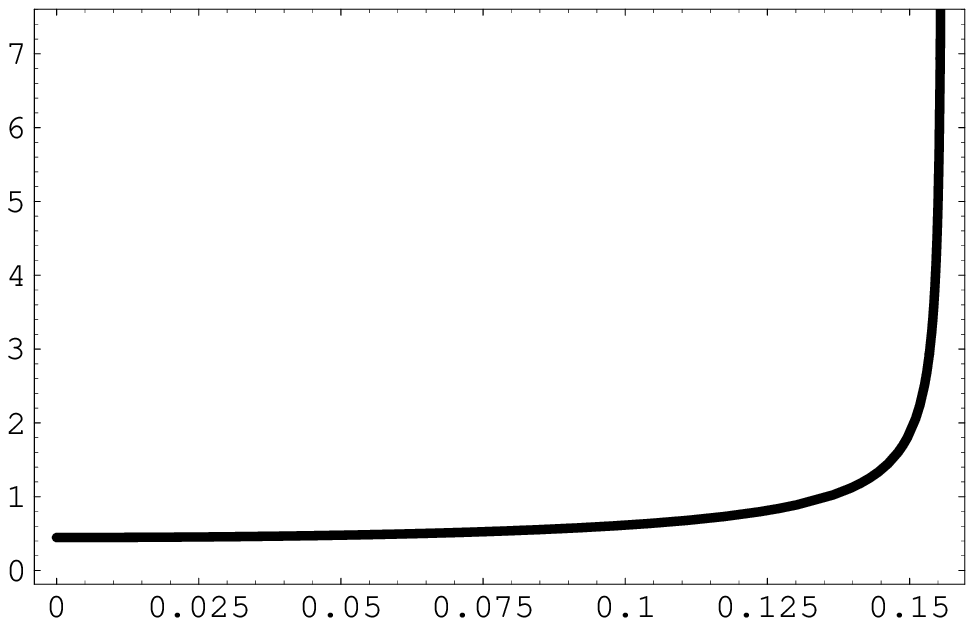,height=5cm,width=7cm}\\
\hspace{3cm} (a) & \hspace{3.1cm} (b) 
\end{tabular}
\caption{{\em (a): plot of decay ratios $A$ (lower curve) and $B$ (middle curve)
as a function of $\xi \in [0,\frac{1}{5}]$. (b): plot of the decay ratio $C$ for 
$\xi \in [0,\xi_c)$.}}
\label{decayratio}
\end{figure}

\noindent
Notice that at $\xi =\xi_{c_1} \simeq 0.1558...$ the $5$-particle has its mass exactly equal to $2 m_2$. 
This makes the amplitude ratio $C$ divergent at this point. For $\xi > \xi_c$, the decay process 
$B_5 \rightarrow B_2 + B_2$ is, instead, obviously forbidden. The plot of the quantities $A,B$ and $C$ 
in the range $0 < \xi < \frac{1}{5}$ (for the first two) and for $0 < \xi < \xi_{c_1}$ for the latter, 
is shown in Figure \ref{decayratio}. 

\begin{figure}[h]
\begin{tabular}{p{8cm}p{8cm}}
\psfig{figure=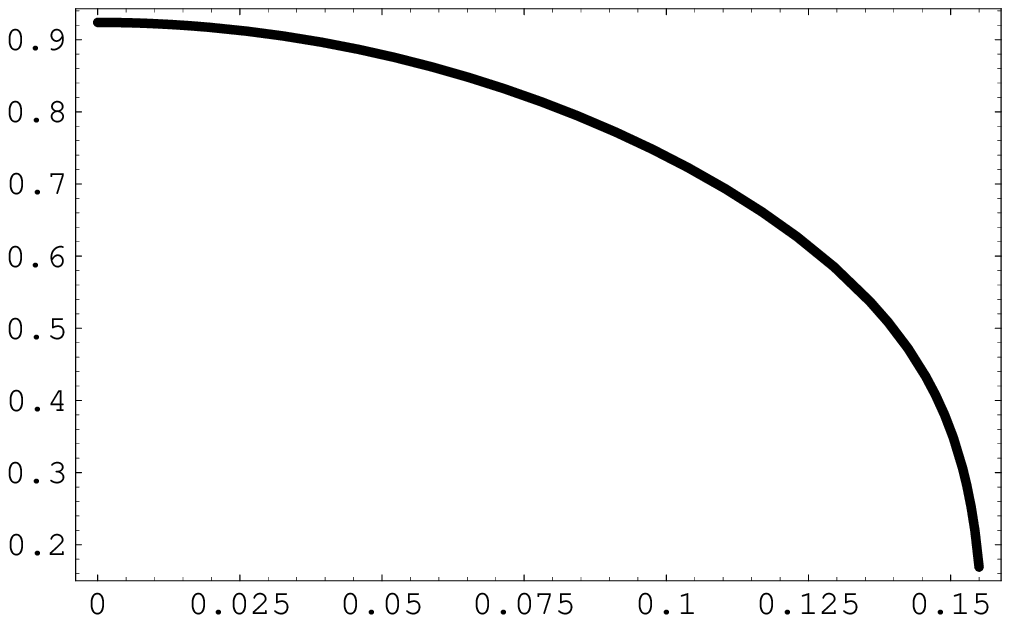,
height=5cm,width=7cm} \vspace{0.2cm}&
\psfig{figure=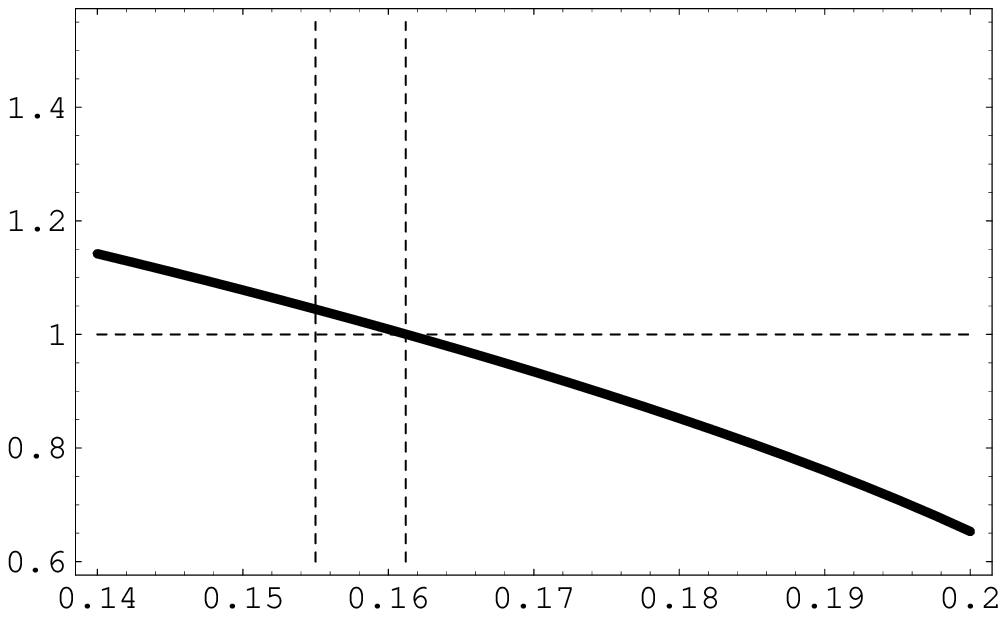,height=5cm,width=7cm}\\
\hspace{3cm} (a) & \hspace{3.1cm} (b) 
\end{tabular}
\caption{{\em (a): plot of the ratio of the life-time of the particles $B_5$ and 
$B_3$ for $\xi < \xi_{c_1}$. (b): plot of the quantity $(A+B)^{-1}$ in the 
interval $\xi \in [0.14,0.2]$, with the window of the values of $\xi$ where $B_5$ is 
more stable than $B_3$}}
\label{decayratio2}
\end{figure}

Since $\Gamma_t^c \,=\,\sum_{a,b} \Gamma^c_{a,b}$ is the inverse of the life-time of the particle $B_c$, 
from Figure \ref{decayratio2}.a one can see that the higher particle $B_5$ is only 
slightly more stable than the lower particle $B_3$ for $\xi < \xi_c$. It is only around the critical 
value $\xi_c$ that there is an enhancement of $C$ and, correspondingly, the life-time of $B_5$ 
becomes much smaller than the one of $B_3$. For $\xi > \xi_c$, the ratio of the 
total life-time of the particles $B_5$ and $B_3$ is given only by $(A + B)^{-1}$. Plotting this 
quantity as a function of $\xi$, one discoveres that there is a narrow window of values of $\xi$, 
given by the interval $[\xi_{c_1},\xi_{c_2}]$, with $\xi_{c_2} \simeq 0.1612..$, where the 
life-time of $B_5$ is larger than the one of $B_3$ (Figura \ref{decayratio2}.b). This counter-intuitive 
behavior of the life-time ratios is a simple consequence of the peculiar properties of the phase space 
in two dimensions. However, in the decay processes analysed in \cite{decay}, where the transition 
amplitudes were exactly computed, the tendency of the higher particles to be more stable than the 
lower ones was shown to be further enhanced by the dynamics. It would be interesting to study whether 
this is also the case of $\varphi^4$ theory, using, perhaps, the numerical methods 
introduced in \cite{ungheresi1}.  

\vspace{5mm}

\normalsize

Let us now comment the general scenario emerging from the semiclassical analysis.  
One could be obviously suspicious about the conclusion of the absence of the bound states for 
$\lambda > \lambda_c$, with the critical value $\lambda_c$ given in eq.\,(\ref{criticalg}): after all, this 
value is not infinitesimal and it might be, in fact, very well out of the realm of validity of the 
semiclassical approximation. This is a legitimate suspect 
and, in the absence of an exact solution which either confirms or disproves it, it is fair to say that the 
spectrum of the theory in such a strong coupling regime essentially remains an open question. Nevertheless, 
we would like to draw the attention on the following arguments in favour of the semiclassical analysis: 
\begin{enumerate}
\item Concerning $\lambda_c$, in other models the critical value of the coupling is definitely not 
small either. In Sine-Gordon for instance, the critical value is given by $\beta^2_c = 4 \pi$. 
\item The possibility of having bound states in this theory is related to the cubic interaction 
into the potential (\ref{potentialphi4}). This provides, in fact, an effective attractive interaction 
between the particles which, for low values of $\lambda/m^2$, overcomes the repulsive interaction 
given by the $\varphi^4$ term. However, the coefficient in front of $\varphi^3$ scales as $\sqrt{\lambda}$, 
whereas the one in front of $\varphi^4$ as $\lambda$. Therefore, it seems plausible that there should 
exist a sufficient large value of $\lambda/m^2$ where the repulsive force prevails on the attractive one. 
\item There is an additional little piece of information encoded in the semiclassical formula of 
the poles (\ref{polesphi4}) which can be further exploited. Namely, suppose that the adimensional 
coupling constant $g$ is very small, so that both the formula of the poles and the corresponding one 
of the masses (\ref{massphi4}) can be trusted. These expressions show that, by increasing $g$, the 
common tendency of all particles is to move {\em forward} the threshold of the two kink state and to 
decay afterward. If nothing will occur to stop this motion of the poles by increasing the coupling 
constant, the actual scenario of the theory will be then the one predicted by the semiclassical 
analysis, i.e. the existence of a critical value $g_c$ (surely different from the one given by the 
semiclassical approximation) but with no neutral particles beyond it. 
\item Finally, there is an exact mapping (which will be discussed in Section \ref{SGsection}) between 
the kinks of $\varphi^4$ and those of the Sine-Gordon model. Hence, one could argue that, as there is a 
critical value of the coupling in Sine-Gordon model in order to have bound states, the same should also 
happen for $\varphi^4$ theory.
\end{enumerate}

\subsection{Simple but useful observations}\label{Simplebutuseful}
In this section we will discuss some general features of the semiclassical methods which will be 
useful in the study of other models.

We will firstly present an equivalent way to derive the Fourier transform of the kink solution. 
To simplify the notation, let's get rid of all possible constants and consider the 
Fourier transform of the derivative of the kink solution, expressed as 
\EQ
G(k) \,=\,\int_{-\infty}^{\infty} d x \,e^{i k x} \frac{1}{\cosh^2 x} \,\,\,. 
\EN 
We split the integral in two terms
\EQ
G(k) \,=\,\int_{-\infty}^0 dx \,e^{i k x} \,\frac{1}{\cosh^2 x} 
+ \int_0^{\infty} dx \,e^{i k x} \,\frac{1}{\cosh^2 x} 
\,\,\,,
\label{FT}
\EN 
and we use the following series expansion of the integrand, valid on the entire real axis (except the origin)  
\EQ
\frac{1}{\cosh^2 x} \,=\,4 \,\sum_{n=1}^{\infty} (-1)^{n+1} n \,e^{-2 n |x|} \,\,\,.
\EN 
Substituting this expression into (\ref{FT}) and computing each integral, we have 
\EQ
G(k) \,=\,4 i\sum_{n=1}^{\infty} (-1)^{n+1} n \left[-\frac{1}{k - 2n} + \frac{1}{k + 2n }\right] 
\,\,\,. 
\EN 
Obviously it coincides with the exact result, $G(k) \,=\,\pi k/\sinh\frac{\pi}{2} k$, but this derivation 
permits to easily interpret the physical origin of each pole. In fact, changing $k$ to the original variable
in the crossed channel, $k \rightarrow (i \pi - \theta)/\xi$, we see that the poles which determine the bound 
states at the vacuum $\mid a \rangle$ are only those relative to the exponential behaviour of the kink 
solution at $x \rightarrow -\infty$. This is precisely the point where the classical kink solution takes 
values on the vacuum $\mid a \rangle$. In the case of $\varphi^4$, the kink and the antikink are the same 
function (up to a minus sign) and therefore they have the same exponential approach at $x = -\infty$ at both 
vacua $\mid \pm 1 \rangle$. Mathematically speaking, this is the reason for the coincidence of the 
bound state spectrum on each of them: this does not necessarily happens in other cases, as 
we will see in the next section, for instance. 

The second comment concerns the behavior of the kink solution near the minima of the potential. 
In the case of $\varphi^4$, expressing the kink solution as 
\EQ
\varphi(x) \,=\,\frac{m}{\sqrt \lambda} \,\tanh\left[\frac{m \,x}{\sqrt 2}\right] \,=\,
\frac{m}{\sqrt \lambda}\,\,\frac{e^{\sqrt{2} \,x} -1}{e^{\sqrt{2}\,x} + 1} \,\,\,,
\EN 
and expanding around $x = -\infty$, we have 
\EQ
\varphi(t) \,=\,-\frac{m}{\sqrt \lambda}\,\left[1 - 2 t + 2 t^2 - 2 t^3 + \cdots 2 \,(-1)^n t^n \cdots \right] 
\,\,\,,
\EN 
where $t = \exp[\sqrt{2} x]$. Hence, all the sub-leading terms are exponential factors, with exponents which 
are multiple of the first one. Is this a general feature of the kink solutions of any theory? The answer is 
positive. To prove it, consider the equation (\ref{kinkequation}) 
\EQ
\frac{d\varphi}{d x} \,=\,\sqrt{2 U(\varphi)} \,\,\,, 
\label{kinkequation2}
\EN 
in the limit in which $x \rightarrow - \infty$ and $\varphi \rightarrow \varphi_a$ (the same reasoning 
can be done, as well, in the other limit $x \rightarrow +\infty$ and $\varphi \rightarrow \varphi_b$). 
To simplify the expression, let's make the shift $\eta(x) = \varphi(x) - \varphi_a$. Assuming 
that the potential $U(\varphi)$ is regular near $\varphi_a$, we can expand the right hand side of 
(\ref{kinkequation2}) in power of $\eta$ 
\EQ
\frac{d\eta}{dx} \,=\,\alpha_1 \eta + \alpha_2 \eta^2 + \alpha_3 \eta^3 + \cdots 
\label{difeq}
\EN 
It is now easy to see that the nature of the solution strongly depends on the presence or on the absence 
of the first term, which express the square root of the curvature of the potential 
$U(\varphi)$ at $\varphi_a$. 
In fact, if $\alpha_1 = \omega \neq 0$, there is an exponential approach to the minimum, with all sub-leading 
terms given by multiples of the same exponential. To show this, let's introduce $t = e^{\omega x}$ and 
express $\eta$ in power series of $t$, $\eta(t) \,=\,\sum_{n=1}^{\infty} \mu_n \,t^n $. Substituting this 
expression into (\ref{difeq}), we have the following recursive equations for the coefficients $\mu_n$ 
\begin{eqnarray}
\sum_{n=1}^{\infty} n \,\mu_n \,t^n  \,& = & \ \,\sum_{n=1} \mu_n \,t^n + 
\frac{\alpha_2}{\omega} \,\sum_{n=1}^{\infty} \left(\sum_{k=1}^n \mu_{n-k} \mu_k\right) t^n \nonumber\\ 
&& + \frac{\alpha_3}{\omega} \,\sum_{n=1}^{\infty} \left(\sum_{k_i;k_1+k_2+k_3 = n} 
\mu_{k_1} \,\mu_{k_2} \,\mu_{k_3}\right) t^n + 
\cdots 
\end{eqnarray}
which iteratively permit to determine uniquely all of them. Explicitly, with the normalization $\mu_1 =1$, 
we have 
\begin{eqnarray}
2 \mu_2 & = & \mu_2 + \frac{\alpha_2}{\omega} \mu_1 ^2 \,\,\, \\
3 \mu_3 & = & \mu_3 + 2 \,\frac{\alpha_2}{\omega} \mu_1 \,\mu_2 + \frac{\alpha_3}{\omega} \,\mu_1^3 
\nonumber \\
\cdots & = & \cdots \nonumber 
\end{eqnarray}
By summoning to well known theorems of uniqueness of the solution of the differential equation 
(\ref{kinkequation2}), the determination of these coefficients uniquely defines the kink configuration 
near the minimum. 

When $\omega = 0$, the approach to the minimum is instead no longer exponential but is developped 
through a power-law. 
For instance, if the first non zero coefficient is $\alpha_2$, by posing $x = - 1/(\alpha_2 t)$,  
$\eta(t) \,=\,\sum_{n=1}^{\infty} \gamma_n t^n$ and substituting into (\ref{difeq}), we have the 
recursive equations for the coefficients $\gamma_n$ 
\begin{eqnarray}
\sum_{n=1} n \gamma_n \,t^{n+1} & \,=\,& \sum_{n=1}^{\infty} \left(\sum_{k=1}^n \gamma_{n-k} 
\,\gamma_k \right)\, t^{n} \nonumber \\
&& + \frac{\alpha_3}{\alpha_2} \,\sum_{n=1}^{\infty} \left(\sum_{k_i;k_1+k_2+k_3 = n} 
\gamma_{k_1} \,\gamma_{k_2} \,\gamma_{k_3}\right) t^n + 
\cdots 
\end{eqnarray}
which, as before, iteratively permit to fix all of them. 

The fact that the approach to the minimum of the kink solutions is always through multiples of the 
same exponential (when the curvature $\omega$ at the minimum is different from zero) implies 
that the Fourier transform of the kink solution has poles regularly spaced by $\xi_a \equiv 
\frac{\omega}{\pi M_{ab}}$ in the variable $\theta$. If the first of them is within the physical 
strip, the semiclassical mass spectrum derived from the formula (\ref{remarkable1}) near the vacuum 
$\mid a \,\rangle$ has therefore the universal form 
\EQ
m_n \,=\,2 M_{ab} \,\sin\left(n\,\frac{\pi\,\xi_a}{2}\right) \,\,\,. 
\EN 
As we have previously discussed, this means that, according to the value of $\xi_a$, 
we can have only the following situations at the vacuum $\mid a \,\rangle$: (a) no bound state if
$ \xi_a > 1$; (b)  one particle if $\frac{1}{2} < \xi_a < 1$; (c) two particles 
if $\frac{1}{3} < \xi_a < \frac{1}{2}$; (d) $\left[\frac{1}{\xi_a}\right]$ particles 
if $\xi_a < \frac{1}{3}$, although only the first two are stable, the others being 
resonances. So, semiclassically, each vacuum of the theory cannot have more than two stable particles above it. 
Viceversa, if $\omega = 0$, there are no poles in the Fourier transform of the kink and 
therefore there are no neutral particles near the vacuum $\mid a \,\rangle$.

\begin{figure}[h]
\hspace{25mm}
\vspace{3mm}
\psfig{figure=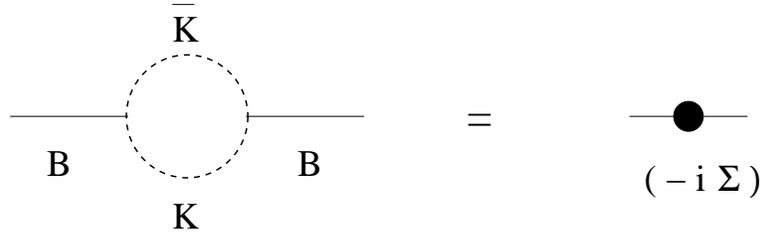,height=3cm,width=10cm}
\vspace{1mm}
\caption{{\em Self-energy due to the loop of the kink-antikink.}}
\label{blog}
\end{figure}

Finally, we would like to comment on the mass formula of the first neutral state, given by 
\EQ
m_1 \,=\,2 M \,\sin\left(\pi \frac{\xi_a}{2}\right) \,\simeq m\, \left[ 1 - \frac{1}{24} 
\left(\frac{m}{M}\right)^2 + \cdots \right] \,\,\,.
\label{primamassa}
\EN 
The first term coincides with the curvature of the minimum. Is there a way to understand the presence
of the subleading correction? Although it is well known (see, for instance, the discussion in \cite{raj})
that this term cannot be correctly reproduced by semiclassical perturbation theory\footnote{The reason 
is related to the intrinsic ambiguity of the normal-ordering procedure, which can arbitrarly alter 
the value of the mass scale.}, it is worth showing a simple calculation which indicates its relation 
with the dynamics of the kinks. To this aim, let's assume that the propagator of the kink (and the anti-kink) 
can be written as 
\EQ
{\cal G}(k) \,\simeq \,\frac{i}{k^2 - M^2} \,\,\,, 
\EN 
while for the propagator of the neutral particle we take  
\EQ
G_0(k) \,\simeq \,\frac{i}{k^2 - m^2} \,\,\,,
\EN 
where $m^2$ is the curvature of the potential at the minimum. If the kink-antikink has $B_1$ as bound state, 
there exists a non-zero $3$ particle coupling $c^1_{K,\bar K}$ and the possibility of a 
virtual process as the one shown in Figure (\ref{blog}), in which the neutral particle $B_1$ splits in a 
couple of kink-antikink and recombines afterward.

\vspace{5mm}

\noindent
The above diagram gives rise to the self-energy of the neutral particle. If $p$ is its momentum, its 
explicit expression is given by 
\EQ
(-i \Sigma(p^2)) \,=\,(i c^1_{K,\bar K})^2 \, J(p^2) \,\,\,,
\EN 
where 
\EQ
J(p^2) \,=\,
\int \frac{d^2 q}{(2\pi)^2} \,\frac{i}{q^2 - M^2 + i \epsilon} \,
\frac{i}{(p-q)^2 - M^2 + i \epsilon} \,\,\,.
\EN 
We are interested in evaluating this expression on mass-shell, i.e. at $p^2 = m^2$. 
The two denominators in $J(p^2)$ can be combined as 
\[
\frac{1}{q^2 - M^2 + i \epsilon} \,\frac{1}{(p-q)^2 - M^2 + i \epsilon} \,=\,
\int_0^1 dx \,\frac{1}{\left[q^2 - 2 x q \cdot p + x p^2 - M^2 + i \epsilon\right]^2}\,\,\,.
\]
Making the change of variable $l = q - x p$ and introducing $\Delta(x) = - x (1 - x) m_1^2 + M^2$, 
we have
\begin{eqnarray}
J(m^2)& \,=\, & \int_0^1 dx \,
\int \frac{d^2 l}{(2 \pi)^2} \frac{1}{(l^2 - \Delta + i \epsilon)^2} \nonumber \\
& = & \,\frac{i}{4 \pi} \int_0^1 \frac{1}{\Delta(x)} \,=\,\frac{i}{\pi} \frac{1}{m \,\sqrt{4 M^2 - m^2}}\,
\arctan\frac{m}{\sqrt{4 M^2 - m^2}} \,\,\,, 
\end{eqnarray}
where the factor $i$ comes from the analytical continuation to the euclidean variables in the 
integral on $l$. In the limit $\xi_a \rightarrow 0$, the mass of the kink is much larger than the 
mass of the neutral particle, so that 
\EQ
J(m^2) \,\simeq \,i\,\frac{1}{4 \,\pi\,M^2 } \,\,\,,
\EN 
and therefore, for the value of the self-energy on mass-shell, we have 
\EQ
(- i \Sigma(m^2)) \,\simeq \,i\,(c^1_{K,\bar K})^2 \,\frac{1}{4 \pi M^2} \,\,\,.
\EN 

\vspace{5mm}
\begin{figure}[h]
\hspace{3mm}
\vspace{3mm}
\psfig{figure=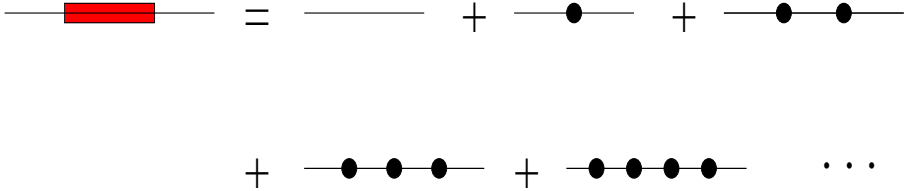,height=2.5cm,width=12cm}
\vspace{1mm}
\caption{{\em Propagator of the scalar particle in the ladder approximation.}}
\label{propagator}
\end{figure}

\noindent
In the ladder approximation shown in Figure \ref{propagator}, the propagator of the neutral particle becomes 
\begin{eqnarray}
& G(k) & \, \simeq \, G_0(k) + G_0(k) (-i \Sigma(m_1^2))\,G_0(k) \\
& & \,\,\,+ G_0(k) (-i \Sigma(m_1^2))\,G_0(k) \,
(-i \Sigma(m_1^2)) \,G_0(k) + \cdots \,=\, \frac{i}{k^2 - m^2 - \Sigma(m_1^2)} \nonumber  
\end{eqnarray}
and, correspondingly, the mass of the particle changes as 
\EQ
m^2 \rightarrow m_1^2 = m^2 + \Sigma(m_1^2) \,=\,m^2 - (c^1_{K,\bar K})^2 \frac{1}{4 \,\pi\,M^2}
\,\,\,.
\EN 
Notice that this correction is always negative, i.e. the presence of the kink tends to decrease the 
value of the mass of the neutral particle, initially expressed by the curvature of the potential. If 
it happens that, by varying the coupling constant, the second term exceeds the first, there is an 
imaginary value in the mass $m_1$. This implies that the particle disappears from the stable part of 
the spectrum. Considering the coefficient $c^1_{K,\bar K}$ as fixed, this occurs for sufficiently large 
values of the coupling constant. Viceversa, decreasing the coupling, the mass of the kink becomes 
larger and the correction gets consequently smaller, making the mass of the neutral particle closer 
to its perturbative value. 

\subsection{Watson's equation and the $S$-matrix}

As a final topic of this section, we now consider the issue of the Watson's equation satisfied by the 
Form Factor. This concerns, in particular, the interesting possibility of extracting the $S$-matrix of 
the kinks, at least in certain regimes of rapidity and coupling constant. The basic idea 
behind the Watson's equation is simply the completeness of the asymptotic states: given a matrix 
element of a local operator 
${\cal G}$ on a given {\em in} state $\mid n \rangle_{in}$, i.e. $F_{in}^{\cal G} \,=\,
\langle 0 \mid {\cal G}(0)\,\mid n\,\rangle_{in}$, one can employ the completeness relation of 
the {\em out} states to get the following relation\footnote{This is a scheleton form of the 
Watson's equation: 
the matrix elements depend on the momenta of the particles and therefore the sum of the intermediate states 
stays also for a multiple integral on these variables.}
\begin{eqnarray}
&& F_{in}^{(n)} \,=\,\langle 0 \mid {\cal G} \mid n \rangle_{in} 
\,=\,\sum_{m} \langle 0 \mid {\cal G} \mid m \rangle_{out} \,
{}_{out}\langle m \mid n \rangle_{in} \nonumber \\
&& \sum_{m} \langle 0 \mid {\cal G} \mid m \rangle_{out} \,S_{n \rightarrow m} \,=\,
\sum_{m} S_{n \rightarrow m} \,F_{out}^{m} \,\,\,.
\label{Watson}
\end{eqnarray}
In this equation $S_{n \rightarrow m}$ is the $S$-matrix amplitude relative to the scattering of the $n$ 
initial particles into $m$. Let's suppose now that the momenta of the incoming particles are so small 
that it is impossible to open higher inelastic channels. Moreover, let's assume that are also absent 
``decay'' processes in lower mass particle states. If this kinematical regime exists, then the 
remaining non-zero scattering amplitudes in (\ref{Watson}) are nothing else but elastic, so that 
the Watson's equations become similar to those employed in the integrable models \cite{ffstructure}, i.e.  
\EQ
\langle 0 \mid {\cal G}(0) \mid n \rangle_{in} \,\simeq \,S_{n \rightarrow n} \,
\langle 0 \mid {\cal G}(0) \mid n \rangle_{out} \,\,\,. 
\EN 
In this case, from the ratio of the $F_{in}^{(n)}/F_{out}^{(n)}$, one could get the elastic part of the 
$S$-matrix $S_{n \rightarrow n}$, an expression obviously valid only in the kinematical region below 
the lowest threshold of the $n$-particle channel. Concerning the two-body $S$-matrix, it is important 
to notice that its elastic part is a pure phase only for real values of $\theta$ below threshold. In the 
following, however, we enforce it to be a phase also for complex values of the rapidity since, anyhow, 
this is the best we can do to obtain an estimate of this quantity. 

Let's follow this suggestion to see whether it would be possible to determine the elastic part of 
the $S$ matrix of the two kink states in the $\varphi^4$ theory. First of all, we have to establish 
that there are no neutral particles $B_n$ with mass $m_n < M$ otherwise, for the non integrability 
of the theory, the ``decay'' channel $\mid K \bar K \rangle \rightarrow 
\mid B_n B_n \rangle$ will always be open, even if the kinks are at rest (here $\bar K$ denotes the anti-kink). 
Since $m_n > m_1$, it is sufficient to impose $m_1 > M$ in order to prevent such decays. This 
gives rise to the following condition on the coupling constant 
\EQ
\sin \frac{\pi \,\xi}{2} \,\geq \,\frac{1}{2} 
\,\,\,\,\,\,\,\,\,\,
i.e.
\,\,\,\,\,\,\,\,\,\,
\xi \geq \frac{1}{3} \,\,\,.
\label{conditionelasticity}
\EN 
Once we are in this range of the coupling constant, the absence of the higher mass thresholds 
is ensured by taking sufficiently small values of the rapidity difference of the two kinks. 

Notice that the Watson's equations are valid irrespectively of the operator ${\cal G}$, an important point 
on which we shall come back later. Taking for granted this insensitivity to the operator ${\cal G}$, then 
we can take the Form Factor of the field $\varphi(x)$ computed in (\ref{FFphi4}). The Form Factor 
of the {\em out} state is simply obtained by substituting in (\ref{FFphi4}) $\theta \rightarrow - \theta$.  
By the ratio of these quantities, we arrive to the putative expression 
\EQ
S_{-a,-a}^{a,a}(\theta) \,=\, S_{a,a}^{-a,-a}(\theta) \equiv {\cal S}(\theta)  
\simeq \frac{\sinh\left(\frac{(i \pi + \theta)}{\xi}\right)}
{\sinh\left(\frac{(i \pi - \theta)}{\xi}\right)}
\,\,\,.
\label{S1phi4}
\EN
If correct, this formula should describe the elastic scattering of the kinks 
\[ \mid K_{-a,a}(\theta_1) 
K_{a,-a}(\theta_2)\rangle \rightarrow \mid K_{-a,a}(\theta_2) K_{a,-a}(\theta_1)\rangle
\]
below their inelastic threshold.

\begin{figure}[h]
\hspace{8mm}
\vspace{10mm}
\psfig{figure=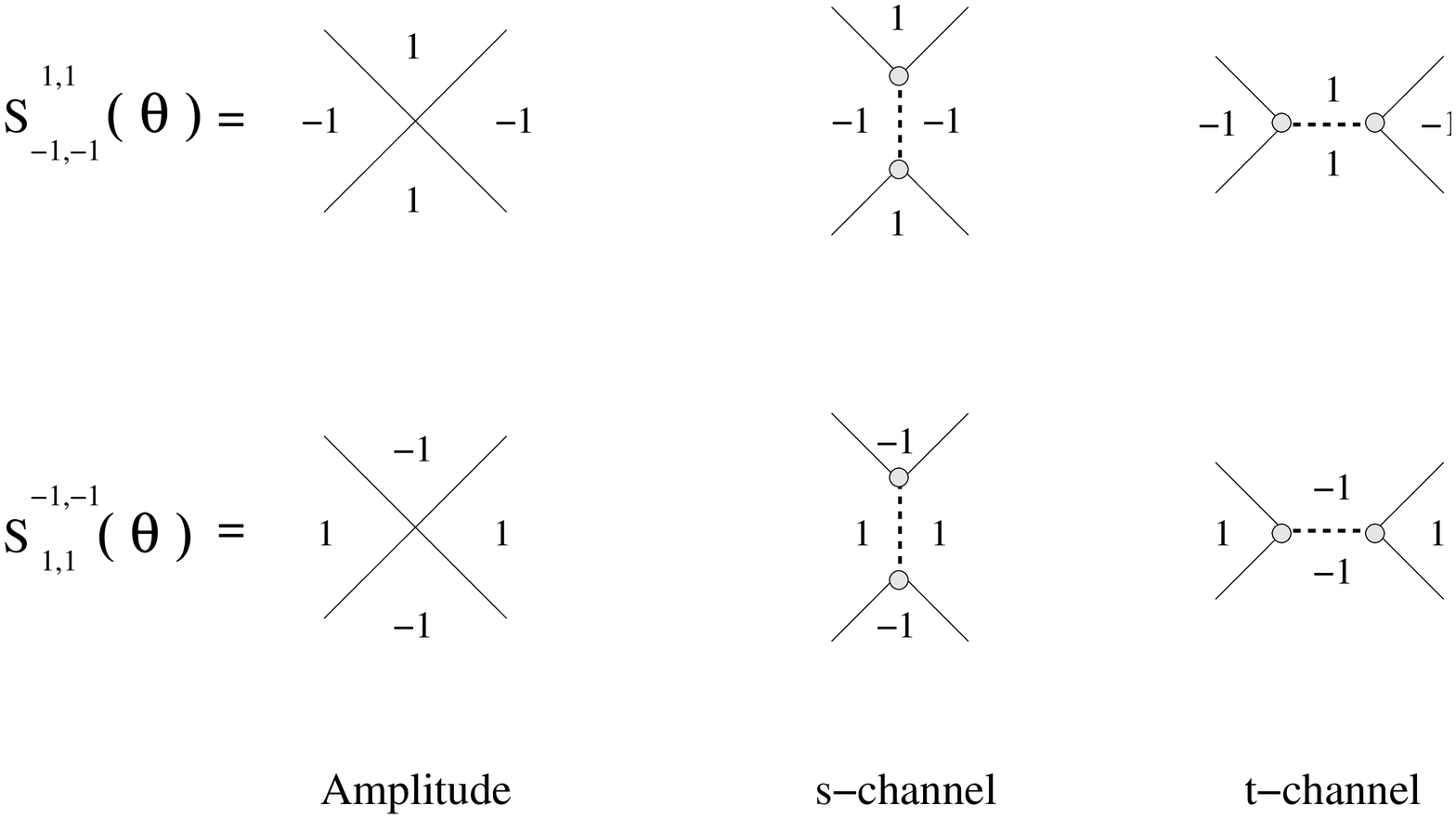,height=8cm,width=12cm}
\vspace{1mm}
\caption{{\em Elastic scattering amplitudes of the kinks and their poles in the s-channel and in the t-channel.
The dots at the vertices are the on-shell $3$-particle couplings.}}
\label{diagram}
\end{figure}

The two amplitudes of the kink scattering can be represented by the diagrams of Figure \ref{diagram}, 
where the indices on the left and on the right are those relative to the initial and final vacua 
respectively, whereas the indices on the top and on the bottom are the other vacua ``visited'' 
during the scattering\footnote{
In this particular theory, it is impossible to change the intermediate vacua in the scattering process: 
this means that the kinks of $\varphi^4$ essentially behave as ordinary particles.}.

\vspace{3mm}

The amplitudes obtained in eq.\,(\ref{S1phi4}) satisfy (by construction) the unitarity equation 
\EQ
{\cal S}(\theta) \,{\cal S}(-\theta) \,=\,1 \,\,\,. 
\label{unitarityphi4}
\EN 
However, they fail to satisfy the crossing relation which is expected for the correct $S$-matrix  
\EQ
S_{-a,-a}^{a,a}(i \pi - \theta) \,=\,S_{a,a}^{-a,-a}(\theta) 
\,\,\,.
\label{crossingphi4}
\EN  
The validity of this relation simply follows by turning of $90^0$ one of the amplitudes of 
Figure \ref{diagram} and comparing with the other. In addition to this problem, the above 
expression for $S$ has another drawback, as it becomes evident by a closer look at its analytic 
structure: even though it has all the poles of the bound states in the $s$-channel correctly 
localised at $\theta = i\pi (1 - n \xi)$, their residue 
\EQ
{\rm Res}\,\,{\cal S}(i\pi (1 - n \xi)) \,=\,- i \,\sin\left(\frac{2\pi}{\xi}\right) \,\,\,,
\EN 
is not always (imaginary) positive as, instead, it should be. In the vicinity of 
these poles the correct $S$-matrix should indeed reduce to 
\EQ
S(\theta) \,=\,i \frac{(g_{-a,a}^n)^2}{\theta -i u_{-a,a}^n} \,\,\,,   
\EN 
and, for unitary theories, the $3$-particle coupling $g_{-a,a}^n$ is real. 

To make a further progress toward the correct identification of the elastic $S$ matrix of the kinks, 
let's now explore the arbitrariness of the operator ${\cal G}(x)$ entering the Watson's equation. 
Semiclassically, for any operator which is a reasonable function $G(\varphi)$ of 
the field $\varphi(x)$, its Form Factor is given by 
\EQ
f_{ab}^{\cal G}\,=\,\langle K_{ab}(\theta_1) \mid G[\varphi(0)] \mid K_{ab}(\theta_2) \,= \,
\int_{-\infty}^{\infty} dx\,e^{i M_{ab}\,\theta \,x}\,G[\varphi_{ab}(x)] \,\,\,.
\label{FFG}
\EN 
To be defined, let's consider a class of operators expressed by power series in $\varphi$
\EQ
{\cal G} \,=\,a_1 \,\varphi + a_2 \,\varphi^2 + \cdots a_n \,\varphi^n + \cdots 
\label{seriesoperator}
\EN  
The regular part of the above Form Factor is obtained in terms of the derivative of the 
function inside the integral which, in a simplified notation, is given by 
\EQ
H(k) \,=\,\int_{-\infty}^{\infty} dx \,e^{i k x} \,\,\frac{dG}{d\varphi} \,\,\left(\frac{d\varphi}{dx}\right) 
\,\,\,,
\EN 
where $\varphi$ denotes here the kink solution. It is easy to see that $H(k)$ has always the same 
poles of the Fourier transform of $\left(\frac{d\varphi}{dx}\right)$ (just using the previous argument 
on the asympotic behavior of the integrand). However, its residues can be arbitrarirly varied by 
changing the coefficients $a_n$ of the expansion (\ref{seriesoperator}). Said in another way, 
at the semiclassical level, the most general expression of the Form Factors for the $\varphi^4$ 
theory is given by 
\EQ
H(k) \,=\, k \,\sum_{n =-\infty}^{\infty} \frac{s_n}{k + 2 n} \,\,\,,
\EN 
where the coefficients $s_n$ are arbitrary numbers. By varying them, the only effect is to change the 
position of the zeros of the function $H(k)$, obviously leaving the position of its poles untouched. 

Reestablishing now the original variable $k \rightarrow (i \pi - \theta)/\xi$ and taking the ratio
of the Form Factors of an operator of the above class 
\EQ
R(\theta) \,=\,\frac{H(i\pi -\theta)}{H(i\pi + \theta)} 
\,\,\,,
\EN 
we see that the resulting function can be {\em any} function which fulfills 
the unitarity equation 
\EQ
R(\theta) R(-\theta) \,=\,1 \,\,\,,
\EN 
and which has the correct poles in the $s$-channel. Arbitrary other poles of $R(\theta)$ may come from 
the zeros of $H(i\pi + \theta)$. In conclusion, the farthest we can go in the use of the semiclassical 
form factors, is to fix the poles of the $S$-matrix in the $s$-channel alone. To find its actual expression, 
one necessarily needs to integrate this information with the others coming from the physical nature 
of the problem at hand.  

For the $\varphi^4$ theory this is easy. In fact, an $S$-matrix which must simultaneously satisfy 
the two equations 
\EQ
\begin{array}{l}
S(\beta) S(-\beta) \,=\,1 
\,\,\,;\\
S(\beta) \,=\,S(i\pi -\theta) \,\,\,,
\end{array}
\EN 
can only be expressed as a product of the elementary functions
\EQ
[ \eta ] \,=\,[ 1-\eta ] \,\equiv \,\frac{\tanh\frac{1}{2} (\theta + i \pi \eta)}
{\tanh\frac{1}{2} (\theta - i \pi \eta)} \,\,\,.
\label{elementary}
\EN 
These functions have two poles, with residues of opposite sign: one at $\theta = i \pi \eta$, 
the other at its crossing symmetric position $\theta = i\pi (1 - \eta)$. Hence, the natural 
proposal for the $S$-matrix of the kinks below their inelastic threshold is  
\EQ
S_{-a,a}^{a,a}(\theta) \,=\,(-1)^{n+1} \prod_{k=1}^n \,[ k \,\xi ] \,\,\,, 
\EN 
an expression which holds for the following values of the coupling constant 
\EQ
\frac{1}{n+1} < \xi < \frac{1}{n} \,\,\,.
\EN
By construction, it satisfies both the unitarity and the crossing symmetry equations. It has  
the right poles in correspondence with those expected from Figure \ref{diagram}. It is also easy to 
check that it has the correct positive residue at {\em all} poles in the $s$-channel (and a negative one at 
all the $t$-channel poles). 

In view of the condition (\ref{conditionelasticity}), the only physical values that $n$ can take in the 
above expression are $n=0,1,2$. Somehow surprisingly, it keeps its validity (for instance, the positivity 
of the residues at all poles) even for other values of $\xi$, where there are decay processes into lighter 
breathers. Notice that, when there are no breathers in the theory ($n = 0$), the $S$-matrix of the kinks 
simply coincides with the one of the Ising model in its low temperature phase, $S = -1$ \cite{Swieca}. 

\subsection{Finite volume}

A way to investigate the spectrum of a quantum field theory is by studying its euclidean version 
on a finite volume, say on an cylinder of width $R$ along the space direction and infinitely long in the 
euclidean time direction $\tau$. Assuming periodic boundary conditions $\varphi(0,\tau) = \varphi(R,\tau)$, 
how the spectrum $E_i(R)$ of the finite-volume Hamiltonian would look like? 

For $\varphi^4$ theory, the answer is as follows. First of all, for dimensional reasons, the energy 
levels can be cast in the form 
\EQ
E_i(R) \,=\,\frac{2\pi}{R} \, e_i(M R) \,\,\,,
\EN 
where $e_i(M R)$ are the {\em scaling functions}\,\footnote{The semiclassical expression of the scaling 
functions of $\varphi^4$ theory with anti-periodic boundary conditions has been studied in \cite{MRSD}.
An issue of a particular interest is their crossover from the conformal regime to the massive behavior.}. 
They depends on the adimensional parameter $M R$, where $M(\lambda)$ is the mass of the kink, which 
we assume to be always finite. Secondly, it is easy to foresee their behavior in two limits,  
$R\rightarrow 0$ and $R \rightarrow \infty$. 

For $R \rightarrow 0$, the theory presents a conformal invariance. The scaling functions becomes then 
$e_i(0) = (2 \Delta_i - c/12)$, where $c=1$ is the central charge of the bosonic 
theory, whereas $\Delta_i$ are the conformal dimensions of the various conformal fields present 
in this limit. In the other case, $R \rightarrow \infty$, the theory displays instead its massive behavior. 
Therefore all levels go to a multi-particle state, with a mass gap ${\cal M}$
given by the sum of the masses of the excitations entering this state. They can be either breathers 
or kinks (for periodic boundary conditions, there are only couples of kink-antikink states). 
Taking into account a possibile bulk energy term $e_0(\lambda)$, the energy levels are then 
expected to go asymptotically as 
\EQ
E_i(R) \,\simeq \, \epsilon_0 \, R + {\cal M}_i 
\,\,\,\,\,\,\,\,\,
,
\,\,\,\,\,\,\,\,\,
R \rightarrow \infty
\EN 

\begin{figure}[h]
\hspace{15mm}
\vspace{10mm}
\psfig{figure=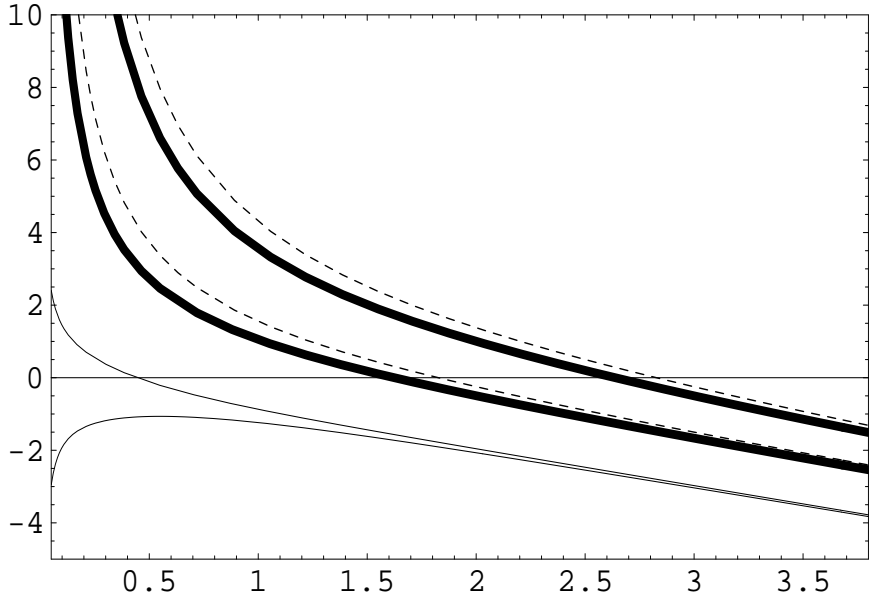,height=8cm,width=12cm}
\vspace{1mm}
\caption{{\em Energy levels of the bound states, as functions of $MR$, for the symmetric well potential in 
a finite volume with periodic boundary conditions.}}
\label{truncationphi4}
\end{figure}

At a finite volume, however, there is a {\em finite} energy barrier (order $R$) between the two vacua and  
they will be in contact each other through the tunneling of the kink states. Correspondingly, the 
energies of the vacua, together with all the energies of the excitations above them, have 
an asymptotical exponential splitting: they come in pairs and become doubly degenerate only in the 
infinite volume limit. A typical outcoming of this circumstance is shown in Figura \ref{truncationphi4}.   

To check whether the the semiclassical prediction is correct or not, it would be sufficient to study the 
movement of the energy levels by increasing $\lambda$. If the prediction of a critical value beyond which there 
are non longer bound states is correct, one should observe a progressive approach of all their energy lines toward 
the two-kink threshold $2 M$ and their disappearance into the continuum once $\lambda > \lambda_c$.  
  
\section{Asymmetric wells}\label{phi6section}

In order to have a polynomial potential with two asymmetric wells, one must necessarily employ 
higher powers than $\varphi^4$. The simplest example of such a potential is obtained with a 
polynomial of maximum power $\varphi^6$, and this is the example discussed here. Apart from its simplicity, 
the $\varphi^6$ theory is relevant for the class of universality of the Tricritical Ising Model. As 
we can see, the information available on this model will turn out to be a nice confirmation of the semiclassical 
scenario. . 

\begin{figure}[h]
\hspace{45mm}
\vspace{10mm}
\psfig{figure=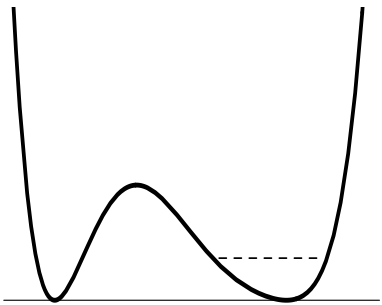,height=4cm,width=5cm}
\vspace{1mm}
\caption{{\em Example of $\varphi^6$ potential with two asymmetric wells and a bound state only on one of them.}}
\label{potential6}
\end{figure}

A class of potentials which may present two asymmetric wells is given by 
\EQ
U(\varphi) \,=\,\frac{\lambda}{2}\,\left(\varphi + a\frac{m}{\sqrt{\lambda}}\right)^2 \,
\left(\varphi - b\frac{m}{\sqrt{\lambda}}\right)^2 \,\left(\varphi^2 + c \frac{m^2}{\lambda}\right) \,\,\,, 
\label{phi6}
\EN 
with $a,b,c$ all positive numbers. To simplify the notation, it is convenient to use the 
dimensionless quantities obtained by rescaling the coordinate as $x^{\mu} \rightarrow  m x^{\mu}$ 
and the field as $\varphi(x) \rightarrow \sqrt{\lambda}/m \varphi(x)$. In this way the 
lagrangian of the model becomes 
\EQ
{\cal L} \,=\,\frac{m^6}{\lambda^2} \left[\frac{1}{2} (\partial \varphi)^2 - 
\frac{1}{2} (\varphi+a)^2 (\varphi-b)^2 
(\varphi^2 + c) \right]\,\,\,.
\label{newphi6}
\EN 
The minima of this potential are localised at $\varphi_0^{(0)} = - a$ and $\varphi_1^{(0)} = b$ and 
the corresponding ground states will be denoted by $\mid 0 \,\rangle$ and $\mid 1 \,\rangle$. The 
curvature of the potential at these points is given by 
\EQ
\begin{array}{lll}
U''(-a) & \equiv & \omega^2_0 = (a+b)^2 (a^2 + c) \,\,\,;\\
U''(b) & \equiv & \omega^2_1 = (a+b)^2 (b^2 + c)\,\,\,.
\end{array}
\label{curvature}
\EN 
For $ a \neq b$, we have two asymmetric wells, as shown in Figure \ref{potential6}. To be 
definite, let's assume that the curvature at the vacuum $\mid 0\,\rangle$ is higher than the 
one at the vacuum $\mid 1\,\rangle$, i.e. $a > b$.

\vspace{3mm}

The problem we would like to examine is whether the spectrum of the neutral particles $\mid B \,\rangle_{s}$ 
($ s = 0,1$) may be different at the two vacua, in particular, whether it would be possible that one of them 
(say $\mid 0 \rangle$) has no neutral excitations, whereas the other has just one neutral 
particle. The ordinary perturbation theory shows that both vacua has neutral excitations, although 
with different value of their mass: 
\EQ
m^{(0)} \,= \,(a+b) \sqrt{2 \, (a^2 + c)} 
\,\,\,\,\,\,\,
,
\,\,\,\,\,\,\,
m^{(1)} \,=\, (a+b) \sqrt{2 \, (b^2 + c)} 
\,\,\,.
\label{baremasses}
\EN 

\begin{figure}[h]
\hspace{30mm}
\vspace{1mm}
\psfig{figure=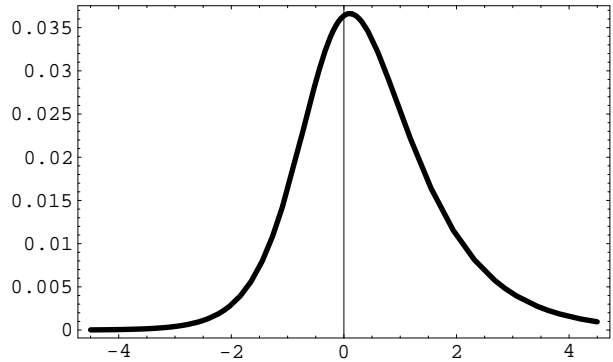,height=5cm,width=8cm}
\vspace{1mm}
\caption{{\em Typical shape of $\left(\frac{d \varphi}{dx}\right)_{01}$, obtained by a numerical solution of eq.\,
(\ref{kinkphi6}).}}
\label{asymmsol}
\end{figure}

Let's see, instead, what is the semiclassical scenario. The kink equation is given in this case by 
\EQ
\frac{d\varphi}{d x} \,=\,\pm (\varphi + a) (\varphi - b) \,\sqrt{\varphi^2 + c}\,\,\,.
\label{kinkphi6}
\EN 
We will not attempt to solve exactly this equation but we can present nevertheless its main features. 
The kink solution interpolates between the values $-a$ (at $ x = -\infty$) and $b$ (at $x = +\infty$). The 
anti-kink solution does viceversa, but with an important difference: its behaviour at $x = -\infty$ 
is different from the one of the kink. As a matter of fact, the behaviour at $x = - \infty$ of the kink 
is always equal to the behaviour at $x = +\infty$ of the anti-kink (and viceversa), but the two vacua are 
approached, in this theory, differently. This is explicitly shown in Figure \ref{asymmsol} and proved  
in the following.


Let us consider the limit $x \rightarrow - \infty$ of the kink solution. For these large values of $x$, 
we can approximate eq.\,(\ref{kinkphi6}) by substituting, in the second and in the 
third term of the right-hand side, $\varphi \simeq -a$, with the result 
\EQ
\left(\frac{d\varphi}{dx}\right)_{0,1} \simeq (\varphi + a) (a + b) \sqrt{a^2 + c} 
\,\,\,\,\,
,
\,\,\,\,\,
x \rightarrow - \infty
\EN 
This gives rise to the following exponential approach to the vacuum $\mid 0 \rangle$ 
\EQ
\varphi_{0,1}(x) \simeq -a + A \exp(\omega_0 x) 
\,\,\,\,\,\,\,
,
\,\,\,\,\,\,\,
x \rightarrow - \infty 
\EN 
where $A > 0$ is a arbitrary costant (its actual value can be fixed by properly solving the non-linear 
differential equation). To extract the behavior at $x \rightarrow -\infty$ of the anti-kink, we 
substitute this time $\varphi \simeq b$ into the first and third term of the right hand side of 
(\ref{kinkphi6}), so that 
\EQ
\left(\frac{d\varphi}{d x}\right)_{1,0} \simeq (\varphi - b) (a + b) \sqrt{b^2 + c} 
\,\,\,\,\,\,
,
\,\,\,\,\,\,
x \rightarrow - \infty
\EN 
This ends up in the following exponential approach to the vacuum $\mid 1 \rangle$ 
\EQ
\varphi_{1,0}(x) \simeq b - B \exp(\omega_1 x) 
\,\,\,\,\,\,\,\,
,
\,\,\,\,\,\,\,\,
x \rightarrow - \infty
\EN 
where $B > 0$ is another constant. Since $\omega_0 \neq \omega_1$, the asymptotic behaviour of the two solutions 
gives rise to the following poles in their Fourier transform 
\begin{eqnarray}
{\cal F}(\varphi_{0,1}) & \rightarrow & \frac{A}{\omega_0 + i k} \nonumber \\
& & \label{polephi6}\\
{\cal F}(\varphi_{1,0}) & \rightarrow & \frac{-B}{\omega_1 + i k} \nonumber 
\end{eqnarray}
In order to locate the pole in $\theta$, we shall reintroduce the correct units. Assuming to have solved  
the differential equation (\ref{kinkphi6}), the integral of its energy density gives the common mass of the kink 
and the anti-kink. In terms of the constants in front of the Lagrangian (\ref{newphi6}), its value is given by 
\EQ
M\,=\,\frac{m^5}{\lambda^2} \,\alpha \,\,\,,
\EN 
where $\alpha$ is a number (typically of order $1$), coming from the integral of the adimensional energy 
density (\ref{integralmass}). Hence, the first pole\footnote{In order to determine the others, one should 
look for the subleading exponential terms of the solutions.} of the Fourier transform of the kink and the 
antikink solution are localised at 
\begin{eqnarray}
\theta^{(0)} \,& \simeq & \,i\pi \left( 1 - \omega_0 \,\frac{m}{\pi M}\right) = i\pi 
\left(1 - \omega_0 \,\frac{\lambda^2}{\alpha m^4}\right)
\nonumber \\
& & \\
\theta^{(1)} \,& \simeq & \,i\pi \left( 1- \omega_1 \,\frac{m}{\pi M}\right) 
\,=\, i\pi \left(1 - \omega_1 \,\frac{\lambda^2}{\alpha m^4}\right) \nonumber 
\end{eqnarray}
If we now choose the coupling constant in the range 
\EQ
\frac{1}{\omega_0} < \frac{\lambda^2}{m^4} < \frac{1}{\omega_1}    \,\,\,,
\label{range}
\EN 
the first pole will be out of the physical sheet whereas the second will still remain inside it!  
Hence, the theory will have only one neutral bound state, localised at the vacuum $\mid 1 \,\rangle$. 
This result may be expressed by saying that the appearance of a bound state depends on the 
order in which the topological excitations are arranged: an antikink-kink configuration gives rise 
to a bound state whereas a kink-antikink does not. 

Finally, notice that the value of the adimensional coupling constant can be chosen so that the mass 
of the bound state around the vacuum $\mid 1 \,\rangle$ becomes equal to mass of the kink. 
This happens when 
\EQ
\frac{\lambda^2}{m^4} \,=\,\frac{\alpha}{3 \omega_1} \,\,\,.
\EN  

Strange as it may appear, the semiclassical scenario is well confirmed by an explicit example. 
This is provided by the exact scattering theory of the Tricritical Ising Model perturbed by its sub-leading 
magnetization. Firstly discovered through a numerical analysis of the spectrum of this model \cite{LMC}, its 
scattering theory has been discussed later in \cite{CKM}. It involves several amplitudes but, for our purposes, 
it is enough to focus only on those given below. With the same meaning of the diagrams as in the 
previous section, their exact expression is given by 

\vspace{10mm}

\begin{picture}(190,40)
\thicklines
\put(60,0){\line(1,1){30}}
\put(60,30){\line(1,-1){30}}
\put(60,15){\makebox(0,0){$0$}}
\put(90,15){\makebox(0,0){$0$}}
\put(75,0){\makebox(0,0){$1$}}
\put(75,30){\makebox(0,0){$1$}}
\put(100,15){\makebox(0,0)[l]{$\displaystyle{\hs =\hs S_{00}^{11}(\theta)
\hs =\hs \frac{i}{2}\hs 
S_0(\theta) 
\hs\sinh\left(\frac{9}{5} \theta -i \frac{\pi}{5}\right)}$}}
\end{picture}
\vspace{10mm}

\begin{picture}(190,40)
\thicklines
\put(60,0){\line(1,1){30}}
\put(60,30){\line(1,-1){30}}
\put(60,15){\makebox(0,0){$1$}}
\put(90,15){\makebox(0,0){$1$}}
\put(75,0){\makebox(0,0){$0$}}
\put(75,30){\makebox(0,0){$0$}}
\put(100,15){\makebox(0,0)[l]{$\displaystyle{\hs =\hs S_{11}^{00}(\theta)
\hs =\hs -\frac{i}{2}\hs S_0(\theta) \hs 
\frac{\sin\left(\frac{\pi}{5}\right)}{\sin\left(\frac{2\pi}{5}\right)} 
\hs\sinh\left(\frac{9}{5} \theta +i \frac{2\pi}{5}\right)}$}}
\end{picture}

\vspace{10mm}

\noindent
The function $S_0(\theta)$ ensures the unitarity condition of the whole set of amplitudes and 
it is given by 
\begin{eqnarray}
&& S_0(\theta) \,=\, -\frac{ w\left(\theta,-\frac{1}{5}\right)
w\left(\theta,\frac{1}{10}\right)
w\left(\theta,\frac{3}{10}\right)
t\left(\theta,\frac{2}{9}\right)
t\left(\theta,-\frac{8}{9}\right) t\left(\theta,\frac{7}{9}\right) 
t\left(\theta,-\frac{1}{9}\right)}
{\sinh\frac{9}{10}(\theta-i\pi)\,\sinh\frac{9}{10}\left(\theta-\frac{2\pi i}{3}\right)}
\nonumber\\
\label{SO}
\end{eqnarray}
where 
\[
w(\theta,x)=\frac{\sinh\left(\frac{9}{10} \theta +i\pi x\right)}
{\sinh\left(\frac{9}{10} \theta -i\pi x\right)}\hs\hs\hs\hs\hs ;
\hs\hs\hs\hs
t(\theta,x)=\frac{\sinh\hs \frac{1}{2}(\theta +i\pi x)}
{\sinh\hs \frac{1}{2}(\theta -i\pi x)} \hs\, .
\]
The structure of poles and zeros of the $S$-matrix of this problem is quite rich but, 
on the physical sheet, $0\leq {\rm Im} \hs\theta\leq i \pi$, the only poles of the $S$-matrix are 
located at $\theta=\frac{2\pi i}{3} $ and $\theta=\frac{i\pi}{3}$.

\vspace{3mm}

\begin{figure}[h]
\hspace{5mm}
\vspace{10mm}
\psfig{figure=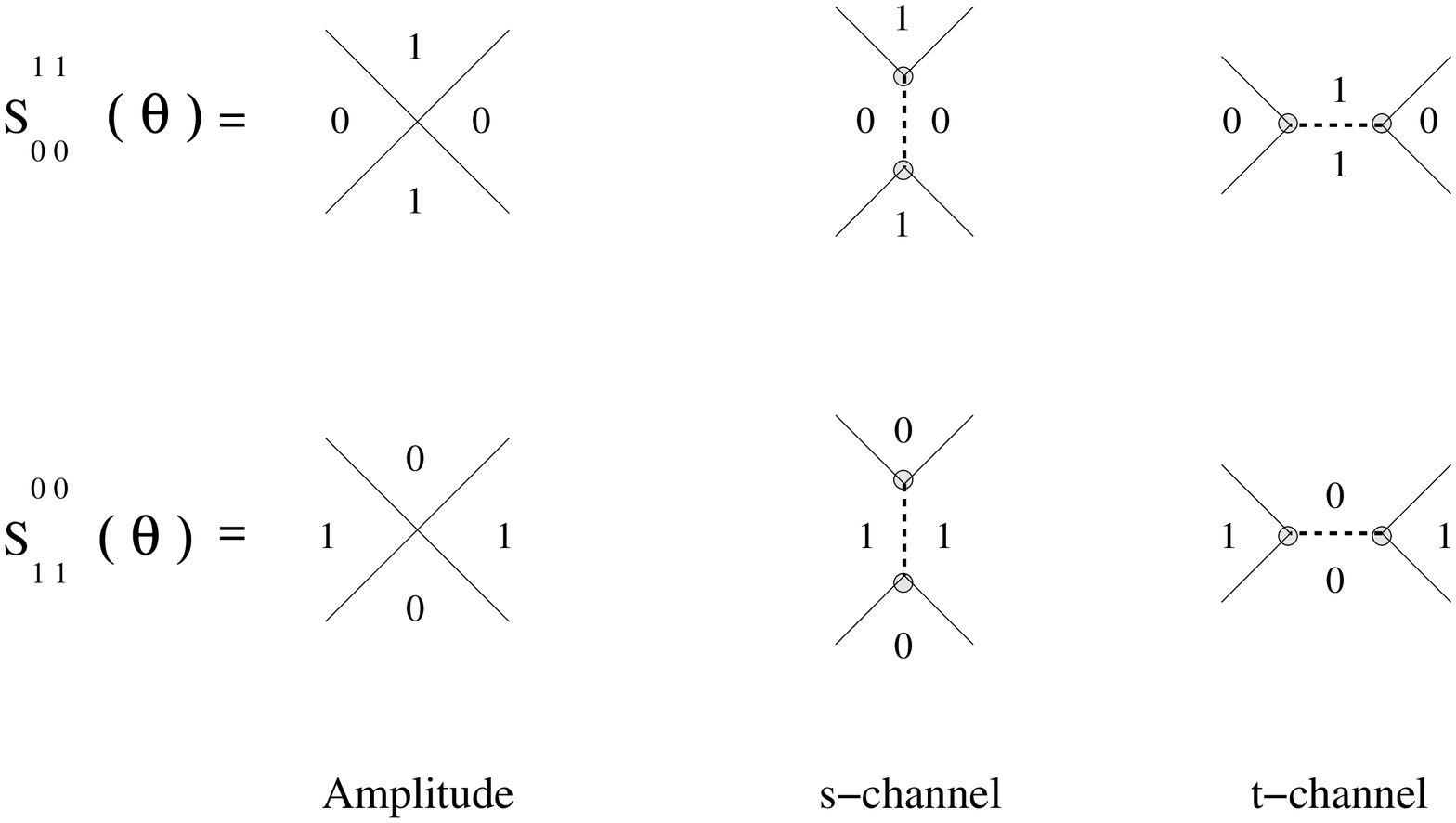,height=8cm,width=12cm}
\vspace{1mm}
\caption{{\em Elastic scattering amplitudes of the kinks in an asymmetric wells potential 
and their intermediate states in the s-channel and in the t-channel.}}
\label{diagramtim}
\end{figure}
\noindent
The first pole corresponds to 
a bound state in the $s$-channel whereas the second one is the singularity due to the particle 
exchanged in the crossed $t$-channel. However, the residues of the two amplitudes at 
$\theta=\frac{2\pi i}{3}$ are quite different! In fact, for the first amplitude we have  
\EQ
\mbox{Res}_{\theta=\frac{2\pi i}{3}} \hs\hs S_{00}^{11}(\theta) = 0 
\hs; 
\EN 
while for the second 
\EQ
\mbox{Res}_{\theta=\frac{2\pi i}{3}} \hs
\hs S_{11}^{00}(\theta) = i \hs 
\frac{s\left(\frac{2}{5}\right)}{s\left(\frac{1}{5}\right)} 
\hs \omega \hs ,
\EN 
where 
\EQ
\omega =\frac{5}{9} \hs \frac{
s\left(\frac{1}{5}\right)
s\left(\frac{1}{10}\right)
s\left(\frac{4}{9}\right)
s\left(\frac{1}{9}\right)
s^2\left(\frac{5}{18}\right)}
{s\left(\frac{3}{10}\right)
s\left(\frac{1}{18}\right)
s\left(\frac{7}{18}\right)
s^2\left(\frac{2}{9}\right)} \hs\hs , 
\EN
($s(x)\equiv \sin(\pi x)$). Hence, in the $s$-channel of the amplitude $S_{00}^{11}$, there is no 
bound state related to the vacuum $\mid 0 \,\rangle$: its only singularity comes from the 
bound state on the vacuum $\mid 1 \,\rangle $, exchanged in the $t$-channel. In the amplitude 
$S_{11}^{00}$ the situation is reverted (the two amplitudes are related by crossing): there is the 
$s$-channel singularity due to the bound state present on the vacuum $\mid 1 \,\rangle$ while the 
one of the $t$-channel is absent. This is easily seen in Figure \ref{diagramtim}, where the 
original amplitudes are streched along the vertical direction ($s$-channel) and along the horizontal 
one ($t$-channel). 

\begin{figure}[h]
\hspace{25mm}
\vspace{10mm}
\psfig{figure=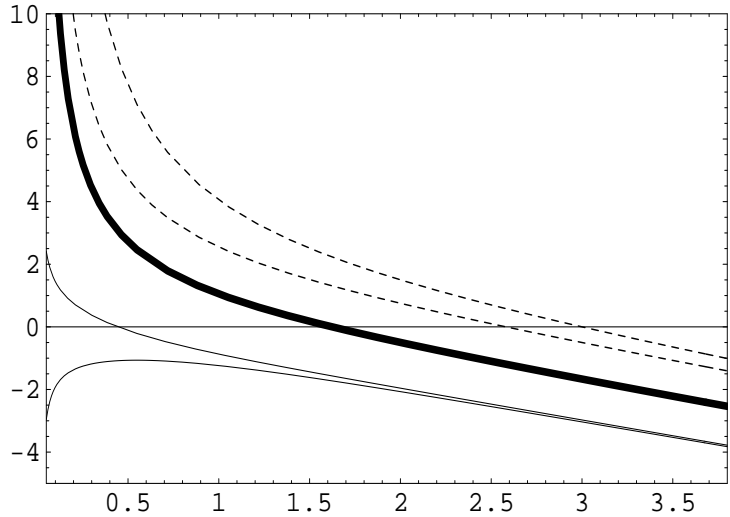,height=6.9cm,width=10cm}
\vspace{1mm}
\caption{{\em Energy levels of the asymmetric well potential in a finite volume with periodic boundary 
conditions.}}
\label{truncation}
\end{figure}

A simple way to check the above scenario would be to study the finite volume energy spectrum with 
periodic boundary conditions for the field $\varphi(x,\tau)$ at the edge of the cylinder of 
width $R$, as was done, in fact, for the Tricritical Ising Model in \cite{LMC}. At a finite $R$ 
the energies of the two vacua are exponentially splitted through the tunnelling process of the kinks. 
However, this does not occur for their excitation. If we consider,  for simplicity, the case of only 
one bound state, its energy level is a single, isolated curve placed between the vacua energy and 
the threshold lines (Figura \ref{truncation}). This situation has to be contrasted with the one 
shown in Figura \ref{truncationphi4}, related to a potential with two symmetric wells.

\section{Sine-Gordon model in the semiclassical limit}\label{SGsection}

In this section we will use the exact solution of the Sine-Gordon model in order to check 
the semiclassical approximation and to learn some important lessons from this comparison. The 
potential is given in this case by 
\EQ
U(\varphi) \,=\,\frac{m^2}{\beta^2} (1 - \cos \beta\varphi) \,\,\,.
\label{SGpotential}
\EN 
There is an infinite number of minima, $\varphi_n^{(0)} = 2 \pi n/$, which correspond to the 
quantum vacua $\mid n \,\rangle$. For their equivalence, one can choose to study the 
excitations on one of them, say the vacuum $\mid 0 \,\rangle$. 

\subsection{Exact scattering theory}

\begin{figure}[t]
\hspace{5mm}
\vspace{10mm}
\psfig{figure=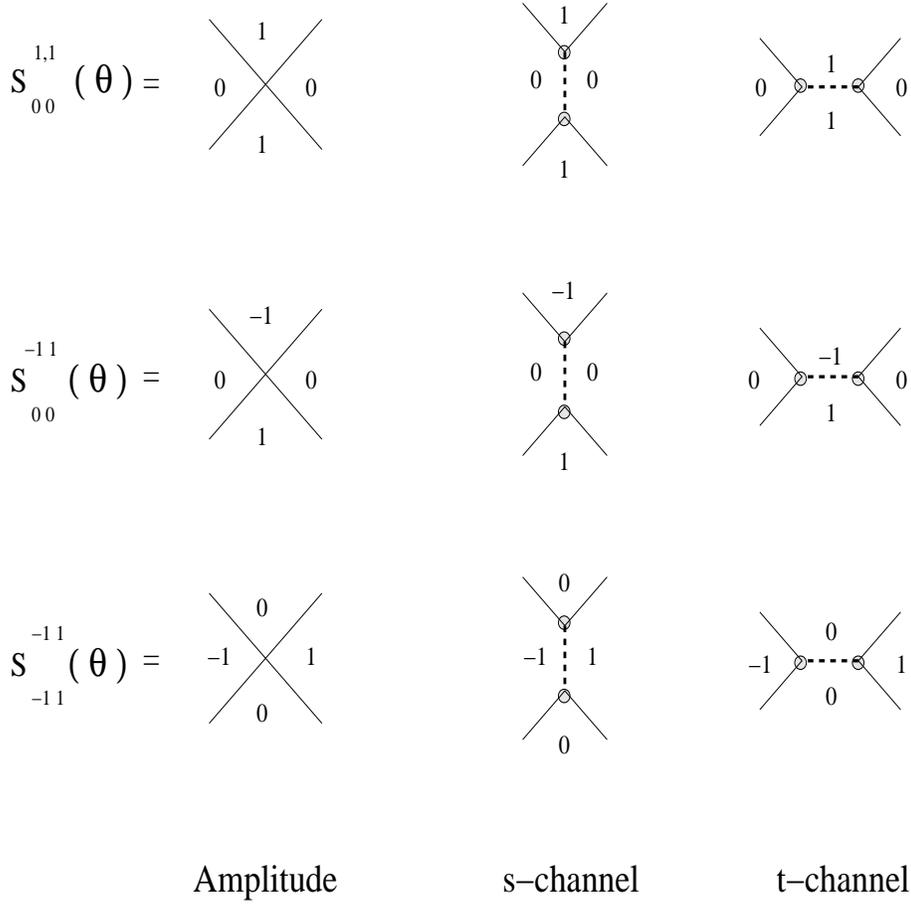,height=12cm,width=12cm}
\vspace{1mm}
\caption{{\em Elastic scattering amplitudes of the kinks and their poles in the s-channel and in the t-channel.}}
\label{diagramSG}
\end{figure}

The exact scattering theory of this model has been discussed in \cite{zamzam} and it will 
be briefly summarised below. The kink of the Sine-Gordon model interpolate between the 
vacuum $\mid 0\,\rangle$ and its neighbouring ones $\mid \pm 1 \, \rangle$, and their 
scattering processes are described by the amplitudes 
\begin{eqnarray}
&& \mid K_{0,a}(\theta_1) K_{a,0}(\theta_2) \rangle  \,=\, S_{00}^{aa}(\theta) \,\mid K_{0,a}(\theta_2) 
K_{a,0}(\theta_1) \rangle + S_{00}^{a,-a}(\theta) \,\mid K_{0,-a}(\theta_2) K_{-a,0}(\theta_1) \rangle 
\nonumber\\
&& \mid K_{a,0}(\theta_1) K_{0,-a}(\theta_2) \rangle  \,=\, S_{a,-a}^{00}(\theta) \,
\mid K_{a,0}(\theta_2) K_{0,-a}(\theta_1)\rangle \label{scatteringSG} 
\end{eqnarray}
with $a = \pm 1$. In the neutral kink-antikink channel, the amplitude $S_{00}^{aa}(\theta) = S_R(\theta)$ 
describes their reflection process while $S_{00}^{a,-a}(\theta) = S_T(\theta)$ describes their transmission. 
In the kink-kink scattering there is only the transmission amplitude $S_{a,-a}^{00}(\theta) = S(\theta)$. 
The reason of this terminology stays in the identification of the states, due to the equivalence of the 
various vacua: for instance, the kink $\mid K_{-1,0} \rangle$ must be identified with $\mid K_{0,1} \rangle$ 
and similar identification can be also established for the others. The above amplitudes, represented as 
in Figure \ref{diagramSG}, satisfy the unitarity equations 
\EQ
\begin{array}{l}
S_R(\theta) \, S_R(-\theta) + S_T(\theta) \, S_T(-\theta) \,=\, 1 \,\,\,;\\
S_R(\theta) \, S_T(-\theta) + S_T(\theta) \, S_R(-\theta) \,=\, 0 \,\,\,;\\
S(\theta) \, S(-\theta) \,=\,1 \,\,\,,
\end{array}
\label{unitaritySG}
\EN 
and the crossing symmetry relations 
\EQ
S_R(i\pi - \theta) \,=\,S_R(\theta) 
\,\,\,\,\,\,\,\,\,
;
\,\,\,\,\,\,\,\,\,
S_T(i\pi - \theta) \,=\,S(\theta) 
\,\,\,.
\label{crossingSG}
\EN    
Their closed solution is given by \cite{zamzam}
\EQ
S_T(\theta)  \,=\, \frac{\sinh\frac{\pi}{\xi} \theta}{\sinh\frac{\pi}{\xi}(i\pi - \theta)}\,
\,S(\theta) 
\,\,\,\,\,\,\,\,\,\,
;
\,\,\,\,\,\,\,\,\,\,
S_R(\theta) \,=\,\frac{i \sin\frac{\pi^2}{\xi}}{\sinh\frac{\pi}{\xi}(i\pi -\theta)} \,\,S(\theta) \,\,\,,
\EN 
where $\xi$ is the so-called renormalised coupling constant  
\EQ
\xi \,=\,\frac{\beta^2}{8\pi} \,\frac{1}{1-\frac{\beta^2}{8\pi}} \,\,\,, 
\label{xiSG}
\EN 
while the amplitude $S(\theta)$ is given by 
\EQ
S(\theta) \,=\,-\exp\left[-i \int_0^{\infty} \frac{dt}{t} \frac{\sinh\frac{t}{2}(\pi-\xi)}
{\sinh\frac{\xi t}{2} \,\,\cosh\frac{\pi t}{2}} \,\,\sin\theta t \right] 
\,\,\,.
\EN 
The pole of $S(\theta)$ are all outside the physical sheet, as it can be read from its equivalent 
infinite-product representation 
\begin{eqnarray}
S(\theta) & \,=\,& \prod_{n=0}^{\infty}
\frac{\Gamma\left((n+1) \frac{\xi}{2\pi} + i \frac{\theta}{2 \pi}\right)
\,
\Gamma\left(\frac{1}{2} +(n+1) \frac{\xi}{2\pi} - i \frac{\theta}{2 \pi}\right)
} 
{\Gamma\left((n+1) \frac{\xi}{2\pi} - i \frac{\theta}{2 \pi}\right)
\,
\Gamma\left(\frac{1}{2} + (n+1) \frac{\xi}{2\pi} + i \frac{\theta}{2 \pi}\right)} \nonumber \\
& & \,\,\,\times  
\frac{\Gamma\left(1 + n \frac{\xi}{2\pi} + i \frac{\theta}{2 \pi}\right)
\,
\Gamma\left(\frac{1}{2} + n \frac{\xi}{2\pi} - i \frac{\theta}{2 \pi}\right)
}
{
\Gamma\left(1+ n \frac{\xi}{2\pi} - i \frac{\theta}{2 \pi}\right)
\,
\Gamma\left(\frac{1}{2} + n \frac{\xi}{2\pi} + i \frac{\theta}{2 \pi}\right)
}\,\,\,.
\label{infiniteproduct}
\end{eqnarray}

\noindent
Hence, the bound states can be obtained from the poles inside the physical strip of the amplitudes 
$S_R$ and $S_T$. Since they are located at 
\EQ
\theta = i \pi (1 - k \,\xi)  
\,\,\,\,\,\,\,
,
\,\,\,\,\,\,\,
k=1,2,\ldots 
\label{poleSG}
\EN  
$\xi$ must be less than $1$ in order to have a bound state. This leads to the critical 
value of the coupling constant given by    
\EQ
\beta^2_c = 4 \pi \,\,\,, 
\EN 
beyond which, there are no bound states. For $\beta < \beta_c$, their number $N$ is given by 
$N = \left[\frac{\pi}{\xi}\right]$. Calling $M$ the mass of the kink, their mass is expressed 
as  
\EQ
m_n \,=\,2 M \,\sin\left(n \,\frac{\xi}{2}\right) 
\,\,\,\,\,\,\,\,
,
\,\,\,\,\,\,\,\,
n = 1,2,\ldots N 
\label{massSG}
\EN 
Obviously, the stability of the particles with mass $m_n > 2 m_1$ is ensured in this case by the integrability of 
the model. Concerning the residues of the amplitude $S_R$ and $S_T$ in the $s$-channel, they are alternatevely 
equal or opposite, depending on $n$ 
\EQ
{\rm Res}\,\,S_R[i\pi(1 - n \xi)] \,=\,(-1)^n \,{\rm Res}\,\,S_T[i\pi (1  - n \xi)] \,=\,
(-1)^{n+1} \frac{\xi}{\pi} \,\sin\frac{\pi^2}{\xi} \, S[i \pi( 1 -  n \xi)] \,\,\,.
\label{residueSG}
\EN 
To appreciate the meaning of this result, let's introduce the combinations which diagonalise 
the $S$-matrix and which display, in this case, also the charge conjugation symmetry of the kinks 
\begin{eqnarray}
&& S^{(+)}(\theta) \,=\,S_R(\theta) + S_T(\theta) \,=\,\frac{\sinh\frac{\pi}{2 \xi}(i \pi + \theta)}
{\sinh\frac{\pi}{2 \xi}(i \pi - \theta)} \,\,S(\theta) \,\,\,,\nonumber \\
&& \label{SplusSminus}\\
&& S^{(-)}(\theta) \,=\, \,S_R(\theta) - S_T(\theta) \,=\,\frac{\cosh\frac{\pi}{2 \xi}(i \pi + \theta)}
{\cosh\frac{\pi}{2 \xi}(i \pi - \theta)} \,\,S(\theta) \,\,\,.\nonumber 
\end{eqnarray} 
In view of (\ref{residueSG}), the bound states with $n$ odd appear as poles only in $S^{(-)}(\theta)$ 
whereas those with $n$ even appear as poles only in $S^{(+)}(\theta)$. The two sets couple, respectively, 
to the combination of the kink states 
\[
\mid K_{0,1} K_{0,1} \rangle \,\,\pm \,\, \mid K_{0,-1} K_{-1,0} \rangle \,\,\,.
\]
The above result permits to check explicitly the non-degeneracy of each mass level (\ref{massSG}). 
As a matter of fact, the equal or opposite value of the residues of $S_R$ and $S_T$ is a general 
feature, which is valid each time that a vacuum state is in communication with two symmetric neighbouring 
vacua. It is worth spending few words on its derivation and on its possible generalization. 

\vspace{5mm}

\begin{figure}[ht]
\hspace{38mm}
\vspace{10mm}
\psfig{figure=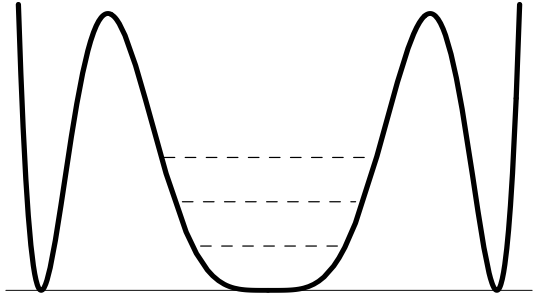,height=4cm,width=7cm}
\vspace{1mm}
\caption{{\em Portion of a potential with three neighbouring vacua, two of them symmetrically placed 
with respect to the central one, which supports a certain number of bound states. }}
\label{3wells}
\end{figure}

\subsection{Residue relations}

Consider a potential which has, locally, a situation like the one shown in Figure \ref{3wells}.
The vacuum in the middle, which we denote by $\mid 0 \,\rangle $, exchanges kinks with the two neighbouring 
ones, here labelled by $\mid \pm 1 \,\rangle$. The latter vacua are equal between each other, but not 
necessarily equal to the central one. This means that the kinks  $\mid K_{01} \rangle$ and $\mid K_{0,-1} \rangle$ 
(as well as their anti-kinks) have the same masses, and that $\mid K_{01} K_{01} \rangle$ has 
the same properties of $\mid K_{0,-1} K_{-1,0} \rangle$. In this situation, even if the theory may be 
not-integrable, one can define an (elastic) reflection and transmission amplitudes for the scattering 
of the above states, just as in the case of Sine-Gordon. Obviously, in the non-integrable case, 
this picture will only be valid in the kinematical region below their inelastic threshold (and also 
with the lowest mass of the bound states higher than the mass of the kink themselves). 

\vspace{3mm}

\begin{figure}[h]
\hspace{15mm}
\vspace{10mm}
\psfig{figure=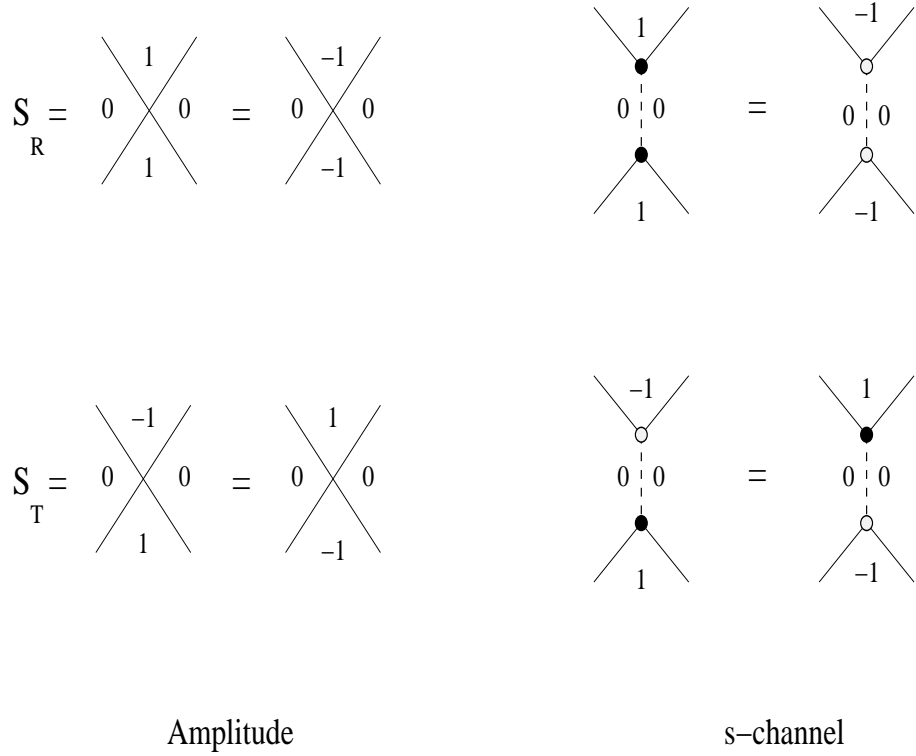,height=10cm,width=12cm}
\vspace{1mm}
\caption{{\em Reflection and transmission amplitudes for the kinks $\mid K_{0 a} K_{a 0} \rangle$ ($a = \pm 1$). 
The black dot represents $g_+^{(n)}$ while the white one $g_-^{(n)}$.}}
\label{3welldiagram}
\end{figure}

Suppose now that the vacuum $\mid 0 \,\rangle$ has certain bound states $\mid B_n\, \rangle_0$. 
Let's call $g_+^{(n)}$ and $g_-^{(n)}$ the coupling to the $n$-th bound state to the kinks  
$\mid K_{01} K_{10}\rangle$ and $\mid K_{0,-1} K_{-1,0} \rangle$, respectively. The bound states 
show up in the poles of both $S_R$ and $S_T$.  Streching the corresponding amplitudes along the $s$-channel, 
the residue in $S_T$ is proportional to $(g_-^{(n)} g_+^{(n)})$, whereas the residue in $S_R$ is 
proportional either to $(g_+^{(n)})^2$ or to $(g_-^{(n)})^2$, depending on which amplitude 
one is looking at (Figure \ref{3welldiagram}). However, for the equivalence of the two vacua $\mid \pm 1\,\rangle$, 
the last two quantities must be equal. Hence 
\EQ
g_+^{(n)} \,=\,\pm \, g_-^{(n)} \,\,\,,
\label{importantrelation}
\EN 
i.e. the residues of the $S_R$ and $S_T$ are either equal or opposite. Correspondingly, at the pole, 
the even, or the odd combination of the amplitudes, becomes a one-dimensional projector: 
the neutral particle state is not degenerate, as we already know. 

\begin{figure}[ht]
\hspace{38mm}
\vspace{10mm}
\psfig{figure=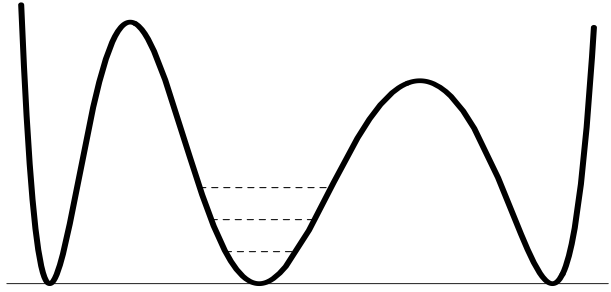,height=3.9cm,width=8cm}
\vspace{1mm}
\caption{{\em Portion of a potential with three neighbouring vacua of different shape but with kinks of equal mass 
interpolating between them.}}
\label{def3wells}
\end{figure}

The reasoning can be further generalised in the case of three neighbouring vacua, one different from the other, 
but with kinks of the same masses. This is not impossible: suppose, in fact, that we deform the potential of 
Figure \ref{3welldiagram} but always keeping equal the integrals (\ref{integralmass}) of the 
right and left kink. Notice that, by using the equation (\ref{kinkequation}), $\epsilon_{ab}(x)$ can be 
equivalently expressed as 
\[
\epsilon_{ab}(x) \,=\,\left\{\begin{array}{l}
\left(\frac{d \varphi_{ab}}{d x}\right)^2 \,\,\,,\\
U(\varphi_{ab}) \,\,\,.
\end{array}
\right.
\]
So, we are simply looking for a potential which supports two kink configurations $\varphi_{0,\pm 1}(x)$ that, 
although different, satisfy however
\EQ
\int_{-\infty}^{+\infty} \left(\frac{d \varphi_{01}}{d x}\right)^2 \,dx \,=\,\int_{-\infty}^{+\infty}
\left(\frac{d \varphi_{0,-1}}{d x}\right)^2 \, dx \,\,\,.
\EN 
There is, of course, no mathematical obstacles in fulfilling this request\footnote{Such a potential can be 
found, for instance, by employing the deformation procedure described in \cite{Bazeia}.}. Concerning the shape 
of the potential, it may look like the one shown in Figura \ref{def3wells}.  
 
\vspace{5mm}

Under this deformation, the masses of the bound states may change but their values can never cross, 
for the non-degeneracy of these particle. The kinks, on the contrary, are still degenerate and therefore  
they can mix each other under the scattering processes. Due to their asymmetric shape, the elastic part 
of their scattering is described, in this case, by $3$ amplitudes. Using the short notation 
$K_{\pm}$ and $\overline K_{\pm}$ to denote the left and right kink/antikinks, we have 
\EQ
\begin{array}{l}
\mid \overline K_-(\theta_1)  K_{-}(\theta_2) \rangle\, \,=\, S_1(\theta) \, 
\mid \overline K_-(\theta_2)  K_-(\theta_1) \rangle  \, + \, 
S_2(\theta) \, \mid K_{+}(\theta_2) \overline K_{+}(\theta_1) \rangle  \,\,\,,\\
\\
 \mid K_+(\theta_1) \overline K_{+}(\theta_2) \rangle\, \,=\, S_2(\theta) \, 
\mid \overline K_-(\theta_2)  K_-(\theta_1) \rangle  \, + \, 
S_3(\theta) \, \mid K_{+}(\theta_2) \overline K_{+}(\theta_1) \rangle  \,\,\,. 
\end{array}
\EN
Calling as before $g_-^{(n)}$ and $g_+^{(n)}$ the on-shell coupling of the left and right kinks to the 
bound states, the residue $r_2$ of $S_2$ will be proportional to $(g_-^{(n)} g_+^{(n)})$, whereas those of 
$S_1$ and $S_3$ ($r_1$ and $r_3$) will be proportional to $(g_-^{(n)})^2$ and $(g_+^{(n)})^2$, respectively. 
However, the residues $r_1$ and $r_3$ are not equal in this case, although they satisfy the relation 
\EQ
r_1 \, r_3 \,=\,(r_2)^2 \,\,\,.
\label{residuerelation}
\EN 
Diagonalising the above $S$-matrix, we have two different phase shifts 
\EQ
e^{i \delta_{\pm}(\theta)} \,=\,\frac{1}{2} \left(S_1 + S_3 \pm \sqrt{(S_1 - S_3)^2 + 4 (S_2)^2} \right) 
\,\,\,.
\EN 
Accordingly to the relative sign of $g_+^{(n)}$ and $g_-^{(n)}$, one of the two will have a vanishing 
residue of the pole of the bound state, as it is easily checked by using eq.\,(\ref{residuerelation}). 
The other amplitude will act, then, as a one-dimensional projector.  

An explicit example of an asymmetric 3-vacua kink scattering is provided, for instance, by the 
following set of functions 
\begin{eqnarray}
S_1(\theta) & \,=\,& \frac{i}{2} \,S_0(\theta) \,
\frac{\sin\left(\frac{\pi}{5}\right)}{\sin\left(\frac{2\pi}{5}\right)}\,
\sinh\left(\frac{9}{5}\theta - \frac{2 \pi}{5} i \right) \,\,\,,\nonumber \\ 
S_2(\theta) & \,=\,& - \,\frac{i}{2} \,S_0(\theta) \, \left(\frac{\sin\left(\frac{\pi}{5}\right)}
{\sin\left(\frac{2\pi}{5}\right)}\right)^{1/2} \,\sinh\frac{9}{5}\theta \,\,\,,\\
S_3(\theta) & \,=\,& - \,\frac{i}{2} \,S_0(\theta) \,
\frac{\sin\left(\frac{\pi}{5}\right)}{\sin\left(\frac{2\pi}{5}\right)}\,
\sinh\left(\frac{9}{5}\theta + \frac{2 \pi}{5} i \right) \,\,\,,\nonumber 
\end{eqnarray}
where $S_0(\theta)$ is the same of eq.\,(\ref{SO}) and ensures the unitarity condition of these amplitudes. It is 
easy to see that, at $\theta = 2 \pi i/3$, the residues of $S_1$ and $S_2$ are different, although they 
satisfy the relation (\ref{residuerelation}).

\subsection{Semiclassical analysis} 

Before proceeding with the analysis of the classical kink configurations of the Sine-Gordon theory, 
let's underline one fact that will be important later. Namely, if we take the limit $\xi \rightarrow 0$ 
in the exact scattering amplitudes, the $S$-matrix develops an essential singularity in $\xi$. Moreover, 
it is no longer a meromorphic function of $\theta$. The term responsable for this singular behavior is 
$S_0(\theta)$, with the breaking of the analiticity of this function due to the accumulation of its 
infinite number of poles and zeros when $\xi \rightarrow 0$. In this limit, its explicit expression 
can be obtained by using 
the infinite product representation (\ref{infiniteproduct}). It reads as follows \cite{zamzam,JW,Korepin}
\EQ
S(\theta) \,=\,\exp\left[\frac{\pi}{\xi}\,\int_0^{\pi} \log\left[\frac{e^{\theta - i \eta} +1}{e^{\theta} + 
e^{-i \eta}}\right]\,d\eta \right] 
\,\,\,\,\,\,\,\,\,
,
\,\,\,\,\,\,\,\,\,
\xi \rightarrow 0 \,\,\,.
\label{Sgoestozero}
\EN

Turning now the attention to the semi-classical data of the kinks, the solutions of (\ref{kinkequation}) 
are given by 
\begin{equation}\label{SGkinkinf}
\phi_{a,a \pm 1}(x)\,=\,\pm \frac{4}{\beta}\arctan(e^{mx})\,\,\,,
\end{equation}
with the usual identification of the vacua, modulo $2\pi/\beta$. Since the kink and 
anti-kink solutions are equal (up to a sign), the semiclassical spectrum on $\mid 0 \,\rangle$ 
and on its neighrouring vacua are the same, as it is expected from their equivalence. The classical 
mass of the soliton is given by $\frac{8 m}{\beta^{2}}$. Including its first quantum corrections, 
one has 
\EQ
M\,=\,\frac{8 \,m}{\beta^2} - \frac{m}{\pi} \,=\,\frac{m}{\pi \xi} \,\,\,.
\label{masskinkSG}
\EN 

\vspace{3mm}
\noindent
Let's now compute the semiclassical Form Factor of an even and an odd operator. For the even one, we 
can choose, for instance, $G_+ = {\cal A}_+\,\cos [\beta \varphi(x)]$ 
\begin{eqnarray}
f^{(+)}(\theta) & \,=\,& {\cal A}_+\,\langle K_{01}(\theta_1) \mid \cos[\beta\varphi(0)] \mid K_{01} \rangle \,=\,
{\cal A}_+ \,\int_{-\infty}^{\infty} dx \,e^{i M \,\theta\,x}\,\cos[\beta\varphi_{01}(x)] \nonumber \\
& = & - \,\pi \,{\cal A}_+\,\left(\frac{M}{m}\right) \,\,\frac{\frac{\theta}{2}}{\sinh\frac{\theta}{2 \,\xi}} 
\,\,\,,
\end{eqnarray}
while for the odd one, $G_- = {\cal A}_-\,\sin[\beta\varphi(x)]$
\begin{eqnarray}
f^{(-)}(\theta) & \,=\,& {\cal A}_-\, \langle K_{01}(\theta_1) \mid \sin[\beta\varphi(0)] \mid K_{01} \rangle \,=\,
{\cal A}_- \,\int_{-\infty}^{\infty} dx \,e^{i M \,\theta\,x}\,\sin[\beta\varphi_{01}(x)] \nonumber \\
& = & - i \,\pi \,{\cal A}_-\,\left(\frac{M}{m}\right) \,\,\frac{\frac{\theta}{2}}{\cosh\frac{\theta}{2 \,\xi}} 
\,\,\,.
\end{eqnarray}  
In the final expressions we have isolated the term $\theta/2$. The reason comes from a close comparison 
with the exact expression of the Form Factors of the above operators, which is available in the literature 
\cite{ffstructure}. By choosing a proper normalization ${\cal A}_{\pm}$ of the operators, they can be 
written as 
\begin{eqnarray} 
f^{(+)}(\theta) \,=\,\frac{\theta}{2} \,\frac{1}{\sinh\frac{\theta}{2 \,\xi}}
&\,\,\, \longleftrightarrow \,\,\, & 
f_{\rm{exact}}^{(+)}(\theta)\,=\,\sinh\frac{\theta}{2} \,\frac{1}{\sinh\frac{\theta}{2 \, \xi}} \,
F_{\rm{min}}(\theta) \nonumber \\
& & \\
f^{(-)}(\theta) \,=\,\frac{\theta}{2} \,\frac{1}{\cosh\frac{\theta}{2 \,\xi}}
& \,\,\, \longleftrightarrow \,\,\, & 
f_{\rm{exact}}^{(+)}(\theta)\,=\,\sinh\frac{\theta}{2} \,\frac{1}{\cosh\frac{\theta}{2 \, \xi}} \,
F_{\rm{min}}(\theta) \nonumber 
\end{eqnarray}
where 
\EQ
F_{\rm {min}}(\theta) \,=\,\exp\left[\int_0^{\infty} \frac{dt}{t}\,
\frac{\sinh\frac{t}{2}(1-\xi)}{\sinh\frac{\xi t}{2}\,\cosh\frac{t}{2}}
\,\frac{\sin^2\frac{\theta t}{2 \pi}}{\sinh t} \,\right] \,\,\,.
\EN 
We see that, a part of the function $F_{\rm min}(\theta)$, on which we are going to comment soon, the 
remaining expressions can be made equal by substituting $\theta/2 \rightarrow \sinh\theta/2$, an operation 
which is definetly permitted within the semiclassical approximation. 

\begin{figure}[h]
\hspace{18mm}
\vspace{0mm}
\psfig{figure=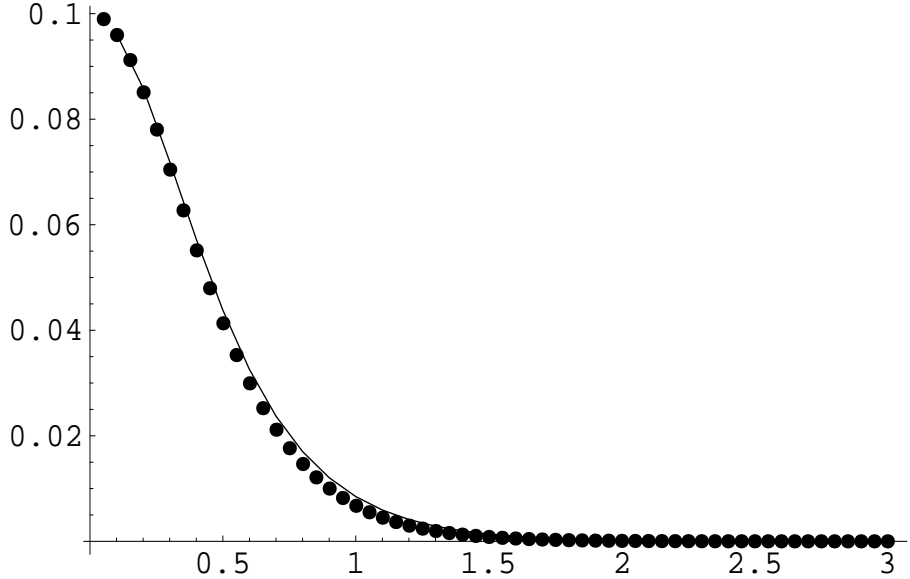,height=8cm,width=12cm}
\vspace{1mm}
\caption{{\em Comparison at $\xi =0.3$, between the exact form factor of $\cos\beta \varphi(x)$ (continous line) 
and its semiclassical expression (dotted line).}}
\label{comparison}
\end{figure}

From a numerical point of view, as far as $\xi < 1$, the above substitution is really harmless. In fact, 
at fixed $\xi$, the function $F_{\rm min}(\theta)$ asympotically goes as 
\EQ
F_{\rm min}(\theta) \,\simeq \,\exp\left[\frac{\theta}{4} \,\frac{(1 - \xi)}{\xi}\,\right] 
\,\,\,\,\,\,\,
,
\,\,\,\,\,\,\,
\theta \rightarrow \infty
\EN 
and, therefore, for $\xi < 1$, the ratio $F_{\rm min}(\theta)/\sinh\frac{\theta}{2\xi}$ becomes negligible 
before one has the possibility to appreciate the difference between the term $\theta/2$ and $\sinh(\theta/2)$: 
plotting together the semiclassical and the exact expressions, one can hardly distinguish them on the 
entire infinite range of $\theta$ (Figure \ref{comparison}).

\vspace{5mm}

One may have noticed that the form factors $f^{(\pm)}(\theta)$ of the operators 
$\cos\beta(\varphi)$ and $\cos\beta(\varphi)$ has a finite value at $\theta = 0$. The reason 
is that they are local fields with respect to the kinks. The same is true for all integer harmonics 
$\cos(n \beta\varphi(x))$ and $\sin(n \beta\varphi(x))$. In fact, the semilocal index $\gamma_{\alpha}$ 
of the exponential operator $e^{i\alpha \varphi(x)}$ with respect to the kink is \cite{DM}
\EQ
\gamma_{\alpha} \,=\,\frac{\alpha}{\beta} \,\,\,. 
\label{semilocal}
\EN 
For the form factor $g^{\alpha}(\theta) = \langle K(\theta_1) \mid e^{i \alpha \varphi(0)} \mid K(\theta_2)$ 
of these operators, $\gamma$ is the quantity that rules their residue at $\theta = 0$   
\EQ
{\rm Res}_{\theta = 0} \,g^{\alpha}(\theta) \,=\,(1 - e^{2 \pi i \gamma_{\alpha}})\,\,\langle 0 \mid e^{i \alpha 
\varphi(0)} \mid 0 \rangle \,\,\,.
\label{residuenonl}
\EN  
However, taking the operators $\cos\frac{\beta}{2} \varphi(x)$ and $\sin\frac{\beta}{2} \varphi$, it is easy 
to see that the form factor of the first operator should have a pole at $\theta = 0$, whereas the other should not. 
This is, indeed, well confirmed by the semiclassical expression of these quantities. For the first, we have in fact 
\begin{eqnarray}
f^{(+)}_{\frac{1}{2}}(\theta) & \,=\,& \langle K_{01}(\theta_1) \mid \cos\left[\frac{\beta}{2}\varphi(0)\right] 
\mid K_{01} \rangle \,=\,
\,\int_{-\infty}^{\infty} dx \,e^{i M \,\theta\,x}\,\cos\left[\frac{\beta}{2}\varphi_{01}(x)\right] \nonumber \\
& = & - i\, \pi\,\,\frac{M}{m} \,\frac{1}{\sinh\frac{\pi \,M \,\theta}{2 m}} \,=\,
- \,\frac{i}{\xi} \,\,\frac{1}{\sinh\frac{\theta}{2 \,\xi}} 
\,\,\,, 
\end{eqnarray}
while for the second 
\begin{eqnarray}
f^{(-)}_{\frac{1}{2}}(\theta) & \,=\,& \langle K_{01}(\theta_1) \mid \sin\left[\frac{\beta}{2}\varphi(0)\right] 
\mid K_{01} \rangle \,=\, \,\int_{-\infty}^{\infty} dx \,e^{i M \,\theta\,x}\,\sin\left[\frac{\beta}{2}
\varphi_{01}(x) \right] \nonumber \\
& = &  \,\pi\,\,\frac{M}{m} \,\frac{1}{\cosh\frac{\pi \,M \,\theta}{2 m}} \,=\,
\,\frac{1}{\xi} \,\,\frac{1}{\cosh\frac{\theta}{2 \,\xi}} 
\,\,\,.
\end{eqnarray}

Concerning the bound states, making the analytic continuation $\theta \rightarrow i \pi - \theta$, 
it is easy to see that  one obtains the exact spectrum of 
the bound states (\ref{massSG}), with $n$ even, from the poles of the form factors of the even operators, 
as for instance $F^{(+)}(\theta) = f^{(+)}(i \pi - \theta)$,. Viceversa, from the poles of the odd operator 
form factors, like $F^{(-)}(\theta) = f^{(-)}(i \pi - \theta)$, one obtains the exact mass of the particles 
with $n$ odd. This happens because the neutral bound states are eigenvectors of the charge conjugation of the model, 
with eigenvalues $(-1)^n$ and, therefore, the two sets couple only to operators with the same parity. 

Let's now discuss the issue of the $S$-matrix. Taking, for instance, the ratios of the 
semiclassical form factors of the operators $\cos\frac{\beta}{2} \varphi$ and $\sin\frac{\beta}{2} \varphi$ 
computed $ i \pi \mp \theta$, one obtains 
\begin{eqnarray}
&& {\cal S}^{(+)} \,=\,\frac{F^{(+)}_{\frac{1}{2}}(i \pi - \theta)}{F^{(+)}_{\frac{1}{2}}(i \pi + \theta)} \,=\,
\frac{\sinh\frac{\pi}{2 \xi}(i \pi + \theta)}
{\sinh\frac{\pi}{2 \xi}(i \pi - \theta)} \,\,\,,\\
&& {\cal S}^{(-)} \,=\,\frac{F^{(-)}_{\frac{1}{2}}(i \pi - \theta)}{F^{(-)}_{\frac{1}{2}}(i \pi + \theta)} \,=\,
\frac{\cosh\frac{\pi}{2 \xi}(i \pi + \theta)}
{\cosh\frac{\pi}{2 \xi}(i \pi - \theta)} \,\,\,. \nonumber 
\label{attemptSmatrix}
\end{eqnarray}
Comparing these results with eq.\,(\ref{SplusSminus}), we see that they are remarkably close to the exact expressions 
but, nevertheless, they miss the important term $S(\beta)$. This is related to the problem, previously 
discussed, of the impossibility of recovering the correct crossing symmetry of the $S$ by using the 
semiclassical Form Factors of a generic operator. In this case, however, this problem seems to get 
even worse for the mathematical nature of the function which is missing, eq.\,(\ref{Sgoestozero}). 
In fact, this is a function with a branch cut in $\theta$, that can only be obtained by an accumulation of 
infinite number of poles and zeros. Together with its essential singularity at $\xi = 0$, its 
non-analytic behavior makes $S(\theta)$ completly invisible to the semiclassical formula (\ref{remarkable1}) 
we are using. Saying differently, the term $S(\theta)$ could only be recovered by the ratio of the 
functions $F_{\rm min}(i\pi \pm \theta)$. By the exact solution of theory \cite{ffstructure}, 
we have indeed  
\EQ
F_{\rm min}(i\pi + \theta) \,=\,S(\theta) \,F_{\rm min}(i\pi -\theta) 
\,\,\,\,\,\,\,\,
,
\,\,\,\,\,\,\,\,
F_{\rm min}(\theta) \,=\, F_{\rm min}(- \theta) \,\,\,.
\EN 
However, in order to obtain the expression (\ref{Sgoestozero}), such function should behave, 
for $\xi \rightarrow 0$, as 
\EQ
F_{\rm min}(\theta) \,=\,\exp\left[\frac{\pi}{\xi} \,\sum_{k=1}^{\infty} k\,\,\int_0^{\pi} 
d \eta \,\log\left[\frac{1 + \left(\frac{\theta}{\pi (2 k+2 - \eta)}\right)^2}
{1 + \left(\frac{\theta}{\pi ( 2 k + 2 + \eta)}\right)^2}\right]\right] \,\,\,.
\EN 
It is easy to convince oneself that the semiclassical formula (\ref{remarkable1}) can never produce such 
an expression for the form factor of any operator $G(\varphi(x))$ which is an analytic function of 
the field $\varphi(x)$. 

In the end, we would like to close this section with a remark that may lighten the bound states 
of $\varphi^4$ theory, analysed in Section \ref{phi4section}. To do so, it is convenient to rescale 
the coordinates $x^{\mu} \rightarrow m x^{\mu}$ and the field $\varphi \rightarrow \beta \varphi$ of the Sine-Gordon 
theory, so that its Lagrangian becomes 
\EQ
{\cal L} \,=\,\frac{m^2}{\beta^2} \left[\frac{1}{2} (\partial \varphi)^2 + (\cos\phi -1)\right] \,\,\,.
\label{LSG}
\EN 
Under the substitution 
\EQ
\varphi_{\pm}(x,t) \,=\,\pi \pm 4 \arctan\Phi(x,t) \,\,\,,
\EN 
the Lagrangian of the Sine-Gordon model becomes 
\EQ
{\cal L} \,=\,\frac{m^2}{\beta^2} \,\frac{1}{(1 + \Phi^2)^2} \,
\left[\frac{1}{2} (\partial \Phi)^2 - \frac{1}{8} (\Phi^2 -1)^2 \right] \,\,\,.
\EN 
The static solutions of this Lagrangian coincide with those of $\varphi^4$, once 
the coupling constants of the two theories are related as 
\EQ
\frac{\lambda}{m^2} \,=\,\frac{\beta}{2} \,\,\,. 
\EN 
By inserting in this relation the critical value of the coupling of the Sine-Gordon, 
$\beta_c = \sqrt{4 \pi}$, we may get a different estimate of the critical value of $\varphi^4$ 
theory 
\EQ
\frac{\lambda_c}{m^2} \,=\,\sqrt{\pi} \,=\, 1.77245...
\EN 
This value is of the same order of magnitude of (\ref{criticalg}), previously obtained.  
It would have been, probably, too ambitious to search a better agreement between the two values. 
After all, even though the static solutions of the two models are similar, their time dependent 
solutions are instead rather different (the Sine-Gordon has an integrable dynamics, whereas $\varphi^3$ does not).

\section{Double Sine-Gordon wells done doubly well}\label{DSGsection}

\begin{figure}[h]
\begin{tabular}{p{8cm}p{8cm}}
\psfig{figure=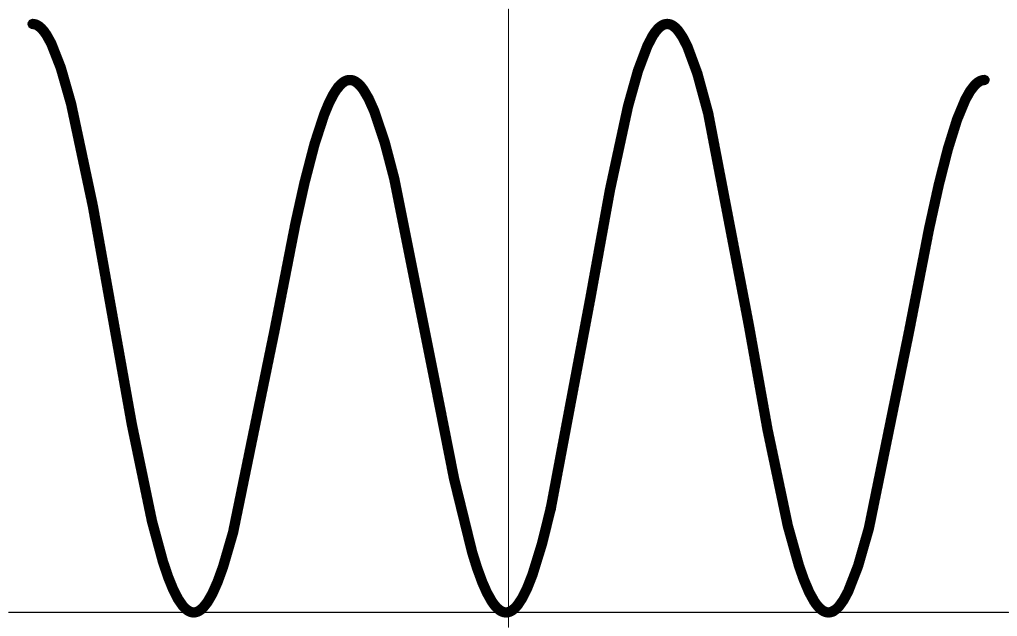,
height=4cm,width=6cm} \vspace{0.2cm}&
\psfig{figure=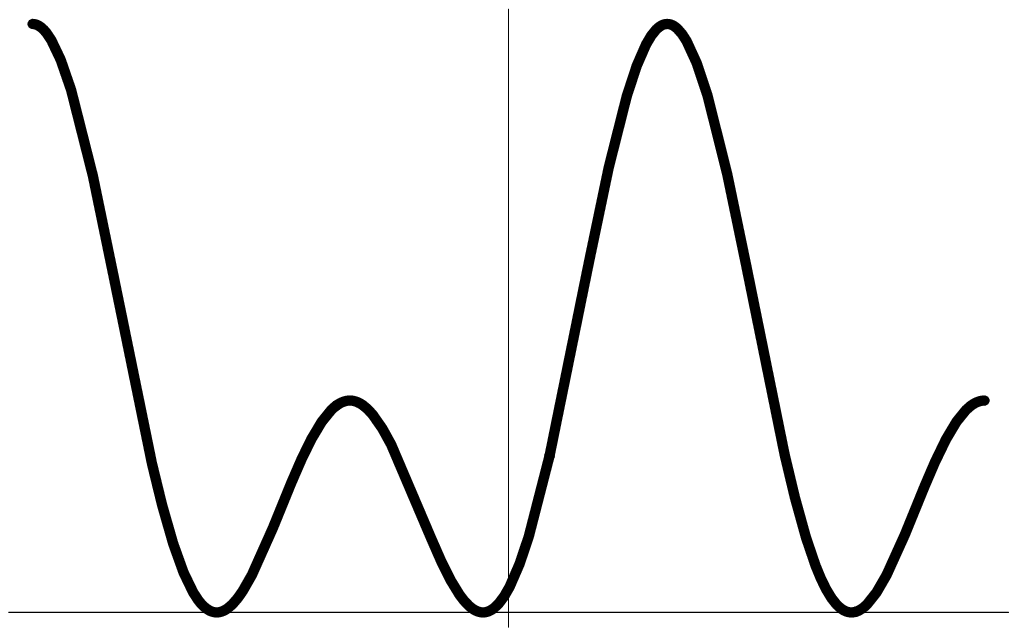,height=4cm,width=6cm}
\end{tabular}
\caption{{\em Shapes of the Double Sine Gordon potential by varying $\lambda$.}}
\label{figdeltapi2}
\end{figure}

In this section we consider the situation of a vacuum state in communication with 
two neighboring ones through kinks of different masses. A prototype of this situation 
is given by the Double Sine-Gordon model, with potential given by 
\EQ
V(\varphi) \,=\,\label{pot1}
-\frac{\mu}{\beta^{2}}\,\cos\beta\,\varphi -
\frac{\lambda}{\beta^{2}} \,\cos\left(\frac{\beta}{2}\,
\varphi+\delta\right) + C\;. 
\end{equation} 
By choosing 
\EQ
\delta \,= \,\frac{\pi}{2}
\,\,\,\,\,\,\,\,
,
\,\,\,\,\,\,\,\,
C \,=\,\frac{1}{\beta^2} \left(\mu + \frac{\lambda^2}{8 \mu}\right) \,\,\,,
\EN 
and by varying $\lambda$, the shape of the potential changes as shown in Figure \ref{figdeltapi2}. 

\footnotesize

\vspace{3mm}

\normalsize

There are two regions, qualitatively different, in the space of
parameters, the first given by $0 < \lambda < 4 \mu$ and the
second given by $\lambda > 4 \mu$. They are separated by the value
$\lambda = 4 \mu$ (where the curvature at the minima vanishes) which can be 
identified as a phase transition point \cite{DM}.

Let's focus our analysis in the coupling constant region where
$\lambda < 4 \mu$. Switching on $\lambda$, there are several effects 
on the potential: the original minima of the Sine-Gordon, located at
$\varphi_{\rm{min}} = 0,\,\frac{2\pi}{\beta}\;
(\rm{mod}\,\frac{4\pi}{\beta})$, remain degenerate but they move to
$\varphi_{\rm{min}} \,=\, -\varphi_{0},\,\frac{2\pi}{\beta} +
\varphi_{0}\;(\rm{mod}\,\frac{4\pi}{\beta})$, with
$\varphi_{0} \,=\,\frac{2}{\beta}\arcsin\frac{\lambda}{4\mu}$. The different minima 
has, however, the same curvature, given by 
\begin{equation}\label{curvaturedeltapi2}
m^{2}\,=\, \mu-\frac{1}{16}\,\frac{\lambda^{2}}{\mu} \;.
\end{equation}
In correspondance with the above shifts, there are two different types of kinks, 
one called\lq\lq large kink" and interpolating through the higher barrier
between $-\phi_{0}$ and $\frac{2\pi}{\beta} + \phi_{0}$, the other
called \lq\lq small kink" and interpolating through the lower
barrier between $-\frac{2\pi}{\beta} + \phi_{0}$ and $- \varphi_{0}$. 
Their classical expressions were explicitly given in the reference 
\cite{peyrard}
\begin{equation}\label{largekink}
\varphi_{L}(x)\,=\, \frac{\pi}{\beta}+
\frac{4}{\beta}\arctan\left[\sqrt{\frac{
4\mu+\lambda}{4\mu-\lambda}}\,\tanh\left(
\frac{m}{2}\,x\right)\right]\qquad(\rm{mod}\; \frac{4\pi}{\beta})\;,
\end{equation}
\begin{equation}\label{smallkink}
\varphi_{S}(x)\,=\,-\frac{\pi}{\beta} +
\frac{4}{\beta}\arctan\left[\sqrt{\frac{ 4\mu - \lambda}{4\mu +
\lambda}}\,\tanh\left(\frac{m}{2}
\,x\right)\right]\qquad(\rm{mod}\; \frac{4\pi}{\beta})\;.
\end{equation}
For the following considerations, we can neglect the periodic structure of 
this potential\footnote{The periodicity of the potential obviously implies that the 
following analysis applied as well to any other vacuum of this theory.} and concentrate 
our attention only on the three vacua around the origin, here denoted by $\mid 0 \,\rangle$ 
(the one near the origin) and $\mid \pm 1\,\rangle$ (the other two). Around the 
vacuum $\mid 0\,\rangle $, the admitted quantum kink states are 
\[
\mid L\,\rangle \,=\, \mid K_{0,1}\,\rangle
\,\,\,\,\,
\makebox{and}
\,\,\,\,\, 
\mid \overline S \,\rangle
\,=\, \mid K_{0,-1}\, \rangle \,\,\,,
\]
together with the corresponding antikink states 
$\mid \overline L \,\rangle =\mid K_{1,0} \,\rangle$ and
$\mid S \,\rangle = \mid K_{-1,0}\,\rangle$. The topological
charges of these kinks are different, and given by 
\EQ
\begin{array}{c}
Q_{L} = - Q_{\overline L} = 1+\frac{\beta\phi_{0}}{\pi}\;,\\
Q_{S} = -Q_{\overline S} = 1-\frac{\beta\phi_{0}}{\pi}\;. \\
\end{array}
\label{topologicalcharge} 
\EN 
The classical masses of the large and small kink get splitted when we switch 
on $\lambda$, and their exact value can be easily computed
\begin{equation}\label{largesmallkinkmass}
M_{L,S}\,=\,\frac{8\,m}{\beta^{2}}
\left\{1\pm\frac{\lambda}{\sqrt{16\,\mu^{2}-\lambda^{2}}}
\left(\frac{\pi}{2}\pm\arcsin\frac{\lambda}{4\mu}\right)\right\} \,\,\,.
\end{equation}
The expansion of this formula for small $\lambda$ is given by
\begin{equation}\label{largesmallkinkfirst}
M_{L,S}\;{\mathrel{\mathop{\kern0pt\longrightarrow}
\limits_{\lambda\to 0 }}}\;
\frac{8\sqrt{\mu}}{\beta^{2}}\pm\frac{\lambda}{\beta^{2}}\,
\frac{\pi}{\sqrt{\mu}}+O(\lambda^{2})\;,
\end{equation}
where the first term is the classical mass of the unperturbed Sine-Gordon kink. 

\subsection{The embarassment of the riches}

Let's now apply the semiclassical formula (\ref{remarkable1}) for  
obtaining the spectrum of the neutral particles at the vacuum $\mid 0\,\rangle$. 
For the form factor $F^{\varphi}_{L\bar L}(\theta) = f_{LL}^{\varphi}(i\pi -\theta)$ 
of the large kink (\ref{largekink}) (with $\lambda < 4 \mu$) we have
\begin{equation}
F^{\varphi}_{L\bar{L}}(\theta)\,=\,
i\,\frac{4\pi}{\beta}\,
\frac{1}{i\pi-\theta}\;\frac{\sinh\left[\alpha
\,\frac{M_{L}}{m}\,(i\pi-\theta)\right]}{\sinh\left[\pi\,
\frac{M_{L}}{m}\,(i\pi-\theta)\right]}\;,
\end{equation}
where
$$
\alpha\,=\,2\arctan\sqrt{\frac{4\mu+\lambda}{4\mu-\lambda}}\;,
$$
while $m$ and $M_{L}$ are given by (\ref{curvaturedeltapi2}) and
(\ref{largesmallkinkmass}), respectively. For the form factor 
$F^{\varphi}_{S\bar S}(\theta) = f_{SS}^{\varphi}(i \pi -\theta)$ of the 
small kink (\ref{smallkink}) (with $\lambda<4\mu$) we have instead 
\begin{equation}
F^{\varphi}_{\bar{S} S}(\theta)\,=\,- 
i\,\frac{4\pi}{\beta}\,
\frac{1}{i\pi-\theta}\;\frac{\sinh\left[\alpha
\,\frac{M_{S}}{m}\,(i\pi-\theta)\right]}{\sinh\left[\pi\,
\frac{M_{S}}{m}\,(i\pi-\theta)\right]}\;,
\end{equation}
where
$$
\alpha\,=\,2\arctan\sqrt{\frac{4\mu-\lambda}{4\mu+\lambda}}\;,
$$
while $m$ and $M_{S}$ are given by (\ref{curvaturedeltapi2}) and
(\ref{largesmallkinkmass}), respectively. 

\begin{figure}[h]
\begin{tabular}{p{8cm}p{8cm}}
\psfig{figure=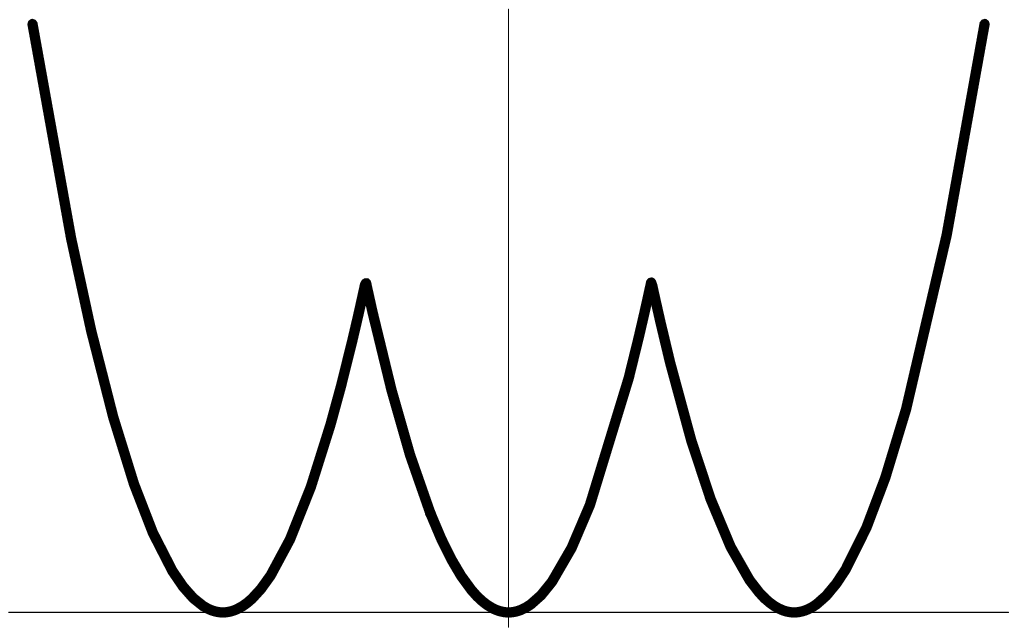,
height=4cm,width=6cm} \vspace{0.2cm}&
\psfig{figure=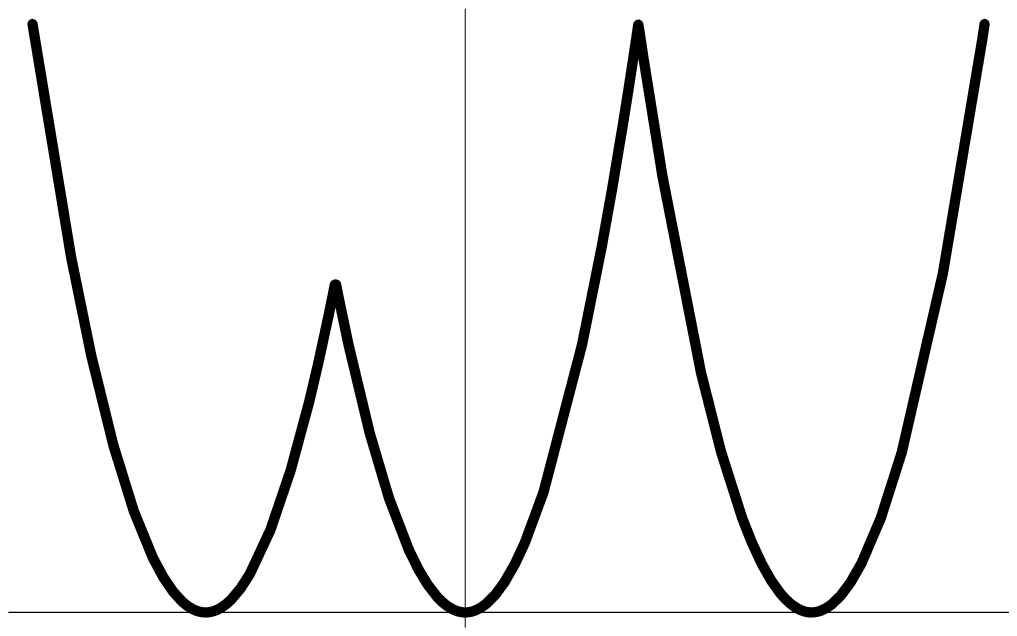,height=4cm,width=6cm}
\end{tabular}
\caption{{\em Shape of $V(\varphi)$ with $a=b$ (left hand side) and with $ a > b$ (right hand side).}}
\label{3parabole}
\end{figure}

By looking at the poles of these expressions within the physical strip, 
it seems that there are two towers of neutral particles at the vacuum 
$\mid 0 \,\rangle$: the one coming from the bound states of the $\mid L\,\bar L\,\rangle$ 
\begin{equation}
\label{largebound}
m_{(L)}^{(n)}\,=\, 2 M_{L} \sin\left(n_{L}\,\frac{m}{2M_{L}}\right)
\,\,\,\,\,\,\,
,
\,\,\,\,\,\,\,
0 < n_{L} < \pi\frac{M_{L}}{m}\;,
\end{equation}
the other coming from the bound states of $\mid \bar{S}\,S\,\rangle$
\begin{equation}\label{smallbound}
m_{(S)}^{(n)}\, =\, 2 M_{S} \sin\left(n_{S}\,\frac{m}{2M_{S}}\right)
\,\,\,\,\,\,\,
,
\,\,\,\,\,\,\,
0 < n_{S} < \pi\frac{M_{S}}{m}\;.
\end{equation} 
As a matter of fact, this situation is not peculiar of the Double Sine Gordon model but
it occurs each time that there are kinks of different masses originating from the same 
vacuum. Consider, for instance, a simplified version of a three vacua configuration, realised 
by the potential (Figure \ref{3parabole}) 
\EQ
V(\varphi) \,=\,\frac{m^2}{2}\,\left\{
\begin{array}{lll}
(\varphi + 2 b)^2 &, & \varphi \leq - b \,\,\,;\\
\varphi^2 &, & -b \leq \varphi \leq a \,\,\,; \\
(\varphi - 2 a)^2 &, & \varphi > a \,\,\,.
\end{array}
\right.
\label{3wel}
\EN

The explicit configurations of the long and short kink of this potential are pretty simple
\EQ
\varphi_L(x) \,=\,\left\{
\begin{array}{lll}
a \, e^{m x} & ,& x \leq 0 \,\,\,,\\
2 a - a \, e^{-m x} &, & x \geq 0 
\end{array} 
\right.
\,\,\,\,\,\,\,
,
\,\,\,\,\,\,\,
\varphi_{\bar{S}}(x) \,=\,\left\{
\begin{array}{lll}
-b  \, e^{m x} & ,& x \leq 0 \,\,\,,\\
-2 b + b \, e^{-m x} & , &x \geq 0 
\end{array} 
\right.
\EN
and their classical masses are 
\EQ
M_L \,=\,m\,a^2 
\,\,\,\,\,\,\,
,
\,\,\,\,\,\,\,
M_S \,=\,m\,b^2 \,\,\,.
\EN 
The form factors can be easily computed 
\begin{eqnarray}
F^{\varphi}_{L \bar{L}}(\theta) & \,=\,& \frac{i}{M_L(i \pi - \theta)} \,\left[
\frac{1}{\theta - i \pi (1 - \xi_L)} - \frac{1}{\theta - i \pi (1 + \xi_L)} \right] \,\,\,, \\
F^{\varphi}_{\bar{S} S}(\theta) & \,=\,& -\frac{i}{M_S (i \pi - \theta)} \,\left[
\frac{1}{\theta - i \pi (1 - \xi_S)} - \frac{1}{\theta - i \pi (1 + \xi_S)}\right] \,\,\,, \nonumber 
\end{eqnarray}
where 
\EQ
\xi_L \,=\,\frac{m}{\pi M_L} \,=\,\frac{1}{\pi a^2} 
\,\,\,\,\,\,
,
\,\,\,\,\,\,
\xi_S \,=\,\frac{m}{\pi M_S} \,=\,\frac{1}{\pi b^2} \,\,\,.
\label{xiLS}
\EN 
By looking at the pole in the physical strip of the above amplitudes, it seems then that 
there are {\em two} particles on the vacuum $\mid 0 \,\rangle$, whose masses are expressed by 
the formulas 
\EQ
m_L \,=\,2 m \,a^2 \,\sin\frac{\pi}{2 a^2} 
\,\,\,\,\,\,
,
\,\,\,\,\,\,
m_S \,=\,2 m \,b^2 \,\sin\frac{\pi}{2 b^2} \,\,\,. 
\label{unoedue}
\EN 
When $a = b$, the two masses coincide but, as we discussed in the introduction, the 
corresponding state cannot be degenerate. Hence, one of the two spectra in (\ref{unoedue})
has to be spurious. The same conclusion applies, as well, to the mass formulas (\ref{largebound}) and 
(\ref{smallbound}) of the Double Sine Gordon model\footnote{The argument concerning the non-degeneracy 
of the neutral states invalidates the conclusions previously reached on the spectrum of the Double Sine 
Gordon model \cite{doubleSG}, and it makes somehow obvious the finding of the numerical analysis 
on this model presented in \cite{ungheresi3}.}. But, what went wrong in this case with the semiclassical formula?

To understand the origin of this discrepancy, notice that each kink solution knows only {\em half} 
of the shape of the vacuum state from which it originates: for instance, the long kink starts 
its ``motion'' from the minimum at the origin, but its next values are determined only by the 
shape of the potential on its right. As far as this kink solution is concerned, the shape of the 
potential to the left of the origin could be arbitrarly changed without effecting the behavior of 
this solution. The same considerations also apply to the short kink, which is determined 
only by the shape of the potential on the left of the origin. 

\begin{figure}[t]
\begin{tabular}{p{8cm}p{8cm}}
\psfig{figure=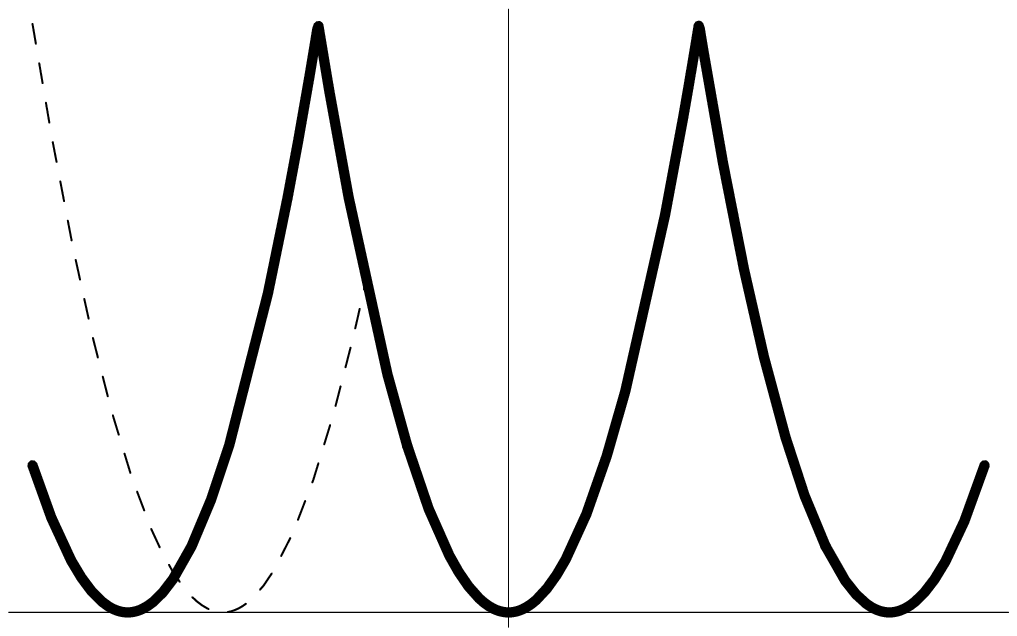,
height=4cm,width=6cm} \vspace{0.2cm}&
\psfig{figure=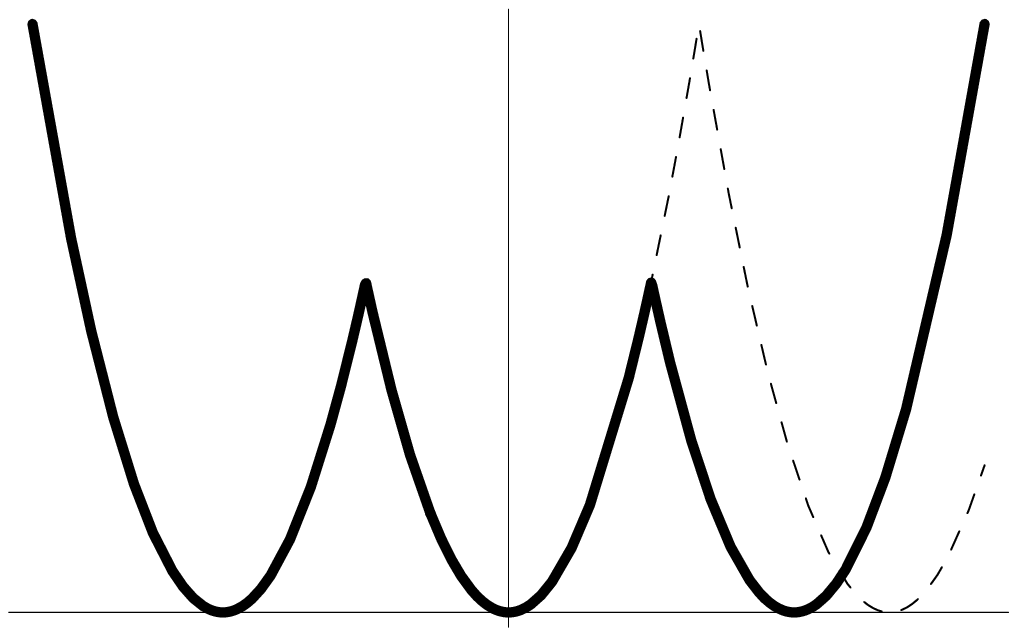,height=4cm,width=6cm} \\
\hspace{2.7cm}(A)  & \hspace{2.7cm} (B) 
\end{tabular}
\caption{{\em Fictitious potentials $V_L(\varphi)$ (A) and $V_S(\varphi)$ (B) 
The dashed line gives the values of the original potential in the regions not seen by the each kink.}}
\label{fictious}
\end{figure}

The above observation means that, when we employ the long-kink solution to extract the mass 
spectrum, it is as we are referring to a potential which is not the {\em actual} one. It is rather 
a potential $U_L(\varphi)$, whose values for $\varphi <0$ are obtained by the specular image of those 
for $\varphi > 0$ of the original potential. Viceversa, when we employ the short-kink solution, 
it is as we are referring to a potential $U_R(\varphi)$, whose values for $\varphi > 0$ are the 
specular image of those for $\varphi < 0$, which determine the short-kink solution. 
For the long and short kinks of the above example, the fictitious potentials $V_L(\varphi)$ and 
$V_S(\varphi)$ reconstructed by the semiclassical solutions are the ones shown in Figure 
\ref{fictious}. Similar fictitious potentials can be extracted, as well, for the Double Sine-Gordon 
model. Hence, no wonder that employing eq.\,(\ref{remarkable1}) in the case of kinks with different 
mass, each of them gives rise to a different spectrum of bound states on the same vacuum. 

Saying the things differently, at the leading order in the coupling constant in which 
the semiclassical form factor (\ref{remarkable1}) was computed, the short and long kink states are 
invisible each other. They start to be aware of the existence of the other only at the next leading order 
in the coupling constant. For instance, in the expression (\ref{completeset}) involving the 
long-kink, the first subleading terms are given by the matrix elements with a couple of 
short kink and antikink state, as 
\EQ
\langle L(\theta_1) \,\mid \varphi(0) \,\mid \, L \,{\bar S} S\, \rangle \,\langle L {\bar S} S \mid \,
\varphi(0) \mid \,L \rangle \ldots \langle L \mid \varphi(0) \mid L(\theta_2) \rangle\,\,\,.
\label{completeset2}
\EN
With $M_L > M_S$, these terms (as well as all the others, obtained by insering more times the 
couples ${\bar S} S$) are always present, no matter which are the values of the external rapidities 
$\theta_1$ and $\theta_2$ of the long kink. In this case, it becomes then rather artificial 
to pin down their presence by simply appealing to the perturbation expansion in the coupling constant. 
Their presence, however, spoil the possibility of obtaining a close differential equation for 
the form factors of the kink, as the one that has led to the semiclassical expression 
(\ref{remarkable1}). In principle, they can be taken in acccount by employing the 
path integral formalism discussed, for instance, in \cite{pathint} but, in practise, this can be 
a rather paintful and ferociously complicated procedure. So, for the purpose of this paper, much 
better to content ourselves with the possibility of identifying the mass spectrum according to 
the heuristic considerations of the next section.

\subsection{The importance of being small}
   
Once it has been clarified the origin of the discrepancy of the spectra coming from the two kink solutions, 
it remains to understand what is the correct spectrum of bound states. Although the exact expression of 
the mass formula has remained elusive to us, we would like to show that the spectrum can be studied, in 
a relatively simple way, at least in two different cases: (a) when the asymmetric kinks have approximately 
the same mass; (b) when the mass of one of them is much smaller than the mass of the other. 

Let's consider first the case (a). This situation can be realised starting by a symmetric configuration of 
the potential (which we assume to be an even function $V(\varphi) = V(-\varphi)$) and slightly deforming it 
by an infinitesimal deformation $\lambda \,\delta V(\varphi)$, with $\delta V(\varphi)$ odd under 
$\varphi \rightarrow -\varphi$. Switching on $\lambda$, the effect of the new term is to decrease one 
the maxima of the potential and to increase the other. This is, for instance, 
what happens in the Double Sine-Gordon for small value of $\lambda$. Under this deformation, the masses of the 
kinks changes as $M_{L,S} \simeq M \pm \lambda {\cal M}$. Denote by $\mid b\,\rangle_L$ and by 
$\mid b\,\rangle_S$ the bound state of the long and the short kinks which, in the unperturbed theory,  
have equal mass. The actual breather of the unperturbed theory is a linear combination of these two 
``half-breathers'' $\mid b\,\rangle_{L,S}$ -- a combination that can be determined by imposing that 
the state is an eigenvector of the parity transformation $P$ ($P^2 =1$). Let's assume, for instance, 
that the bound state is expressed by the combination 
\EQ
\mid B \,\rangle \,=\,\frac{\mid b\,\rangle_L - \mid b\,\rangle_R}{\sqrt{2}} \,\,\,.
\label{truebreather}
\EN 
In the unperturbed potential, its mass can be equivalently written as 
\EQ
m_B \,=\,\frac{1}{2}(m_L + m_S) \,\,\,.
\label{unperturbedmass}
\EN 
Switching now $\delta V$, at the first order in $\lambda$ the state (\ref{truebreather}) does not change. 
The masses $m_L$ and $m_R$ receive, instead, a linear correction of opposite sign, 
\EQ
\delta_{L,S} \simeq\,\pm 
\left(2 \sin\left(\frac{m}{2 M}\right) -\frac{m}{M}\,\cos\left(\frac{m}{2M}\right)\right) \,\,\,. 
\label{linear}
\EN 
Plugging these corrections into (\ref{unperturbedmass}), the mass of the breather remains then unchanged. 
This result matches with the first order Form Factor Perturbation Theory \cite{ungheresi2,ungheresi3}, as 
it can be seen by employing the parity operator $P$
\EQ
\lambda \,\langle \,B\,| \,\delta V(\varphi) \,| B\,\rangle \,=\,
\lambda \,\langle \,B \,| P \left(\,P \,\delta V(\varphi) \,P \right) \,P \,| B\,\rangle \,=\,-\lambda \,
\langle \,B\,| \,\delta V(\varphi) \,| B\,\rangle \,=\,
0 \,\,\,.
\EN   

Let's consider now the case (b), i.e. when one of the kink is much heavier than the other. 
We would like to present a series of arguments in favour of the thesis that, in this circumstance, 
the semiclassical spectrum is {\em essentially} determined by the short-kink solution, 
i.e. 
\begin{equation}\label{actual}
m_{(S)}^{(n)}\, \simeq \, 2 M_{S} \sin\left(n_{S}\,\frac{m}{2M_{S}}\right)
\,\,\,\,\,\,\,
,
\,\,\,\,\,\,\,
0 < n_{S} < \pi\frac{M_{S}}{m}\;,
\end{equation}
with the stable part of the spectrum obtained only for $n_S \leq 2$. The arguments are the following
\begin{enumerate}
\item 
The actual mass of the bound state depends on both $M_S$ and $M_L$. However, repeating the argument presented 
in Section \ref{Simplebutuseful}, the correction induced by the long kink is expected to be suppressed with 
respect to the one of the short kink by the ratio $\left(\frac{M_S}{M_L}\right)^2$. Therefore, making 
heavier the mass of the long kink, it can be forseen that its influence on the mass of the neutral particle 
should becomes less and less relevant. In particular, when $M_L \rightarrow \infty$, the long kink decouples 
from the theory and the masses of the breathers are only determined by the dynamics of the remaining short kink. 
\item 
\begin{figure}[t]
\hspace{60mm}
\vspace{10mm}
\psfig{figure=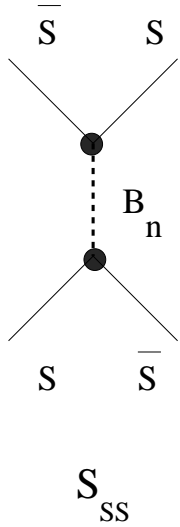,height=6.9cm,width=2.3cm}
\vspace{1mm}
\caption{{\em Elastic scattering amplitude of the short kink-antikink state.}}
\label{shor}
\end{figure}

\vspace{3mm}

Another argument in favour of the formula (\ref{actual}) directly comes from the Watson equations satisfied 
by the form factors of the short and long kink solutions. Let's first consider the scattering process of 
the small kink-antikink state $\mid {\bar S}(\hat\theta_1) S(\hat\theta_2) \,\rangle$. As far as 
we have the inequality    
\EQ
M_S \,\cosh\hat\theta < M_L \,\cosh\theta \,\,\,,
\label{inthr}
\EN 
in the center of mass (defined by $\frac{\theta}{2} \,\equiv \theta_1 = - \theta_2 $ and 
$\frac{\hat\theta}{2} \equiv \hat\theta_1 = - \hat\theta_2$), there is no possibility of converting 
the above state into a $\mid L(\theta_1) \bar{L}(\theta_{2})
\,\rangle$. Therefore the scattering process of this short kink-antikink state can only be 
elastic
\EQ
\mid {\bar S}(\hat\theta_1) \, S(\hat\theta_{2}) \,\rangle \,\,\,
\longrightarrow \,\,\,
\mid {\bar S}(\hat\theta_{1}) \, S(\hat\theta_{2}) \,\rangle \,\,\,.
\label{elas2}
\EN 
The elastic range of $\hat\theta$ obviously enlarges by making the mass of the long kink heavier. In this region, the 
Watson equation which is satisfied by any form factor of the short kink-antikink state becomes   
then 
\EQ
F_{{\bar S} S}^{\cal O}(\hat\theta) \,=\,S_{SS}(\hat\theta) \,F_{{\bar S} S}^{\cal O}(-\hat\theta) \,\,\,,
\label{WS}
\EN 
where $S_{SS}(\hat\theta)$ is the elastic $S$-matrix of the process (\ref{elas2}). Assuming 
that this elastic process leads through the exchange of the scalar particles, we have 
the diagram of Figure \ref{shor}.

Eq.\,(\ref{WS}) implies that the ratio $F_{{\bar S} S}^{\cal O}(\hat\theta)/F_{{\bar S} S}^{\cal O}
(-\hat\theta)$ is a pure phase for the real values of $\hat\theta$ below the inelastic threshold 
given by eq.\,(\ref{inthr}), perfectly in agreement with the semiclassical result. If we now trust the 
semiclassical result of the form factors also for complex values of the rapidity (in particular 
for those concerning the location of their poles), we arrive to the mass spectrum (\ref{actual}). 

Consider now the long kink-antikink scattering. First of all, notice that for the non-integrability 
of the theory, the scattering processes of the state $\mid L(\theta_1) \,\bar{L}(\theta_2)\,\rangle$ 
{\em always} involve, as a final state, $\mid {\bar S}(\hat\theta_1) \,S(\hat\theta_2),\rangle$
\EQ
\mid L(\theta_1) \,\bar{L}(\theta_2) \,\rangle \,\,\,
\longrightarrow \,\,\,
\mid {\bar S}(\hat\theta_1) \,S(\hat\theta_2) \,\rangle \,\,\,.
\label{production}
\EN 
i.e. the production process is always present. 

\vspace{3mm}

\begin{figure}[ht]
\hspace{38mm}
\vspace{10mm}
\psfig{figure=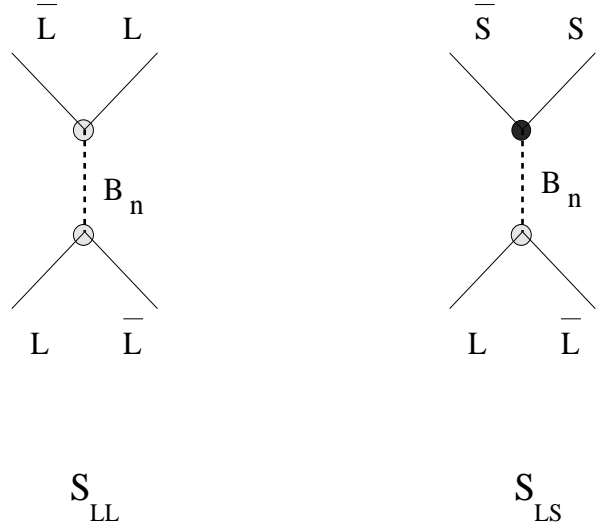,height=6.9cm,width=7.8cm}
\vspace{1mm}
\caption{{\em Elastic and production amplitudes of the long kink-antikink scatterings.}}
\label{long}
\end{figure}

Going to the center of mass, the rapidity of the outcoming small kink-antikink state is determined by 
\EQ
M_S \,\cosh\hat\theta \,=\,M_L \,\cosh\theta \,\,\,.
\EN 
In addition to the production process, in the scattering of the long kink-antikink state, 
there is also its elastic part 
\EQ
\mid L(\theta_1) \,\bar{L}(\theta_2) \,\rangle \,\,\,
\longrightarrow \,\,\,
\mid L(\theta_1) \,\bar{L}(\theta_2) \,\rangle \,\,\,.
\label{elas}
\EN 
For values of $\theta$ below other inelastic thresholds, the Watson equation satisfied 
by the form factors of the long kink-antikink state is 
then 
\EQ
F_{L \bar{L}}^{\cal O}(\theta)\,=\,S_{LL}(\theta) \,F_{L \bar{L}}^{\cal O}(-\theta) + S_{LR}(\theta) 
\,F_{\bar{S} S}^{\cal O}(-\hat\theta) \,\,\,, 
\label{WL}
\EN 
where $S_{LL}(\theta)$ is the $S$-matrix relative to the process (\ref{elas}), whereas $S_{LR}(\theta)$ 
is the one of (\ref{production}). Assuming that, also in this case, the scattering processes are 
dominated by the exchange of the scalar particles, we have the diagrams of Figure \ref{long}. 
In contrast with the short kink case, this time the ratio $F^{\cal O}_{L \bar{L}}
(\theta)/F^{\cal O}_{L \bar{L}}(-\theta)$ can never be a pure phase, not even for real values of $\theta$. 
Hence the semi-classical result cannot be the correct one: in fact, from the ratio of the long kink 
form factors, one always gets a pure phase expression. 

\vspace{3mm}

\begin{figure}[t]
\hspace{43mm}
\vspace{10mm}
\psfig{figure=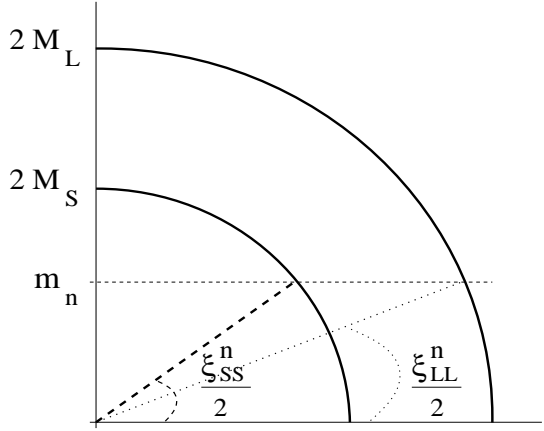,height=5.7cm,width=7.1cm}
\vspace{1mm}
\caption{{\em Relation between the resonance angles of the short and long kink-antikink state.}}
\label{refraction}
\end{figure}

Notice that in the production diagram $\mid L {\bar L}\,\rangle \rightarrow \mid {\bar S} S\,\rangle$ it 
appears the same $3$-particle coupling than in the elastic scattering $\mid {\bar S} S \,\rangle 
\rightarrow \mid {\bar S} S \,\rangle$. This implies that the intermediate particles of the two processes 
are the same. Assuming that their masses are those extracted by the short kink-antikink bound states, we 
can predict where the correct position of the poles of the long kink-antikink form factors should be: 
denoting the position of these poles by $i u_{{\bar S} S}^n = i \pi 
(1 - \xi_{{\bar S} S}^n)$, with 
$\xi_{\bar S S}^n = n \frac{m}{M_S}$, and $i u_{L {\bar L}}^n = i \pi (1 - \xi_{L {\bar L}}^n)$, 
the resonance value $\xi_{L {\bar L}}^n$ is determined by the relation 
\EQ
M_L \,\sin \left(\frac{u_{L {\bar L}}^n}{2}\right) \,=\,
M_R \,\sin \left(\frac{u_{{\bar S} S}^n}{2} \right)
\,\,\,,
\EN 
shown in Figure (\ref{refraction}). It is easy to see that this relation is similar to the law of 
refraction of light between two media of refraction indices $M_L$ and $M_R$.

\item  
In order to extract the spectrum of the neutral particles at a given vacuum $\mid a\,\rangle$, the actual 
thing to do is, of course, to quantize the time-dependent solution nearby the corresponding minimum 
$\varphi_a^{(0)}$ of the potential. To this aim, one initially needs to solve the equation of motion 
\EQ
\frac{\partial^2\varphi}{\partial t^2} - \frac{\partial^2 \varphi}{\partial x^2} \,=\,- 
\frac{dU}{d \varphi} \,\,\,,
\label{fullequation}
\EN 
by requiring the following properties of the solution:
\begin{itemize}
\item To be a periodic function in time, with a certain frequency $\omega$. 
\item To be localised around the minimum $\varphi_a^{(0)}$ of the potential.  
\item To have a finite value of its energy, given by 
\EQ
E\,=\,\int_{-\infty}^{\infty} \,\left[ \frac{1}{2} \left(\frac{\partial \varphi}{\partial t}\right)^2 + 
\frac{1}{2} \left(\frac{\partial \varphi}{\partial x}\right)^2 + U(\varphi) \right] \,dx \,\,\,.
\EN
The finiteness of the energy implies that, at $x \rightarrow \pm \infty$, the solution should necessarily
go to $\varphi_a^{(0)}$, where the potential vanishes. However, we must also require that $E \leq 2 M_S$,
which is the lowest energy threshold in the neutral sector. Without this condition, in fact, the 
asymptotical time evolution of the solution will consist of a state of small kink and small antikink, 
moving a part with respect each other, with a velocity fixed by the excess of energy with 
respect to their rest mass. This is the classical equivalence of a decay process. 
\end{itemize}
Having stated these conditions, the problem of solving eq.\,(\ref{fullequation}) is equivalent to find 
the small oscillation of a string in the funnel of the potential $U(\varphi)$ relative to the 
minimum $\varphi_a^{(0)}$, with its ends (at $x=\pm \infty$) kept fixed at the value of 
its bottom (Figure \ref{funnel}). During its swinging, the string explores and probes the 
shape of the potential nearby the valley of the minimum $\varphi_a^{(0)}$. The oscillations, however,
cannot be too wide, otherwise the motion will violate the energy condition $E < 2 M_S$. 
Notice that, in this analogy, the static kink solutions are nothing else than strings which, 
at $x = -\infty$ are at the bottom of one valley, whereas at $x = \infty$, after overpassing one of 
the peaks of the potential landscape, are in a neighbouring one. 

\vspace{5mm}

\begin{figure}[ht]
\hspace{20mm}
\vspace{10mm}
\psfig{figure=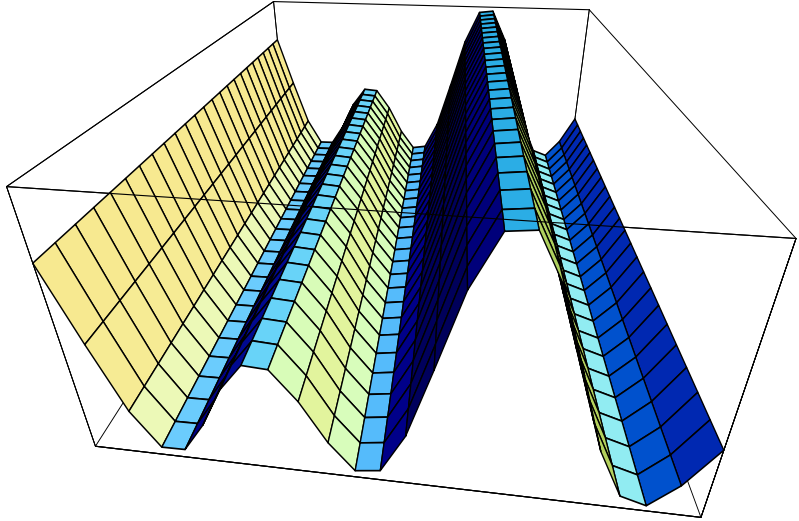,height=6.9cm,width=10.5cm}
\vspace{1mm}
\caption{{\em Landscape of the potential seen by a string oscillating in the central valley, relative to 
the minimum $\varphi_a^{(0)}$.}}
\label{funnel}
\end{figure}
In order to solve (\ref{fullequation}), let's pose $\varphi(x,t) \,=\,\varphi_a^{(0)} + 
\eta(x,t)$ and expand correspondingly the potential near this minimum 
\EQ
U(\varphi_a^{(0)} + \eta) \,=\,\frac{1}{2} \omega_0^2 \,\eta^2 + \frac{\lambda_3}{3} \,\eta^3 + \frac{\lambda_4}{4} 
\,\eta^4 + \frac{\lambda_5}{5}\,\eta^5 + \cdots \,\,\,.
\label{expansionU}
\EN 
Notice that the asymmetry of the potential with respect to its two barriers (i.e. the mere fact that there 
exist the small and the long kink) is encoded in the non-zero coefficients of the odd powers of $\eta$. 
Any smooth, real-valued solution of (\ref{fullequation}) that is periodic in time with frequency $\omega > 0$
has a Fourier representation 
\EQ
\eta(x,t) \,=\,\sum a_n(x) \,\exp(i n \omega\,t) 
\,\,\,\,\,\, 
, 
\,\,\,\,\,\,
a_{-n}(x) \,=\,a_n^*(x) \,\,\,.
\EN 
All the coefficients $a_n(x)$ must be localised, i.e. they should vanish when $x \rightarrow \pm \infty$.  
Substituting this expansion into (\ref{fullequation}) and using eq.\,(\ref{expansionU}), one reaches 
the non-linear equations for $a_n$'s
\EQ
(\partial_x^2 + n^2 \,\omega^2 - \omega_0^2) \,a_n \,=\,
-\,\lambda_3 \,\sum_m \, a_m \,a_{n-m} 
- \,\lambda_4 \,\sum_k \sum_m a_k \, a_m \,\,a_{n-k-m} + \cdots 
\label{nonlinear}
\EN 
Spatially uniform infinitesimal solution of (\ref{fullequation}) oscillate with frequency $\omega_0$, and 
it is possible to prove that (\ref{fullequation}) admits no breather solution for $\omega > \omega_0$ 
\cite{Coron}. Classically, this result is expressing the fact that the actual mass is always smaller 
than the one fixed by the curvature of the potential. Hence, one can focus on the interval 
$0 < \omega < \omega_0$. In order to 
find a small parameter $\epsilon$ to make a reasonable expansion, previous studies of this 
equation\footnote{It is interesting to observe that in \cite{Kruskal,CP}, the authors actually 
argue about the {\em non-existence}, stricly speaking, of a stable solution of eq.\,(\ref{fullequation}). 
However, the stability breaking of the solution occurs through exponentially small terms, invisible 
to any perturbative orders, which correspond to a radiation rate of the perturbative 
solution. Due to the extremely slow decay of these processes, for any practical purpose, one can 
safely ignore this mathematical subtlety, as also admitted by the same authors.} 
(see, for instance, \cite{DHN,Kruskal,CP}) suggest to use 
\EQ
\epsilon \,\equiv \,(1 - \omega^2/\omega_0^2)^{1/2} \,\,\, ,
\EN 
and to introduce a slightly modified asymptotic expansion in terms of the dimensionless variables: 
\EQ
\xi \,=\,\frac{\epsilon \,\omega_0 \,x}{\sqrt{1 + \epsilon^2}} 
\,\,\,\,\,\,
,
\,\,\,\,\,\,
\tau \,=\,\frac{\omega_0 \,t}{\sqrt{1 + \epsilon^2}} \,\,\,.
\EN 
The solution of (\ref{nonlinear}) will be expressed in terms of a series expansion in $\epsilon$ of 
all the terms $a_n$. For instance, in the case of Sine-Gordon model, the breather mode (at the leading 
order in $\epsilon$) 
is given by \cite{raj,CP}
\begin{eqnarray}
\varphi(x,t) & \,=\,& \frac{4}{\beta} \arctan\left(\frac{\epsilon \sin(\omega_0\,t/\sqrt{1+\epsilon^2})}
{\cosh(\epsilon \omega_0\,x/\sqrt{1+\epsilon^2})}\right) \,\,\,\\
& \simeq & \frac{4}{\beta} \,\textrm{sech}\xi \,\,\sin[\tau (1-\epsilon^2/2)] + \cdots 
\,\,\,.\nonumber 
\end{eqnarray}
Once such solution of (\ref{nonlinear}) has been found, the next steps will be to compute its 
classical energy $E$ and its action 
\EQ
W\,=\,\int_{-\infty}^{\infty} dx \,\int_0^{T} \left(\frac{\partial \varphi}{\partial t}\right)^2
\,dt\,\,\,,
\EN 
along one period. Referring once again to the Sine-Gordon model, the results are  
\begin{eqnarray}
E & \,=\,& 2 M \,\left(\epsilon - \frac{\epsilon^3}{3} + \cdots\right) \,\,\,,\\
W & \,=\,& \frac{4\pi M}{m} \left(\epsilon - \frac{\epsilon^3}{3} + \cdots\right) \,\,\,, 
\nonumber 
\end{eqnarray}
where $M$ is the mass of the soliton, alias the heights (per unit length) of the barriers of the 
potential landspace. Eliminating now $\epsilon$, one has 
\EQ
E \,=\,m\left(\frac{W}{2\pi} - \frac{1}{24} \left(\frac{m}{M}\right)^2 \,\left(\frac{W}{2\pi}\right)^3 + 
\cdots \right)\,\,\,.
\EN 
The familiar mass spectrum of the Sine-Gordon model is finally obtained by imposing the 
quantization condition 
\EQ
W \,=\,2\pi n \,\,\,.
\label{quantization}
\EN 
If the same calculations are repeated for $\varphi^4$ theory \cite{DHN,CP}, one finds 
\begin{eqnarray}
\varphi(x,t) & \simeq & \frac{m}{\sqrt\lambda} \left[
\frac{2\epsilon}{\sqrt{3}} \,\textrm{sech}\xi \,\cos\tau - \epsilon^2 \,\textrm{sech}^2\xi  
\right.\nonumber \\ & + & \left. 
\frac{e^2}{3} \,\textrm{sech}^2\xi \,
\cos 2\tau + \frac{\epsilon^3}{6\sqrt{3}} \,\textrm{sech}^4\xi \,\cos 3 \tau + \cdots \right] 
\,\,\,, 
\end{eqnarray}
with the classical energy and the action of the swinging string in 
the valley of this potential given by 
\begin{eqnarray}
E & \,=\,&  M \,\left(2 \epsilon + \frac{37}{27} \frac{\epsilon^3}{3} + \cdots\right) \,\,\,,\\
W & \,=\,& \frac{2\pi M}{\sqrt{2} \,m} \left(2 \epsilon + \frac{46}{27} \frac{\epsilon^3}{3} + 
\cdots \right)\nonumber 
\end{eqnarray}
As in the Sine-Gordon model, also in this case the energy is expressed in terms of the mass of 
the soliton, alias in terms of the height of the barrier of the potential. Eliminating $\epsilon$ 
and imposing the quantizion condition (\ref{quantization}), one arrives, as before, to the series 
expansion of the usual formula $m_n = 2 M \sin\left(n \frac{m}{2 M}\right)$. 

The two examples above should have made clear that, in the general case, the energy and the action 
of the classical solution will be expressed in terms of the lowest heights of the potential, a feature 
which is pretty intuitive. In the case in which these heights are the same (like the Sine-Gordon case), 
there is only one energy scale, given by the mass of the soliton. But, also in the case of $\varphi^4$, 
there is only one energy scale, given by the height of the potential between the two vacua: on the 
left (or on the right) of each of them,  there is in fact an infinite barrier.   

The same one-scale situation is expected to be valid when the potential has a barrier much higher 
than the other: in this case, the small oscillations of the string will be essentially determined by the 
barrier given by the lowest peak (the other barrier wll only induce corrections at higher order in $\epsilon$). 
Correspondingly, the spectrum of the neutral bound states should essentially coincide with the one 
extracted by the lowest kink. 

This discussion should also clarify the reason of the difficulty to find an exact expression 
of the mass spectrum when the kinks have a comparable mass. In this case, in fact, there are 
{\em two scales} in the problem, i.e. $M_L$ and $M_S$, and the energy together with the action 
of the string will be non-trivial functions thereof. To find their expression, at least in a 
concrete example, is an interesting open problem on which we hope to come back in the future. 
From a practical point of view, notice that, when $M_L$ is not much larger than $M_S$, the 
masses obtained by employing either eq.\,(\ref{largebound} or eq.\,(\ref{smallbound}) are always 
very close each other and they can provide an indication on the actual value of the masses of 
the bound states. 

Even though it seems rather difficult, in general, to determine where the crossover from the 
{\em two-scale scenario} to the {\em one-scale scenario} takes place, when $M_L/M_S > 2$ one 
should be relatively safe by taking $M_S$ as the only scale of the problem\footnote{The string 
oscillations are localised in a region order $\epsilon$. It is hard to immagine a string 
oscillating twice higher than the first barrier, whose height for unit length is 
expressed by $M_S$, still keeping its energy lower than $2 M_S$.}. 

\end{enumerate}

\section{The BLLG potentials}
\begin{figure}[h]
\begin{tabular}{p{8cm}p{8cm}}
\psfig{figure=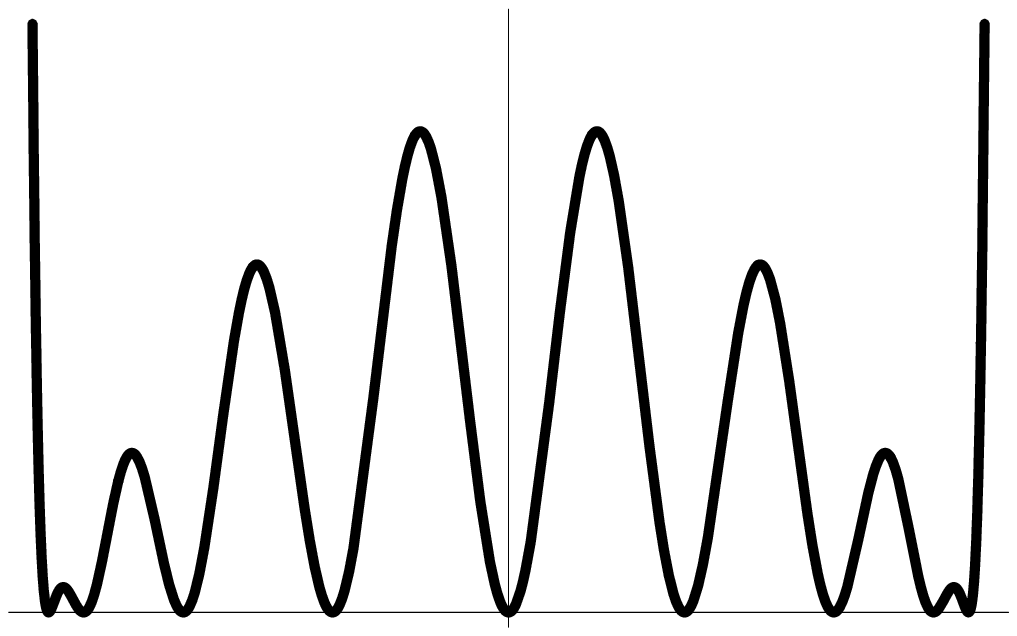,
height=4cm,width=6cm} \vspace{0.2cm}&
\psfig{figure=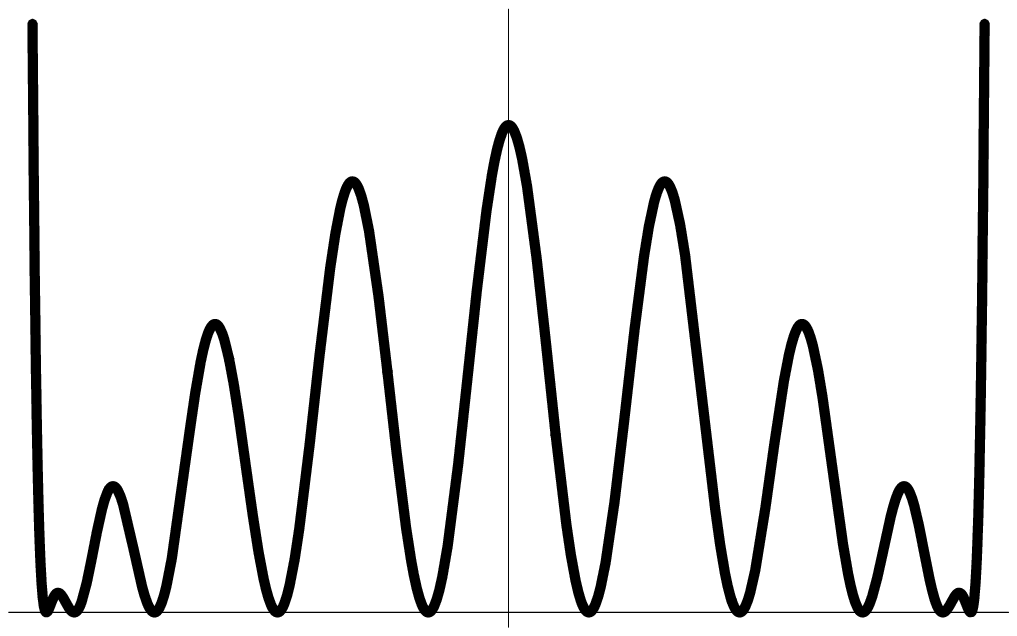,height=4cm,width=6cm} \\
\hspace{2.7cm}(A)  & \hspace{2.7cm} (B) 
\end{tabular}
\caption{{\em BLLG potentials $U_8(\varphi)$ (A) and $U_{9}(\varphi)$ (B).}} 
\label{Baz}
\end{figure}

In this section we will briefly discuss the particle content of an interesting class of potentials, 
introduced by Bazeia et al. in \cite{Bazeia}. These potentials, which can be expressed in terms 
of the Chebyshev polynomials of second kind, are closely related to the $\varphi^4$ potential 
(in its broken phase). In fact, they are obtained from this theory by using the so-called 
deformation procedure, explained in \cite{Bazeia}. In the following we denote them as 
BLLG potentials. In terms of the dimensionless coordinates and the dimensionless field $\varphi$ 
previously used ($x^{\mu} \rightarrow m x^{\mu} $, $\varphi \rightarrow \frac{\sqrt \lambda}{m} 
\varphi$), the Lagrangian of the BLLG models is given by 
\EQ
{\cal L}\,=\,\frac{m^4}{\lambda} \,\left[\frac{1}{2} (\partial_{\mu} \varphi)^2 - U_a(\varphi) \right]
\,\,\,,
\EN 
where 
\begin{eqnarray}
&& U_a(\varphi) \,=\,\frac{1}{2 a^2} (1 - \varphi^2)^2 \,V_{a-1}^2(\varphi) \,\,\, , \\
&& V_{a}(\varphi) \,=\,\frac{\sin[(a+1) \arccos\varphi]}{\sin(\arccos\varphi)} \,\,\,.\nonumber 
\end{eqnarray}
The potentials of these theories fall in two classes, depending whether $a$ is an odd or an even 
integer: those with $a$ odd are like $\varphi^4$ in its broken phase, i.e. with a maximum at the origin, 
whereas those with $a$ even are like $\varphi^6$ potential, with a zero at the origin. Some of these 
potentials are drawn in Figure \ref{Baz}. Following the analysis done in \cite{Bazeia}, let's 
briefly summarise their main properties. 

The minima of the above potentials are localised at 
\EQ
\varphi_k^{(0)} \,=\,\cos\left(\frac{k-1}{a} \pi\right)
\,\,\,\,\,\,\,
,
\,\,\,\,\,\,\,
k=1,2,\ldots,2 a +1 \,\,\,.
\label{zeroBLLG}
\EN 
By taking into account the periodicities $\varphi_k^{(0)} = - \varphi_{a + 2 -k}^{(0)}$ and 
$\varphi_{a+k}^{(0)} = \varphi_{a+2 -k}^{(0)}$, it is easy to see that the number of distinct (positive) 
values are $a+1$, all negative values obtained by reflection. For $a$ even, one has this set of zeros 
\EQ
\{\varphi_1^{(0)}=1,\varphi_2^{(0)},\ldots,\varphi_{\frac{a}{2}}^{(0)},
\varphi_{\frac{a}{2}+1}^{(0)}=0,-\varphi_{\frac{a}{2}}^{(0)},\ldots,-\varphi_2^{(0)},-\varphi_1^{(0)}=-1
\}\,\,\,,
\EN 
whereas for $a$ odd
\EQ
\{\varphi_1^{(0)} = 1,\varphi_2^{(0)},\ldots,\varphi_{\frac{a}{2}}^{(0)},
\varphi_{\frac{a+1}{2}}^{(0)},-\varphi_{\frac{a+1}{2}}^{(0)},\ldots,-\varphi_2^{(0)},-\varphi_1^{(0)}=-1
\}\,\,\,.
\EN
These minima are the vacuum states $\mid k\,\rangle$ ($k =1,2,\ldots,a+1$) of the corresponding 
quantum theory. 
 
The nice thing about these potentials is that, thanks to the deformation procedure, all kink 
solutions are explicitly known. They are expressed as 
\EQ
\varphi_k(x) \,=\,\cos\frac{\psi(x) + (k-1)\pi}{a} \,\,\,,
\EN 
where $k = 1,2,\ldots,a$ and $\psi(x) \in [0,\pi]$ is the principal determination of $\arccos\tanh(x)$. 
These topological configurations interpolate between the vacua $\varphi_{k+1}^{(0)}$ (reached at $x = 
-\infty$) 
and $\varphi_{k}^{(0)}$ (reached at $x = \infty$). The solutions $\varphi_{k+a}(x)$ are the corresponding 
anti-kink configurations. In order to compute their mass, it is useful to introduce the so-called 
super-potential $W_a(\varphi)$ and write the potential $U_a(\varphi)$ as 
\EQ
U_a(\varphi) \,=\,\frac{1}{2} \left(\frac{d W_a}{d \varphi}\right)^2 \,\,\,.
\label{superp}
\EN 
This is always possibile since $U_a(\varphi) \geq 0$. For $a \neq 2$, the explicit expression of 
$W_a(\varphi)$ is 
\begin{eqnarray}
W_a(\varphi) & \,=\,& \frac{1}{a^2 (a^2 - 4)} \left[(a^2 (1-\varphi^2) -2) \cos(a \arccos\varphi) \right. 
\label{wwwa} \\
& & - 2 a \varphi \,\sqrt{1-\varphi^2} \,\sin(a \arccos\varphi) \left. \right] \,\,\,,\nonumber 
\end{eqnarray}
whereas, for $a=2$, 
\EQ
W_2(\varphi) \,=\,\frac{1}{4} \varphi^2 \,(2 - \varphi^2) \,\,\,.
\label{www2}
\EN 
In terms of the super-potential, the mass of the $k$-th kink is given by\footnote{In this formula we have 
restored the original dimensional quantities.}  
\EQ
M_{k} \,=\,\frac{m^3}{\lambda}\,\mid W_a(\varphi_k^{(0)}) - W_a(\varphi_{k+1}^{(0)}) \mid \,\,\,.
\EN 
By using (\ref{www2}), we arrive to the final expression of the kink masses
\EQ
M_k \,=\,\frac{m^3}{\lambda} \,\frac{1}{a^2 (a^2-4)} \,
\left| a^2 \left(\sin^2\left(\frac{(k-1) \pi}{a}\right) + 
\sin^2\left(\frac{k \pi}{a}\right) \,\right) -4 \right|
\,\,\,\,\,\,
,
\,\,\,\,\,\,
a\neq 2 
\EN 
whereas $M_1 \,=\,\frac{1}{4} m^3/\lambda$ for $a=2$. 

\begin{figure}[t]
\begin{tabular}{p{8cm}p{8cm}}
\psfig{figure=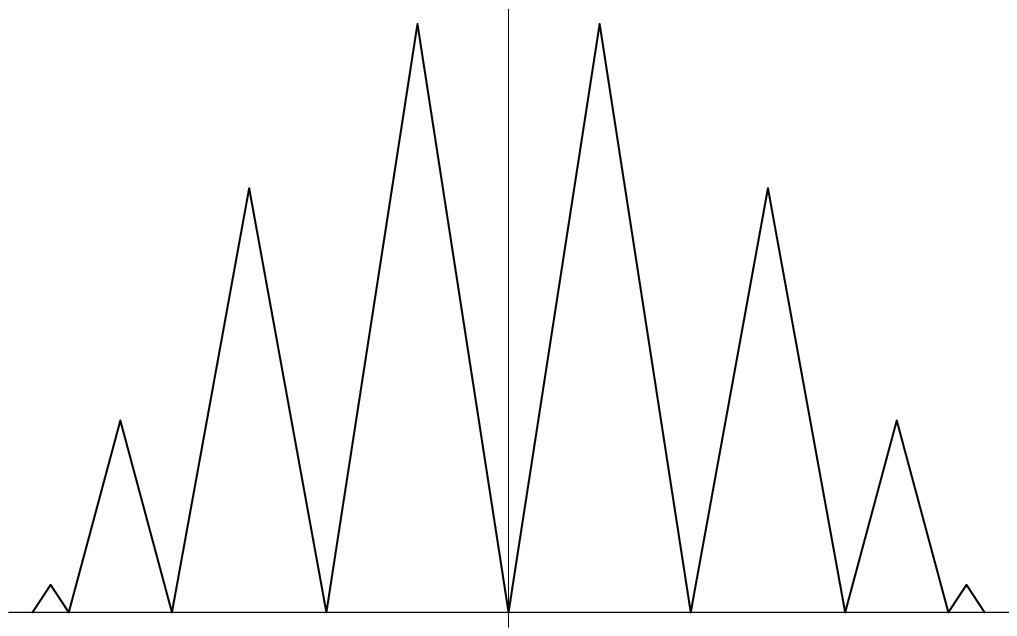,
height=4cm,width=6cm} \vspace{0.2cm}&
\psfig{figure=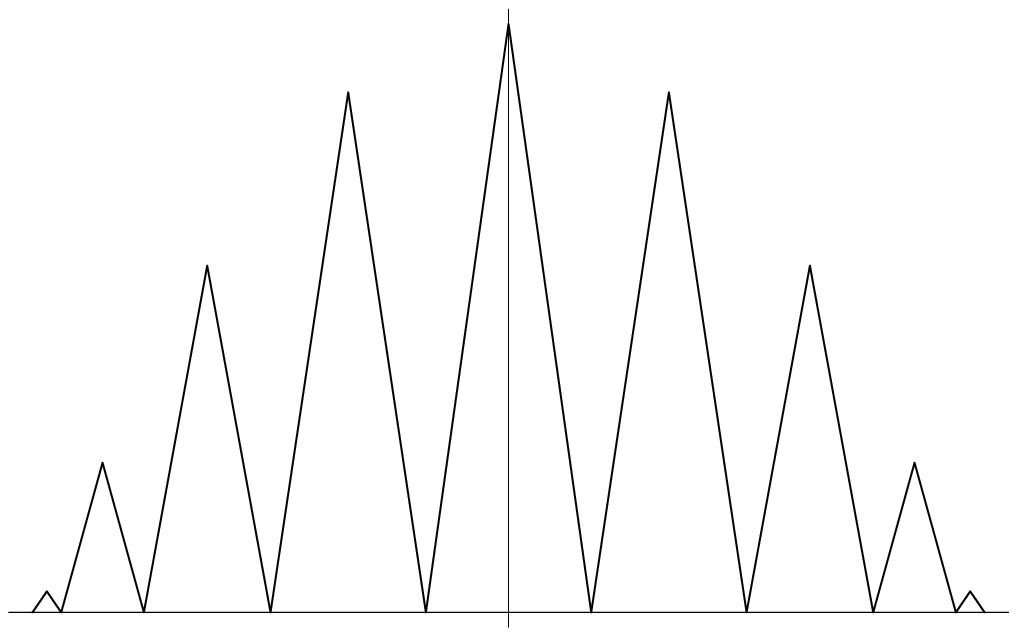,height=4cm,width=6cm} \\
\hspace{2.7cm}(A)  & \hspace{2.7cm} (B) 
\end{tabular}
\caption{{\em Masses of the kinks in the BLLG potentials $U_8(\varphi)$ (A) and $U_{9}(\varphi)$ (B).}} 
\label{massBaz}
\end{figure}

The kink with the lower mass is the one connecting the farthest vacua of the potential, either on the left or 
on the right. The kinks with the higher mass are, instead, the ones related to the vacua nearby the origin. 
Since the masses of the kinks are proportional to the heights of the potential, by making an 
istogram of the their values  -- each of them placed at the middle of the two vacua interpolated 
by the corresponding kink -- one expects to get a scheleton form of the original potential, as it is 
indeed the case, see Figure \ref{massBaz}.  

Let's discuss now the spectrum of the neutral particles at each vacuum $\mid k \,\rangle$. Notice that 
all vacua, except the most external ones, have the same curvature. More precisely, reintroducing the 
correct dimensional units and denoting the curvature by $\omega^2$, one has
\EQ
\omega_1^2 \,=\,\omega_{a+1}^2 \,=\,4 m^2 
\,\,\,\,\,\,\,
,
\,\,\,\,\,\,\,
\omega_k^2 \,=\, m^2
\,\,\,.
\EN 
The perturbative mass of the neutral particles at the vacua $\mid 1 \,\rangle$ and $\mid a+1 \,\rangle$ 
is twice larger than the one at the other vacua. Together with the lowest value of the mass of the kink that 
originates from the most external minima, it is clear than the most sensitive situation for the existence of 
the neutral particles happens at the two farthest vacua. 

Referring to the discussion of the previous section, as far as the mass of the large kink is 
much higher than the mass of the short kink, the spectrum of the neutral particles at the corresponding
vacuum is determined by the smallest one. If these masses are comparable, if one prefers, can use  
instead their average. The conditions on the relative weigths of the masses can be easily 
checked for the BLLG potential. Since, in both cases, it is always the lowest mass that matters, 
we can simplify the notation, at least, by always employing the masses of the smaller kinks. 
To have neutral particles at the vacuum $\mid k\,\rangle$, one has then to check the condition 
\EQ
\xi_k \,=\,\frac{\omega_k}{M_k^{(S)}} < \pi \,\,\,, 
\EN 
where $M_k^{(S)}$ is the mass of the small kink (or antikink) which has the vacuum 
$\mid k\,\rangle$ as asymptotic limit at $x \rightarrow -\infty$. Specializing the above formula
to the vacua $\mid 1\,\rangle$ and $\mid a+1\,\rangle$, one has 
\EQ
\xi_1 \,=\,\frac{2 \,a^2 (a^2 -4)}{4 - a^2 \sin^2\frac{\pi}{a}} \,\,\frac{\lambda}{m^2} \,\,\,. 
\EN 
Concerning the values of $\xi_k$ at the other vacua, notice that for all vacua $\mid k\,\rangle$ to the 
left of the origin, the small kink are always the antikink of $\varphi_k(x)$. For those to the right of 
the origin, the small kink is instead the kink $\varphi_k(x)$ itself. We have 
\begin{eqnarray}
\xi_k & \,=\,& \frac{a^2 (a^2 -4)}{4 - a^2 \left(\sin^2 \frac{(k-1) \pi}{a} 
+ \sin^2\frac{(k-2) \pi}{a}\right)}\,
\,\frac{\lambda}{m^2} 
\,\,\,\,\,\,
,
\,\,\,\,\,\,
k=2,\ldots, \tilde{a} \\
\xi_k & \,=\,& \frac{a^2 (a^2 -4)}{4 - a^2 \left( \sin^2\frac{k \pi}{a} + \sin^2\frac{(k-1)\pi}{a}\right)} \,
\frac{\lambda}{m^2} 
\,\,\,\,\,\,
,
\,\,\,\,\,\,
k= \tilde{a}+1, \ldots,a+1 \nonumber 
\end{eqnarray}
where $\tilde a = \frac{a}{2}+1$ for $a$ even, whereas $\tilde{a} = \frac{a+1}{2}$ for $a$ odd. 
For instance, in the case of the potential $U_9(\varphi)$, with the notation $\tilde\xi = \xi m^2/(\pi \lambda)$, 
one finds 
\EQ 
\begin{array}{ll}
\tilde\xi_1  &\,=\, 725.2.. \\
\tilde\xi_2  &\,=\, 362.6..\\
\tilde\xi_3  &\,=\, 50.98.. \\
\tilde\xi_4  &\,=\, 22.06.. \\
\tilde\xi_5  &\,=\, 14.67..\\
\tilde\xi_6  &\,=\, 14.67..\\
\tilde\xi_7  &\,=\, 22.06..\\
\tilde\xi_8  &\,=\, 50.98..\\
\tilde\xi_9  &\,=\, 362.6..\\
\tilde\xi_{10} & \,=\,725.2..
\end{array}
\EN 
Hence, there are a series of nested equations relative to the bound states on the various vacua: for instance, 
if  
\[
\frac{\lambda}{m^2} \,<\, \tilde\xi_1^{-1}\,\,\,,
\]
there is one particle on the vacua $\mid 1\,\rangle$ and $\mid 10\,\rangle$ and two particles 
on all the others. Increasing the value of $\tilde \xi$ to the interval  
\[
\tilde\xi_1^{-1} \,< \,\frac{\lambda}{m^2} \,< \,\tilde\xi_2^{-1} \,\,\,,
\]
the farthest vacua do not have neutral excitations any longer, the vacua $\mid 2\,\rangle$ and 
$\mid 9\,\rangle$ have one particle, while all the other have two bound states. Increasing 
$\tilde \xi$, there is a progressive emptying of particles on the various vacua, so that, 
when 
\[
\frac{\lambda}{m^2} \,>\, \tilde\xi_5^{-1} \,\,\,,
\]
there are no more neutral particles on all vacua.

\section{Conclusions}

In this paper we have used simple arguments of the semi-classical analysis to investigate the 
spectrum of neutral particles in a quantum field theory with kink excitations. Leaving apart the 
exact values of the quantities extracted by the semiclassical methods, it is perhaps more important 
to underline some general features which have emerged through this analysis. One of them concerns, 
for instance, the existence of a critical value of the coupling constant, beyond which there are 
no neutral bound states. Another result is about the maximum number $n \leq 2$ of neutral particles 
living on a generica vacuum of a non-integrable theory. An additional aspect is the role played by 
the asymmetric vacua and by the asymmetric kinks. 

There are several interesting open problems which deserve a further investigation. An important 
open question is to find the exact mass formula (if it exists) when the asymmetric kinks have a 
comparable value of their masses. This goes together with the problem of finding a convenient way 
of taking into account higher order terms in the form factor expression of the kinks. Another 
challenging aspect concerns the refinement of the analysis of the resonances and of the corresponding 
decay processes. Under this respect, it may be possible that useful insights will come in the future 
from a numerical analysis.

\newpage

\vspace{1cm}
\begin{flushleft}\large
\textbf{Acknowledgements}
\end{flushleft}

I would like to thank G. Delfino, K. Gawedski, J.M. Maillet, D. Serre and V. Riva for interesting discussions. 
I am particularly grateful to M. Peyrard for very useful and enjoyable discussions on solitons. 
This work was done under partial support of the ESF grant INSTANS. It also falls in the scopes 
of the European Commission TMR programme HPRN-CT-2002-00325 (EUCLID).

\end{document}